\documentclass[twocolumn,twocolappendix]{aastex7}

\usepackage{color,soul,comment}
\graphicspath{{./}}

\newcommand{\tgt}{HD\,23514\ }

\accepted{for publication in ApJL }

\shorttitle{Discovery of Volatiles in the HD 23514 Giant Impact Disk}
\shortauthors{Su et al.}

\begin{document}

\title{Discovery of Volatile Gas in the Giant Impact Disk around the 150 Myr old HD\,23514}

\author[orcid=0000-0002-3532-5580]{Kate Y. L. Su}
\affiliation{Space Science Institute, 4750 Walnut Street, Suite 205, Boulder, CO 80301, USA}
\affiliation{Steward Observatory, University of Arizona, 933 N Cherry Avenue, Tucson, AZ 85721--0065, USA}
\email[show]{ksu@spacescience.org}

\author[orcid=0009-0001-9360-2670]{Attila Mo\'or}
\affiliation{Konkoly Observatory, HUN-REN Research Centre for Astronomy and Earth Sciences, MTA Centre of Excellence, Konkoly-Thege Mikl\'os \'ut 15-17, 1121 Budapest, Hungary}
\email[]{moor.attila@csfk.org}

\author[orcid=0000-0001-8184-5547]{Chengyan Xie}
\affiliation{Lunar and Planetary Laboratory, The University of Arizona, Tucson, AZ 85721, USA}
\email[]{cyxie@arizona.edu}

\author[orcid=0000-0001-7962-1683]{Ilaria Pascucci}
\affiliation{Lunar and Planetary Laboratory, The University of Arizona, Tucson, AZ 85721, USA}
\email[]{pascucci@arizona.edu}

\author[orcid=0000-0003-2303-6519]{George H. Rieke}
\affiliation{Steward Observatory, University of Arizona, 933 N Cherry Avenue, Tucson, AZ 85721--0065, USA}
\affiliation{Lunar and Planetary Laboratory, The University of Arizona, Tucson, AZ 85721, USA}
\email[]{grieke@arizona.edu}

\author[orcid=0000-0001-7157-6275]{\'Agnes K\'osp\'al}
\affiliation{Konkoly Observatory, HUN-REN Research Centre for Astronomy and Earth Sciences, MTA Centre of Excellence, Konkoly-Thege Mikl\'os \'ut 15-17, 1121 Budapest, Hungary}
\affiliation{Institute of Physics and Astronomy, ELTE E\"otv\"os Lor\'and University, P\'azm\'any P\'eter s\'et\'any 1/A, 1117 Budapest, Hungary}
\affiliation{Max-Planck-Insitut f\"ur Astronomie, K\"onigstuhl 17, 69117 Heidelberg, Germany}
\email[]{kospal.agnes@csfk.org}

\author[orcid=0000-0001-9064-5598]{Mark C. Wyatt}
\affiliation{Institute of Astronomy, University of Cambridge, Madingley Road, Cambridge CB3 0HA, UK}
\email[]{wyatt@ast.cam.ac.uk}

\author[orcid=0000-0001-6015-646X]{P\'eter \'Abrah\'am}
\affiliation{Konkoly Observatory, HUN-REN Research Centre for Astronomy and Earth Sciences, MTA Centre of Excellence, Konkoly-Thege Mikl\'os \'ut 15-17, 1121 Budapest, Hungary}
\affiliation{Institute of Physics and Astronomy, ELTE E\"otv\"os Lor\'and University, P\'azm\'any P\'eter s\'et\'any 1/A, 1117 Budapest, Hungary}
\affiliation{Institute for Astronomy, University of Vienna,T\"urkenschanzstrasse 17, A-1180 Vienna, Austria}
\email[]{abraham.peter@csfk.org}

\author[orcid=0000-0003-4705-3188]{Luca Matr\`a}
\affiliation{School of Physics, Trinity College Dublin, the University of Dublin, College Green, Dublin 2, Ireland}
\email[]{LMATRA@tcd.ie}

\author[orcid=0009-0003-3696-9673]{Zoe Roumeliotis}
\affiliation{School of Physics, Trinity College Dublin, the University of Dublin, College Green, Dublin 2, Ireland}
\email[]{roumeliz@tcd.ie}

\author[orcid=0000-0003-1526-7587]{D. J. Wilner}
\affiliation{Center for Astrophysics, Harvard \& Smithsonian, 60 Garden Street, Cambridge, MA 02138, USA}
\email[]{dwilner@cfa.harvard.edu}

\begin{abstract}

We report the discovery of CO$_2$ gas emission around HD 23514, an F5V star in the $\sim$150 Myr old Pleiades cluster, hosting one of the rare giant-impact disks with unique mineralogy dominated by silica dust. We show that the dust feature remains stable over several decades, and that the submicron grains, which give rise to the $\sim$9 \micron\ feature, are cospatial with the hot CO$_2$ molecules within the sub-au vicinity of the star. Examining the Spitzer spectrum taken 15 yr earlier, we show that the CO$_2$ emission was also present at 4.3$\sigma$ significance. The existence of tiny silica grains and volatile gas requires special conditions to prevent the rapid loss caused by stellar radiation pressure and photodissociation. We explore several pathways explaining the observed properties and suggest that a past giant impact and/or stripping atmospheric event, involving large bodies with volatile content similar to the carbonaceous chondritic material, can simultaneously explain both the silica and volatile emission. Our discovery provides an important context for the amount of volatiles that a newly formed planet or the largest planetesimals could retain during the giant impact phase in the early solar system evolution.

\end{abstract}


\section{Introduction}
\label{sec:intro}

Young, extremely dusty disks, termed extreme debris disks (EDDs), give new insights into the collisional processes among planetary embryos during the era of terrestrial planet formation. These systems exhibit unique dust mineralogy dominated by thermodynamically altered minerals, likely produced by hypervelocity impacts, analogous to the giant-impact hypothesis for the formation of the Moon. The large amount of warm and highly processed dust definitively tie them to the rapid evolution of dust debris in their terrestrial zones, an environment traditionally thought to be dry and devoid of volatiles. Mid-infrared (mid-IR) monitoring shows that the majority of EDDs exhibit significant variability \citep{meng14,moor21,rieke21_v488per,su23_rzpsc}. In the best studied case of ID8 in NGC 2547, both yearly and monthly semiperiodic variability were observed, which were attributed to the collisional and dynamical evolution of a cloud of escaping boulders and vapor condensates formed in giant-impact events \citep{meng12,su19}. Deep optical transits have also been found in some highly inclined systems where eclipsing dust clumps are found to be as large as the stars \citep{dewit13_rzpsc,gaidos19_hd240779,melis21,su22_hd166}, further validating the scale of such collisions.

\tgt (Cl* Melotte 22 HII 1132), an F5V star at a distance of 138.4$\pm$0.4 pc \citep{gaia_edr3} in the Pleiades cluster, was first discovered to possess an exceptional amount of warm dust and unique dust mineralogy by \citet{rhee08}. The age of the Pleiades, estimated to be $\sim$110--150 Myr \citep{lodieu19_Pleiades_Praesepe_alphaPer}, makes the system particularly interesting because it corresponds to the final building stage of terrestrial planet formation. After that stage, the rate of colossal Moon-forming events is expected to be significantly reduced \citep{quintana16}, consistent with that only $\sim$1\% of the stars younger than a few hundred million years show detectable 12 \micron\ IR excesses above 15\% levels \citep{kennedy_wyatt13}. To put \tgt in perspective, for 100 stars younger than the terrestrial planet formation stage, only one system is like HD 23514. Prior to the end of Spitzer cryogenic mission, \citet{meng12} showed that it is also one of the few systems exhibiting IR variability on timescales of a few years. The system shows no sign of cold (T$\lesssim$40 K) dust and gas, suggesting the disk is compact, likely confined within $\lesssim$10 au \citep{vican16,sullivan22_alma_pleiades}. \tgt also has a wide $\sim$M7 companion at a projected separation of $\sim$360 au and a position angle of $\sim$228\arcdeg\ \citep{rodriguez12_hd23514}, one of the characteristics that some EDDs share \citep{zuckerman15,moor21,moor24_edds_visir}.

The \tgt system is one of the only three EDDs where the Spitzer/IRS spectra show prominent silica dust \citep{rhee08,lisse09,fujiwara12}, similar to those of the impact glasses and melted droplets commonly found in terrestrial impact craters \citep{morlok16_reflectance_impactglasses}, therefore, they are also called giant-impact disks. In the giant-impact hypothesis \citep{canup04_simulations}, the almost fully grown Earth collided with a Mars-size planet called Theia at the end of its formation, and the material from the outer layers of both objects coalesced into the Moon. We can take this well-studied event as a prototype for such impacts. Because the Earth and Moon are almost chemically indistinguishable, Theia was thought to form at a similar distance from the Sun, i.e., dry material with significantly depleted volatile elements -- those that vaporize easily upon heating. Some very young debris disks do contain gas, e.g., $\beta$ Pic \citep{roberge00_betapic_gas,dent14_betapic}, but only typically at very young stages ($\le$ 50 Myr, \citealt{moor17,bonsor23_debrisgas}). Using JWST, we report the truly remarkable detection of hot molecular gas emission in HD\,23514, particularly the CO$_2$ lines have never been detected in debris disks before.

Our observations and basic data reduction are described in Section \ref{sec:data}. In Section \ref{sec:analysis}, we describe the findings in terms of (1) lack of large-scale overall excess over four decades and the dust properties showing stable silica emission compared to the Spitzer data taken 15 yr earlier, and (2) the properties of volatiles (robust detection of CO$_2$ and CO, but tentative detections of H$_2$O and NH$_3$) and their likely existence in the Spitzer IRS data. In Section \ref{sec:discussion}, we show that (1) the silica dust and gas volatiles are likely colocated within the sub-au region where their lifetimes are less than a year due to radiation-pressure blowout and photodissociation, and (2) speculate on the origins of the stable silica dust and that of gas, in which both can be simultaneously explained by invoking a past giant-impact event. The conclusion and future prospect are given in Section \ref{sec:conclusion}.

\begin{figure*}
    \centering
    \includegraphics[width=0.47\linewidth]{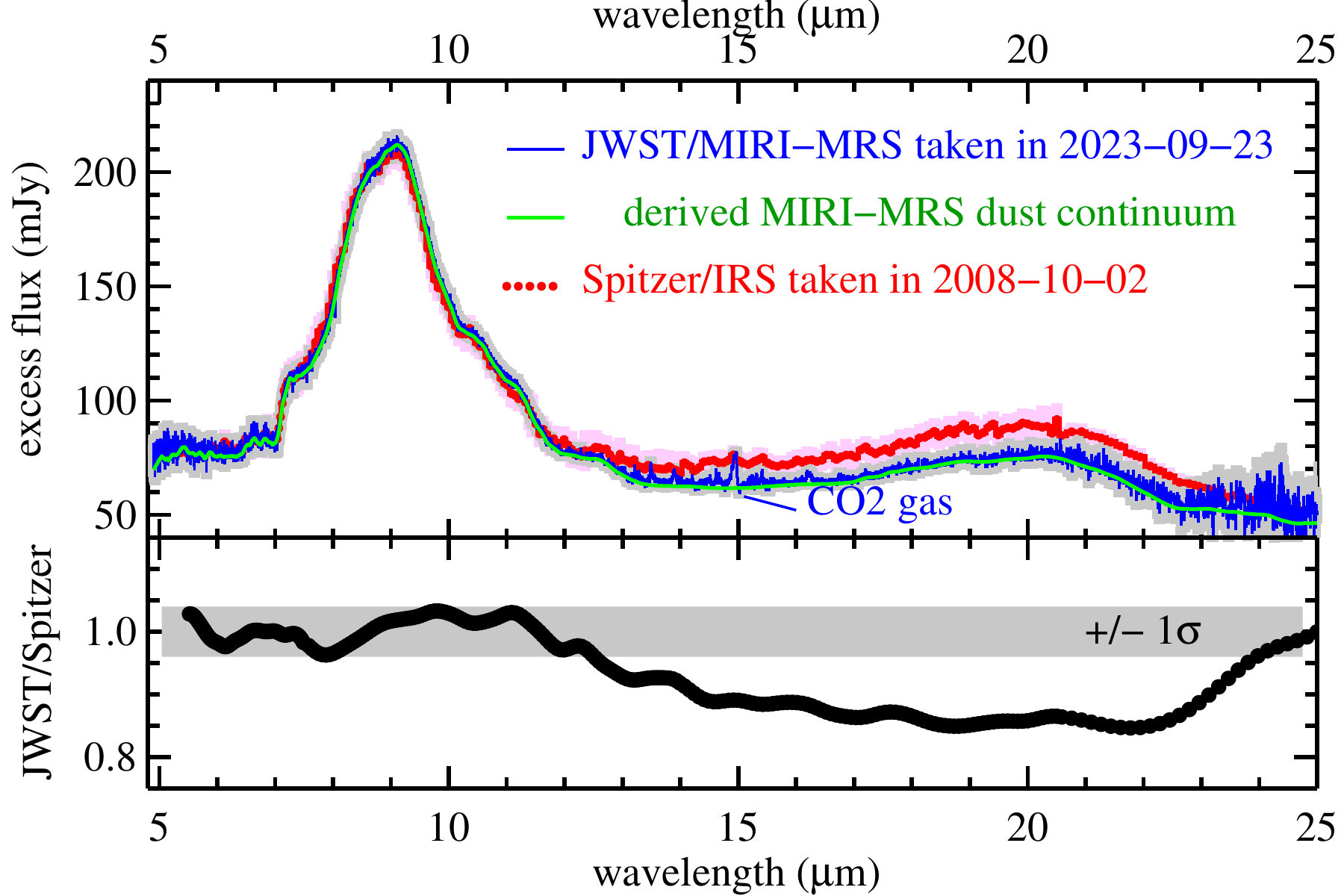}
    \includegraphics[width=0.5\linewidth]{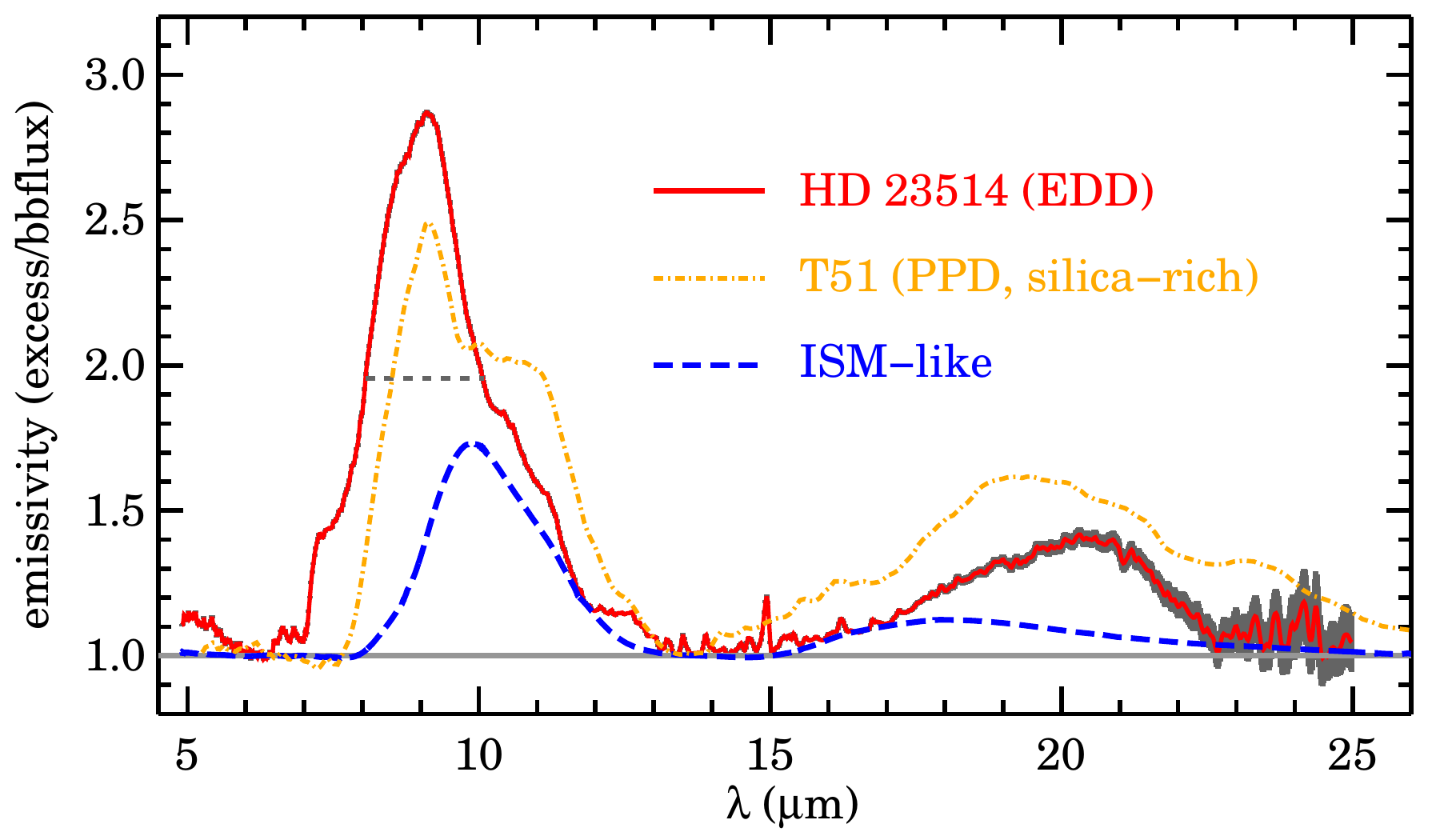}
    \caption{The left panel shows the comparison between the JWST MIRI/MRS and Spitzer/IRS disk spectrum around \tgt taken 15 yr apart with the bottom one showing the flux ratio between the two. The color shaded area indicates the associated $\pm$1$\sigma$ uncertainty. The two spectra agree within $\pm$1$\sigma$ in the silica 9 \micron\ region while in the 20 \micron\ region JWST data appear to be slightly lower by $\sim$3$\sigma$. The thin green line is the estimated dust continuum of the JWST data (for details see Section \ref{sec:gas}). The right panel depicts the derived dust emissivity spectrum in comparison with a silica-rich protoplanetary disk (orange, dot-dashed line), and a typical ISM-like, very-low-crystallinity dust mineralogy (blue dashed line; see Appendix \ref{sec:starsubemi} for details). } 
    \label{fig:mrs_result}
\end{figure*}

\section{Observations and Data Reductions} 
\label{sec:data}

\tgt was observed with MIRI/Medium-Resolution Spectrometer (MRS) on 2023 September 23, as part of the JWST GTO program PID1206 (PI: G. Rieke) to study the dust mineralogy and variability for a handful of EDDs discovered by Spitzer. All MRS channels and subbands were included, covering a wavelength range from 4.9 to 27.9 \micron. Each subband was integrated with 12 groups in a single exposure using the FASTR1 mode with a total integration time of 132 s. A nominal observation setup was used, which includes the target acquisition with a neutral density filter, and a four-point dither pattern optimized for point sources. 

We used the JWST Calibration Pipeline version 1.15 \citep{bushouse24_jwstpipeline} and Calibration Reference Data System context \texttt{pmap\_1298} to reduce the data with default parameter settings and with additional steps for bad pixel self-calibration in the \texttt{calwebb\_spec2} and residual fringe correction in the \texttt{calwebb\_spec3} steps\footnote{see MIRI/MRS Pipeline Notebook at \href{}{https://github.com/spacetelescope/jwst-pipeline-notebooks/blob/main/notebooks/MIRI/MRS/JWPipeNB-MIRI-MRS.ipynb}}. The 12 spectral cubes were inspected visually to identify potential other sources within the integral field unit (IFU) field of view. Only a very faint (by a few hundred times) object, $\sim$2\farcs3 away from \tgt at a position angle of $\sim$148\arcdeg, was found in the medium combined IFU images for the data shortward of 18 \micron. This faint source is unlikely to be the late M-type companion reported by \citet{rodriguez12_hd23514} because of very different position angle and projected distance. Given its faintness, it has no impact in the extracted 1D spectrum. We also examined the target spatial extension by comparing with a calibration star at all observed wavelengths, and concluded that the \tgt system is point-like and not resolved by JWST.

A 1D spectrum was extracted for each of the cubes using the default aperture and sky annulus \citep{law25_mrscalibration} that increased linearly with wavelength. Finally, to improve the continuity of the full spectrum, a small flux shift ($\lesssim$a few percent) was applied to individual spectra using the overlapping wavelength region and lining up with the shortest subband. Because the star contributes significant flux in the short wavelength part of the MIRI/MRS spectrum, we derive the disk emission by subtracting a model photosphere (for details see Appendix \ref{sec:starsubemi}), yielding the spectrum shown in Figure \ref{fig:mrs_result} (left).

\section{Results and Analysis}
\label{sec:analysis}

\subsection{Lack of large-scale variation}
\label{sec:stabledustfeature}

As shown in Figure \ref{fig:mrs_result}, the disk emission (in both flux level and solid-state feature) is remarkably similar to the one taken by Spitzer/IRS 15 yr earlier. The prominent 9 \micron\ silica feature agrees very well within $\pm$5\% (1$\sigma$ uncertainty including the absolute flux calibration from both observatories), while in the broad 20 \micron\ feature region the JWST data are consistently lower by $\sim$15\% compared to the Spitzer one at $\sim$3$\sigma$ significance. This behavior suggests that there is no significant change in terms of dust composition, but the dust temperature is slightly lower in the JWST epoch (by $\sim$20 K, see Appendix \ref{sec:starsubemi}). The \tgt disk is known to display both weekly and yearly stochastic IR variability, at $\sim$10--30\% level in both the 3--5 \micron\ and 24 \micron\ photometry (Appendix \ref{sec:ir_monitoring}). Because there are no dust features in those bands, the variability is likely due to the change in the dust temperature rather than the composition. The overall level of the 10 \micron\ feature has not changed appreciably over a 40 yr baseline from the earliest IRAS 12 \micron\ measurement taken in 1983, AKARI 9 \micron\ taken in 2006, WISE 12 \micron\ measurement taken in 2010 to the latest JWST data taken in 2023 (details see Figure \ref{fig:mid-ir}). Lack of large-scale variation in the 10 \micron\ feature over decades has also been reported for other EDDs like HD\,113766 and HD\,172555 \citep{su20,lisse09,samland25_hd172555}.

\subsection{Dust Properties}
\label{sec:silica}

To characterize solid-state features, we derived the dust emissivity (observed spectrum divided by the featureless dust continuum) and estimated the crystalline mass fraction indices following the method outlined by \citet{watson09} for protoplanetary disks (PPDs). Because debris disk structures generally are not as complex as the PPDs, we use a combination of two blackbody functions to estimate the featureless dust continuum (Appendix \ref{sec:dustemissivity}). The derived emissivity (smoothed to a resolution of $R=500$ for clarity) is shown in Figure \ref{fig:mrs_result} (right). 

\begin{figure*}[t!]
    \includegraphics[width=\linewidth]{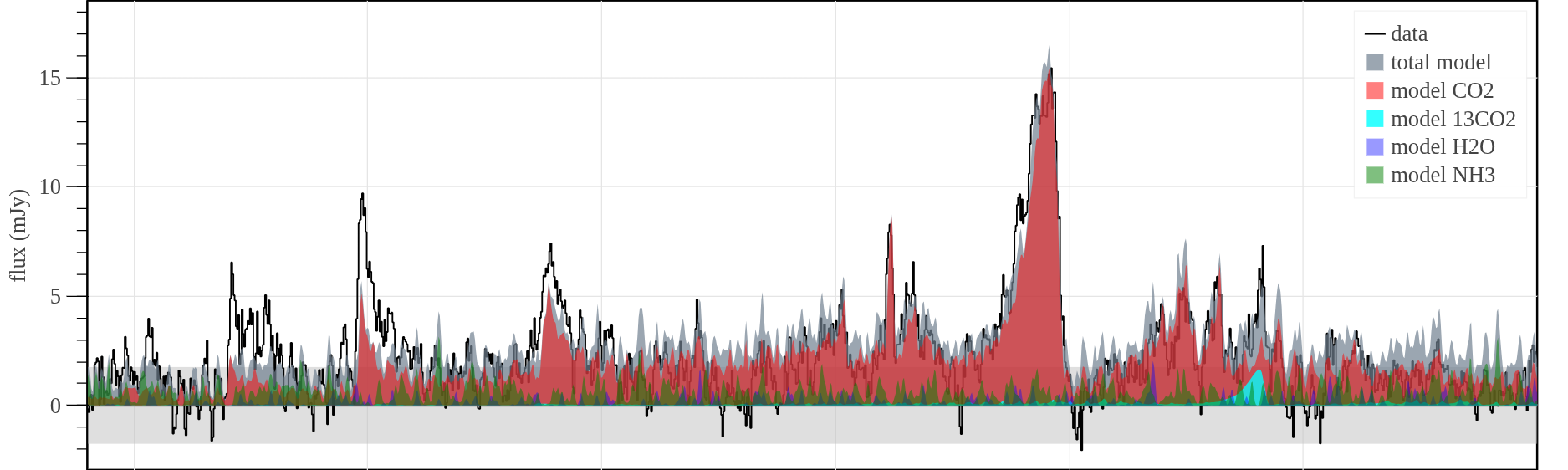}
    \includegraphics[width=\linewidth]{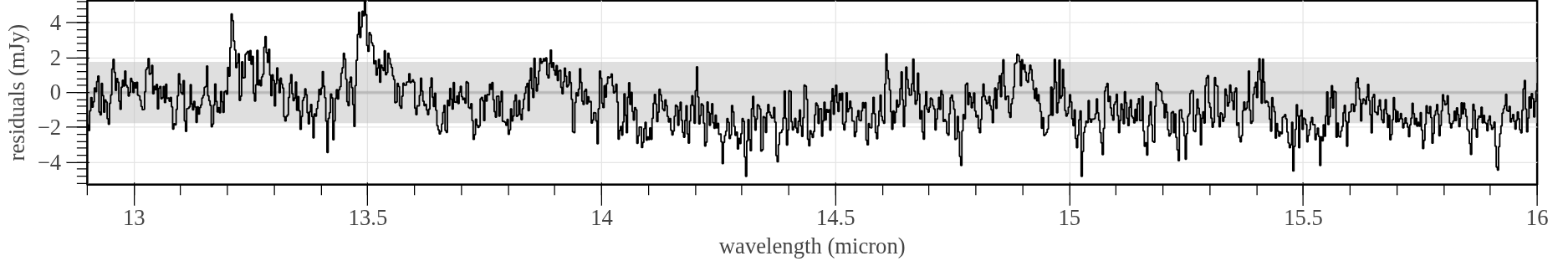}
    \caption{Zoom-in, continuum-subtracted spectrum centered at $\sim$15 \micron\ showing the dominant CO$_2$ Q-, P-, and R-branch molecular lines in the upper panel and the residuals (data $-$ model) in the bottom panel. The black line depicts the data, with the gray horizontal area showing the $\pm$1$\sigma$ estimated in the local spectral region. Filled, color areas show the model emission from four volatile species: CO$_2$, H$_2$O, NH$_3$ and $^{13}$CO$_2$ with the dark gray area showing the model sum. }
    \label{fig:co2model}
\end{figure*}

Mineral features are key in dust composition identifications due to stretching and bending motions in the silicon-oxygen anions. The narrow and strong resonance features from the thermally processed, crystalline silicates are easily distinguishable from the broad features of amorphous grains lacking ordered structure \citep{henning10_astromineralogy}. The mass fraction of crystalline grains present in an IR spectrum is then a good indication of thermal processing in the disk environment \citep{sargent06,juhasz10_herbig}. 
\citet{watson09} defined three crystalline mass fraction indices for the strongest features of pyroxenes ($P_{10}$), olivines ($O_{10}$) and silica ($S_{10}$) in the 10 \micron\ complex, centered at wavelengths 9.21, 11.08, and 12.46 \micron, respectively, by referencing to a pristine (ISM-like, amorphous) silicate feature (for details see \citealt{watson09}). An index value is roughly unity when the emission feature resembles the pristine profile, while an index value above unity reflects increasing prominence of a crystalline signature above the pristine one. The three indices are simple proxies for the amount of crystalline (thermally processed) grains without detailed dust decomposition modeling. Because the $\sim$9 \micron\ silica peak can overlap with the 9.21 \micron\ pyroxene peak under low spectral resolution, the silica index is defined by the narrow feature at 12.5 \micron. In other words, the prominent 9 \micron\ silica feature seen in \tgt is not entirely captured by the silica index ($S_{10}$). Although these indices do not perfectly track the exact level of crystallinity on individual bases, they are good indicators when comparing properties among an ensemble of objects \citep{watson09}.

Using the derived dust emissivity, the 10 \micron\ emission complex in \tgt can be described as having a peak wavelength at 9.1 \micron\ with a full-width-half-maximum (FWHM) of 2.04 \micron, an equivalent width ($W_{10}$) of 4.7$\pm$0.2 \micron, $P_{10}$=2.54, $O_{10}$=0.91, and $S_{10}$=1.82. These crystalline indices suggest that the crystalline dust in \tgt is dominated by both pyroxene and silica with little of olivine type. The crystalline mass fraction is high ($\sim$40\% using the pyroxene index with the empirical trend derived by \citealt{watson09}), which is not common among PPDs. Silica dust has been found in a handful of PPDs, and one example (around a few megayear PPD, T51 (Sz 41)) is also shown in Figure \ref{fig:mrs_result} (right) for comparison. The 10 \micron\ dust emissivity comparison between \tgt and the young star (Appendix \ref{sec:dustemissivity}) suggests that the dust in \tgt has a higher crystallinity and the dominant dust size is smaller than that of the PPD ($\sim$\micron) as evidence by the feature's FWHM.  Furthermore, the $W_{10}$ (an indication of the amount of small grains) of 4.7 is also near the top among the 84 PPDs observed by Spitzer (only three systems have $W_{10}$ over 5 and all have near unity crystalline indices, i.e., low-crystallinity fraction). The highly thermally processed dust mineralogy implies different formation processes. As detailed in Section \ref{sec:stabledustfeature} and Appendix \ref{sec:dustemissivity}, past giant-impact events are the most likely mechanism responsible for the large amount of highly processed crystalline silicates and silica grains in HD\,23514.

\subsection{Surprising Molecular Gas Emission Lines}
\label{sec:gas}

In addition to the prominent silica feature, the high-quality JWST spectrum also reveals surprising gas emission lines. Visual inspection indicates that the lines are dominated by CO$_2$ emission centered at 14.98 \micron\ (Q branch) accompanied by the P and R branches on the short- and long-wavelength sides. To properly characterize the molecular lines and discover other weaker ones, the dust continuum emission needs to be removed from the spectrum. The continuum was determined in an iterative way as detailed in Appendix \ref{sec:gas_model} and shown in the left panel of Figure \ref{fig:mrs_result}. A subset of the continuum-subtracted spectrum centered at $\sim$15 \micron\ is shown in Figure \ref{fig:co2model} while the rest are shown in Figures \ref{fig:comodel} and \ref{fig:hd23514_full_contsub}. 

We used the local standard deviation per channel and subband as the typical rms of the continuum-subtracted MRS spectrum after masking out strong detected lines. Although the overall signal-to-noise ratio (S/N) of the data before continuum subtraction is $>$100 for $\lambda <$20 \micron, unresolved emission lines with fluxes $\lesssim$5 mJy cannot be robustly detected in the current data. Despite the hints of H$_2$O, and NH$_3$ present in the spectrum, only CO$_2$ Q-branch is detected at 10$\sigma$ along with the R and P branches at 3--5$\sigma$, and CO at 2--4$\sigma$. There is no sign of molecular/atomic hydrogen and hydrocarbon lines that are commonly found in PPDs, nor ionized/neutral atomic lines such as [Ne II], [Ar II], [Fe II], [Ni II], [Cl I], and [S I] present in HD 23514. Nondetection of molecular/atomic hydrogen and [Ne II] lines is the best indication that the gas in \tgt is not primordial but secondary \citep{pascucci07}, which is also consistent with the estimated upper limit for the CO-to-H$_2$ mass ratio (Appendix \ref{h2_limits}).

To extract basic properties of the emitting gas, a plane-parallel slab model in local thermodynamic equilibrium (LTE) was adopted using three parameters: the gas emitting area ($A$), column density ($N$), and the gas temperature ($T_{\rm gas}$). The emitting area is often expressed as an equivalent emitting radius ($R_{\rm em}$) where $A = \pi R_{\rm em}^2$. We focused on fitting the CO$_2$ complex in the 13.5--16.3 \micron\ region where the detected lines have the highest S/N. Assuming a turbulence velocity of zero, we find that a slab model with $A=$0.0085 au$^2$ (or $R_{\rm em}=$0.052 au), $N=$1.38$\times10^{18}$cm$^{-2}$, and $T_{\rm gas}=$891 K can reproduce the CO$_2$ emission spectrum well. The associated uncertainties could be large (for details see Appendix \ref{sec:gas_model}), mostly due to the LTE assumption where the inferred temperature is higher, which could lead to an order magnitude lower column density than that in non-LTE \citep{bosman17}.

Due to low S/N in most of the continuum-subtracted spectral region, we simply adopted similar parameters for H$_2$O, NH$_3$ and adjusted them to achieve reasonable fits visually. For CO, the same approach produces much narrower line widths in the limited wavelength range covered by MIRI/MRS, while an extra broadening along with a hotter gas temperature model can also reproduce the data (for details see Appendix \ref{sec:detailed_gas}). Overall, the simple slab models can roughly match the peak flux of the CO lines in both cases, but the hotter one matches the line profiles better. We consider the detections of volatile CO$_2$ and CO is robust, and the H$_2$O and NH$_3$ detections are tentative. Using the LTE slab model, the properties of CO$_2$ are well constrained, but less so for CO and not at all for H$_2$O and NH$_3$. Non-LTE effects are particularly important for the low-density region that is very close to the star. Exploring non-LTE models is beyond the scope of the paper, which needs high S/N data (particularly for many of the weak water lines) and shorter-wavelength coverage between 4.4 and 4.9 \micron\ where the CO lines are brighter. Using LTE models we show that the JWST MIRI spectrum indubitably reveals the presence of hot ($\sim$900 K) gas in the sub-au ($\sim$0.03--0.05 au) region around HD\,23514.

\begin{figure}
    \centering
    \includegraphics[width=\linewidth]{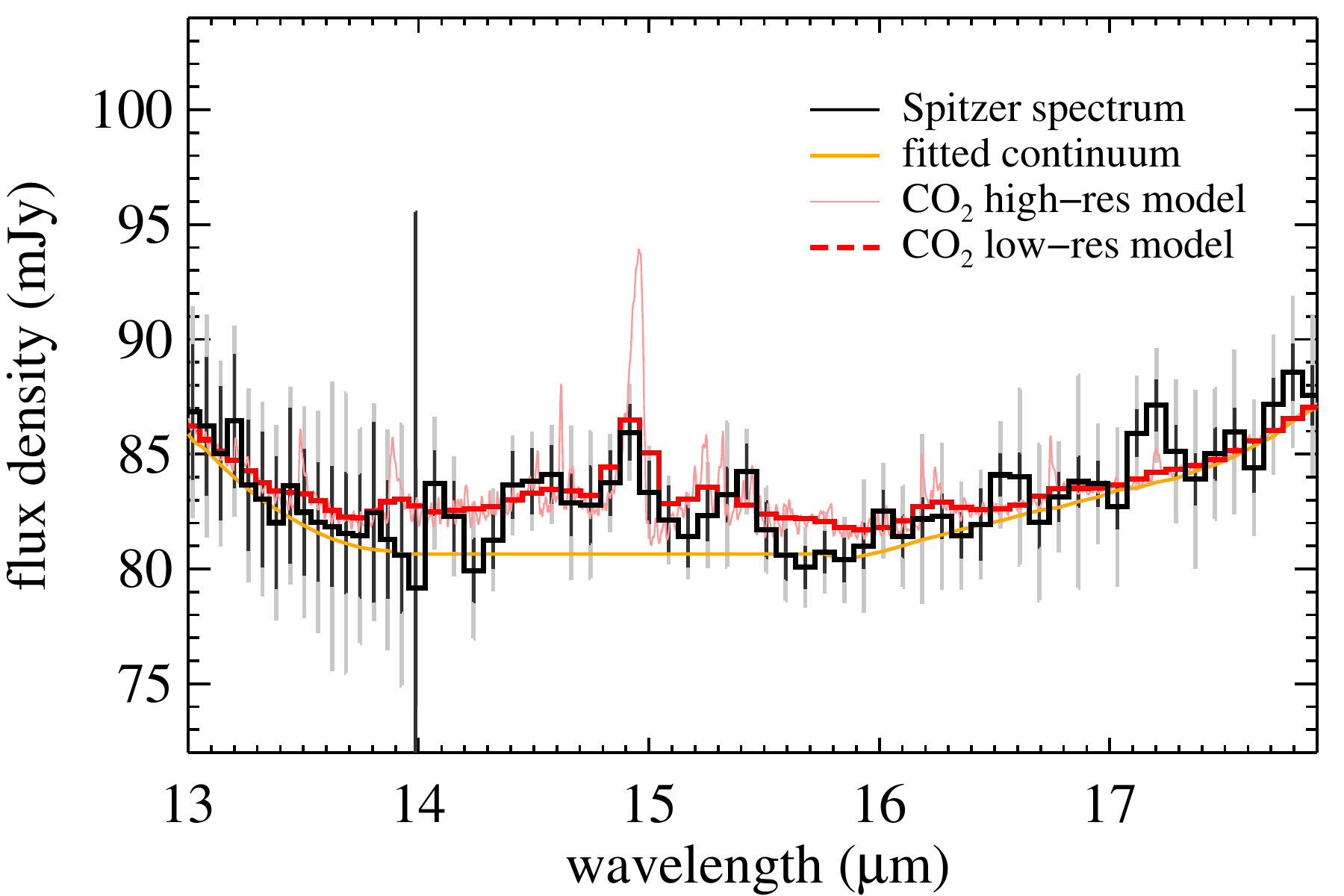}
    \caption{Spitzer IRS low-resolution spectrum of \tgt observed in 2008 centered at the CO$_2$ 15 \micron\ complex, where the data are shown as the black line with 1$\sigma$ rms error showing as dark vertical lines (gray for the combined error). The thick red line is the expected, low-resolution model spectrum if the same level of the CO$_2$ emission detected by JWST (thin red line) was also present. This comparison suggests that the same CO$_2$ emission is presented in the Spitzer spectrum where the brightest Q-branch line is detected at 4.3$\sigma$.}
    \label{fig:co2_irs}
\end{figure}

\subsection{Is CO$_2$ present in the Spitzer/IRS spectrum?}
\label{sec:co2_irs}

Inspired by the discovery of CO$_2$ gas, we went back to examine whether a similar gas emission was also present in the Spitzer IRS spectrum taken 15 yr earlier. We retrieved the calibrated spectrum from the Combined Atlas of Sources with Spitzer IRS Spectra (CASSIS)\footnote{https://cassis.sirtf.com/atlas/} \citep{cassis_ref} site. We used the optimal extraction product best for point sources and adopted the uncertainty including both statistical (rms) and systematic errors. The IRS data shown in Figure \ref{fig:co2_irs} do show a bump near 15 \micron, which was not found in other similar debris systems (for details see Appendix \ref{sec:irs_exam}), suggesting that the bump is likely astrophysical in nature.
We then determined the general continuum by fitting the CO$_2$ complex region using a polynomial function and forcing the flux between 13.9 and 15.8 \micron\ to be relatively flat for a conservative approach as shown in Figure \ref{fig:co2_irs}. We then constructed a low-resolution model spectrum by (1) adding the JWST CO$_2$ model (from Section \ref{sec:gas}) to the continuum, and (2) resampling the wavelength grid to match the Spitzer IRS spectral resolution\footnote{see the IRS Instrument Handbook for details.}. 
As shown in Figure \ref{fig:co2_irs}, the low-resolution model spectrum resembles the observed one well if the same level of the CO$_2$ emission detected by JWST was also present. In this case, the strong CO$_2$ Q-branch was detected at 4.3$\sigma$ statistically (2.5$\sigma$ if using the combined error)  while the others were buried in the noise. In summary, the same level of the hot CO$_2$ emission is also present in the Spitzer data taken 15 yr earlier and remains unchanged between the two epochs.

\section{Discussion} 
\label{sec:discussion}

\subsection{Colocation of silica and hot gas}
\label{sec:sed_model}

The \tgt system has been observed with the near-IR adaptive optics imaging at the 10 m Keck II telescope \citep{rodriguez12_hd23514} and with ALMA at 1.3 mm using a beam size of 1\farcs5 \citep{sullivan22_alma_pleiades}, and neither resolves nor detects the disk. Despite the high degeneracy, the disk extent can be estimated using spectral energy distribution (SED) models because the bulk of disk emission is sensitive to its temperature (i.e., location for a given dust composition). \citet{vican16} characterized the \tgt system as a two-temperature disk with a hot component having a dust temperature ($T_{\rm d}$) of $\sim$1082 K and a warm component with $T_{\rm d}$ of $\sim$168 K, roughly consistent with our estimated temperatures for the featureless dust continuum (750 + 200 K; Appendix \ref{sec:starsubemi}). Given the stellar properties, these temperatures correspond to radial locations of $\sim$0.1 and $\sim$4.8 au, respectively, for blackbody emitters. For imperfect emitters like dust grains, the probable locations are likely to be larger (Figure \ref{fig:radialdistribution_dustTd}). Under the optically thin condition, the contrast/ratio between the prominent 10 and 20 \micron\ dust features is a good proxy for dust temperature, where the larger the ratio the hotter the dust due to Planck emission (Figure \ref{fig:radialdistribution_dustTd}). In other words, the bulk of the dust emission that gives rise to the silica feature most likely comes from the hot rather than the warm component. 

We explored SED models, aiming to constrain the location of the silica dust (Figure \ref{fig:sedmodel}, and for details see Appendix \ref{sec:sedmodels}). Although such models might be degenerate, they provide diagnostic constraints under simplified assumptions. We found that the minimum grain size in the disk needs to be 0.1--0.5 \micron\ in order to reproduce the 10/20 \micron\ feature strengths, which is smaller than the typical blowout size, $\sim$1 \micron\ for \tgt (Appendix \ref{sec:betavalues}). Our SED models suggest that the disk is compact ($\sim$0.1--3 au), and the highly refractory silica grains are mostly confined within the sub-au region, where the temperature is close to their sublimation temperature, in order to reproduce the prominent 9 \micron\ feature and 3--5 \micron\ flux. The system's IR fractional luminosity is 1.76$\times10^{-2}$ with $\sim$70\% coming from the featureless dust continuum. The total dust mass is $\sim$$10^{24}$ g (1.67$\times 10^{-4} M_{\oplus}$) and the feature-producing grains account for $\sim$5\% of the dust mass with only $\sim$1\% coming from tiny silica dust. Based on the model parameters for the gas (Section \ref{sec:gas_model}) and the silica dust, they are likely colocated very close to the star at sub-au distance.

\begin{figure}
    \centering
    \includegraphics[width=\linewidth]{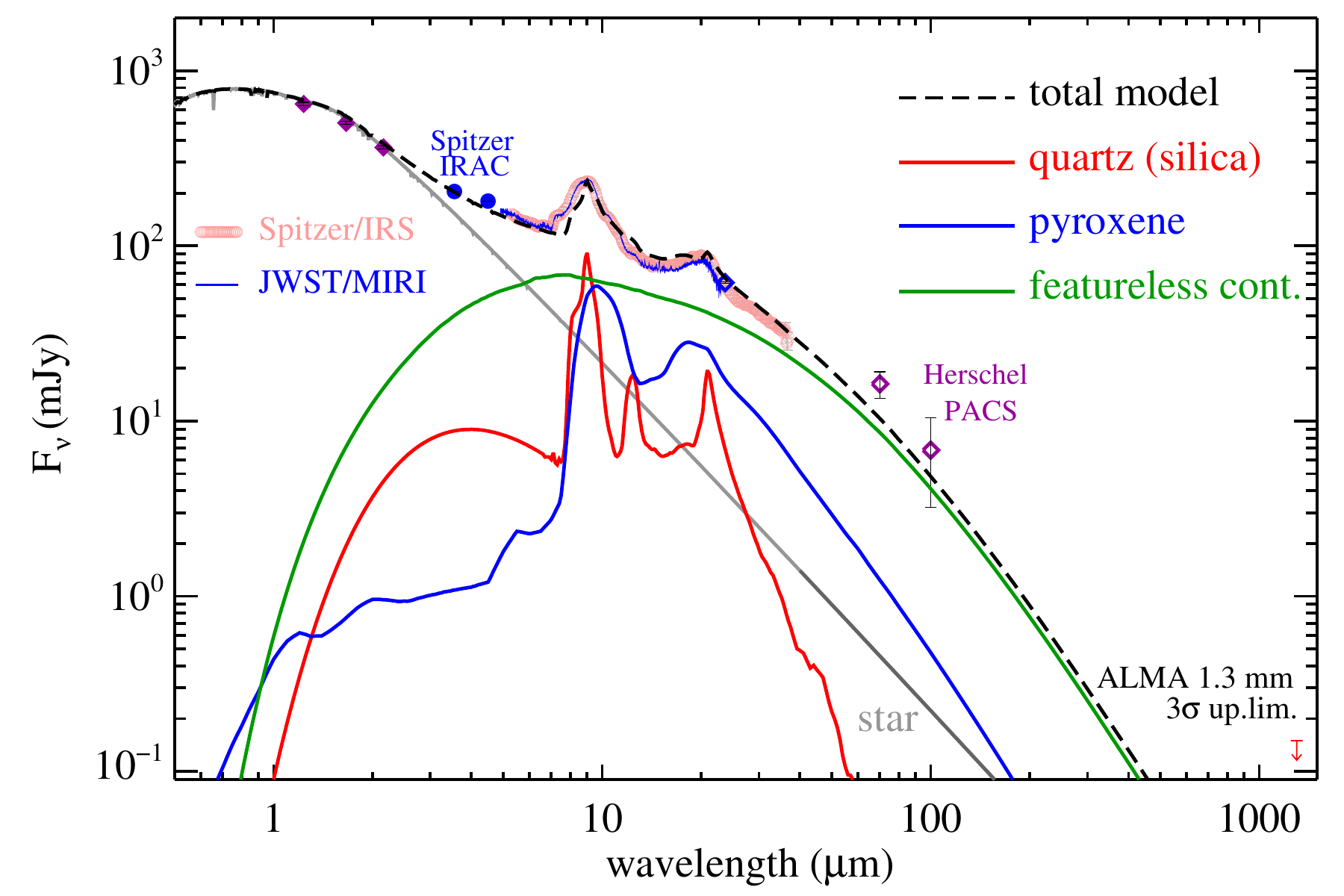}
    \caption{The SED of the \tgt system where the data points come from 2MASS, Spitzer and Herschel photometry along with the mid-IR spectra. Lines show the SED models for a disk in a radial range of 0.1--3 au and a dust composition mixture of silica and pyroxene, and grains lacking insipid features. The prominent 10 and 20 \micron\ features mostly come from optically thin silica and pyroxene grains within sub-au (for details see Section \ref{sec:sed_model} and Appendix \ref{sec:sedmodel_details}). }
    \label{fig:sedmodel}
\end{figure}

\subsection{Origins of the Stable Silica Dust}
\label{stablesilica}

The origin of the large amount of submicron grains in debris disks has been long debated in the literature. The typical radiation-pressure blowout size is $\sim$1 \micron\ around \tgt except for highly porous grains \citep{arnold19}, i.e., the grains that give rise to the prominent mid-IR feature have very short lifetimes (less than a year, Appendix \ref{sec:tau_c}). This implies a very high mass-loss rate (5$\times$10$^{22}$ g\,yr$^{-1}$) if there were no optical depth effects either to prevent them leaving the system by shielding or to replenish them fast enough (e.g., \citealt{thebault19}). If this high rate was sustained through collisional cascades over the age of the system (150 Myr), \tgt would need to be born with a disk that is 35 times of the minimum mass of solar nebula (MMSN), which is not impossible but unlikely. Instead, the large amount of small grains is more likely to be created by transient events where the high rate is only applicable for a limited recent time. The maximum dust production rates in solar system comets can reach $\sim$10$^{13}$ g\,yr$^{-1}$ when they are very close to the Sun. It is then unlikely to sustain the high dust production/loss rate by inward-scattered comets if these grains are continuously being ejected by radiation pressure. 

One easy way to address this dilemma is that all or some portion of the small grains are not subject to radiation-pressure blowout, which can be achieved by certain grain properties or being trapped by other means \citep{rieke16,lebreton13,pearce20_gastrapping}. The prominent $\sim$9 \micron\ feature is thought to come from obsidian or annealed/fused quartz grains \citep{fujiwara12} whose properties make them transparent to stellar photons, therefore not subject to blowout, when they are tiny enough \citep{artymowicz88}. Giant impacts provide perfect conditions for the tiny grain formation. These tiny grains could have been generated either by direct condensation from impact-produced vapor \citep{johnson12a} or by hypervelocity grinding within optically thick clouds of debris produced by violent collisions of large bodies \citep{johnson12b}. In the former case, these silica grains were born tiny; while in the later case, tiny grains were formed through collisional cascades and the optical depth effects within the formation region allow grains that are much smaller than the blowout size to form. 
As a result, the solid-state feature rising from the tiny silica dust would remain stable until they slowly drift inward due to the Poynting-Robertson drag \citep{su20}. The feature would gradually weaken due to dust sublimation once the grains reach the sublimation temperature under timescales of a few hundred years in HD 23514. 

In addition, the giant-impact ejecta would also form a population of planetesimals, maintaining a high level of dustiness and creating temporal variation \citep{watt24_postimpact_edd_evolution} as probed by 3--5 \micron\ photometry. The observed variability levels require a minimal corresponding area change in the dust cross section to be on the order of (1--2.5)$\times$10$^{-3}$ au$^2$ (for details see Appendix \ref{sec:ir_monitoring}). This dust cross section is similar to the emitting area of hot gas, pointing a potential link between the dust and molecular gas production in the sub-au region. Assuming the molecular gas and small dust share the same sub-au region, the estimated gas and dust masses suggest that the small grains could have low ($<$1) Stokes number and coupled with the gas (highly depending on the estimated CO mass, see Appendix \ref{sec:gas_model}). The presence of molecular gas would increase the dust-clearing timescale due to the combination of radiation pressure and gas drag \citep{takeuchi01,kenyon_najita_bromley2016}. 
Exploring grain trapping is beyond the scope of this paper, but it is likely to be an important component to a comprehensive model for the \tgt system, particularly in light of the new discovery of gas in the system.

\subsection{Origins of the Volatile Material}
\label{sec:gas_origin}

\begin{figure}
    \centering
    \includegraphics[width=\linewidth]{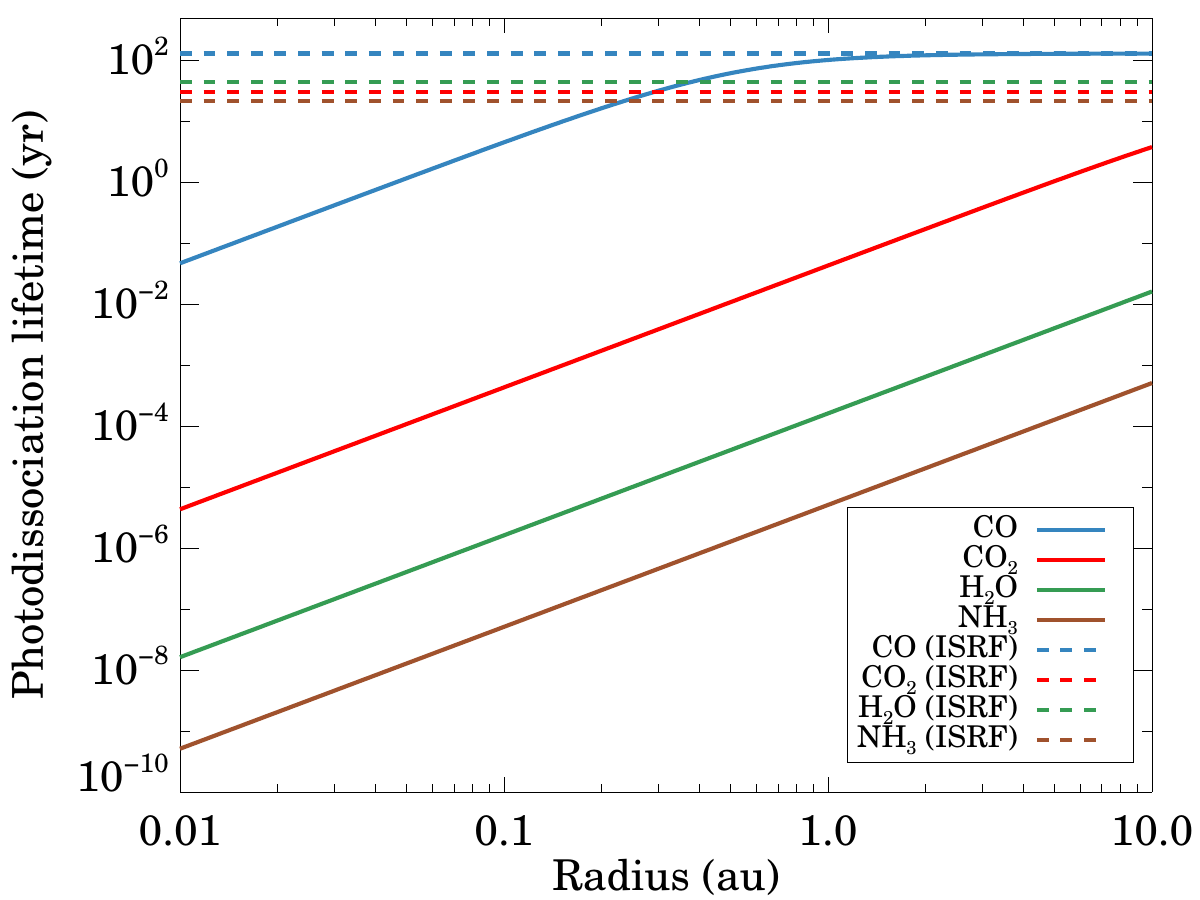}
    \caption{The photodissociation lifetime of unshielded molecules as exposed to both the ultraviolet (UV) photons from \tgt and the interstellar radiation field (ISRF) as a function of stellocentric distance where the dashed lines are those only using ISRF. The UV emission from the star was estimated using the Kurucz model (Appendix \ref{sec:starsub}), which under estimates the amount of UV photons given its youth. The unshielded photodissociation lifetimes shown here should be considered as upper limits.}
    \label{fig:photodisociation_timescale}
\end{figure}

Under the optically thin condition, molecules have very short photodissociation lifetimes, particularly around an F-type star like HD 23514. Figure \ref{fig:photodisociation_timescale} shows the expected unshielded photodissociation timescales for various molecules as a function of stellocentric distance from the star. Volatiles would only survive much less than a year once released within 1 au including additional shielding as detailed in Appendix \ref{sec:pf_shielding}, but CO could persist over $\sim$100 years outside 1 au. The detection of hot molecular gas suggests that these molecules require a continuous replenishment and/or an effective protection/shielding against the strong stellar radiation. The short-lived volatiles in the very close ($<$1 au) vicinity of the star can be produced either in-situ or delivered from the outside 1 au region as discussed below. 

\noindent \textit{In situ formation}\ \  
In the context of the giant-impact hypothesis, molecular gas could be produced by outgassing processes if one or both the impacting bodies are volatile rich. Earth's early atmosphere was thought to form by retention of these outgassing products \citep{abe_matsui85}. Impact experiments and model calculations both show that H$_2$O and CO$_2$ are the major gases produced by outgassing of carbonaceous chondritic material over a wide range of temperatures and pressures (e.g., \citealt{lange_ahrens82,schaefer2010, thompson21_outgassingexperiments}, and references within). Furthermore, shock-recovery experiments targeting a mixture of olivine, iron, and water also reveal preferential formation of ultrafine (tens of nanometers) smoke particles of silicates and metal oxides, facilitated by the role of supercritical water \citep{furukawa07_UPs}. These tiny grains might be ubiquitous in post-ocean-impact events and provide effective UV shielding for volatiles. Alternatively, volatiles could also be released from (1) an atmospheric stripping event as has been suggested for HD\,172555 \citep{schneiderman21_hd172555}, or (2) large remnants of a giant-impact ejecta as long as they are large enough to retain a significant fraction of volatiles. It is then interesting to note that the change in the dust cross section (derived in Appendix \ref{sec:ir_monitoring}) is on the same order as the emitting area of the molecular gas, indicating that collisional activities in the aftermath of giant-impact events could be tied to the formation/protection of molecular gas in the inner region.

\noindent \textit{Delivery from outer region}\ \
This delivery scenario has been widely accepted in both our solar system (comets) and other debris systems (exocomets) -- scattered icy bodies from the outer region releasing volatiles and dust grains as they warm up on their inward migration. In this case, the detected molecular gas might have no direct relationship with the giant-impact-produced small grains observed in the system. For HD 23514, the inward delivery rate might be much lower than that of the debris systems that host Kuiper-belt analogs such as $\beta$ Pic \citep{beust96_betapic_FEBs,matra19b} and $\eta$ Corvi \citep{marino17_etaCrv} because it has a limited volume of cold reservoir for storing icy bodies (lack of a massive Kuiper-belt analog, see Appendix \ref{sec:sedmodel_details}). This scenario also needs a perturber either internally or externally to efficiently scatter icy bodies into the inner region. Ground-based high-contrast imaging found no other sources within 10\arcsec\ around \tgt except for the M7 brown dwarf companion \citep{rodriguez12_hd23514} 2\farcs64 away, while a recent radial velocity (RV) study rules out massive giant planets within 0.1 au \citep{takarada20_RV_Pleiades}. External companions under some favorable conditions can excite the planetesimal population via the eccentric Kozai–Lidov mechanism, leading to inner dust and gas production \citep{naoz16review_kozai,young_wyatt24}.
Without detailed orbital parameters, it is not clear whether the M dwarf companion at $\sim$360 au away is capable of this role. Other challenges in the delivery scenario are: (1) evaporation of icy bodies is expected to occur at a range of distances, which is at odds with the gas concentration within the sub-au region, and (2) the comet delivery rate is expected to be variable based on the solar system observations, which somehow contradicts the tentative CO$_2$ detection in the Spitzer data. The second challenge might be accommodated if scattering of many comets occurs at high efficiency. A future investigation is needed to further explore the delivery model; nonetheless, it remains a possible origin for the molecular gas.

\section{Conclusion }
\label{sec:conclusion}

We have presented a JWST MIRI/MRS 5--28 \micron\ spectrum of HD 23514, an F5V star in the $\sim$150 Myr-old Pleiades cluster, hosting one of the giant-impact disks discovered by Spitzer with unique mineralogy dominated by $\sim$9 \micron\ silica dust. We found that the prominent silica feature remains stable compared to the data taken 15 yr earlier by Spitzer, and that the system lacks large-scale ($>$30\%) variation over a 40-year baseline since the IRAS measurement taken in 1983. Characteristics derived from the observed dust mineralogy and SED modeling suggest that the dust has high crystalline mass fraction preferentially in both pyroxene and silica types with the latter mostly confined within sub-au region. The JWST spectrum also reveals surprising hot molecular gas emission from CO$_2$  and CO with traces of other volatiles such as H$_2$O and NH$_3$, apparently stable as seen in the Spitzer data and colocated with the silica dust. We explored several pathways explaining the observed properties of dust and gas, and found that a past giant-impact event can simultaneously explain both the large amount of highly processed small grains (crystalline silicates and silica) and volatile emission. 

In the preferred giant-impact scenario, at least one of the involved bodies needs to have volatile content similar to the carbonaceous chondritic material in the solar system. Volatiles are expected to form as (1) the outgassing products either directly from the newly formed, molten planet or from the large planetesimal remnants of impact ejecta, or (2) the stripped atmospheric material. Tiny silica grains can form either by direct condensation from the impact-produced vapor or by hypervelocity grinding within pebble-size vapor condensates. If one of the impacting bodies had a surface ocean before the violent collision, the post-ocean-impact event also favors the formation of ultrafine particles, which can also provide effective UV shielding for the volatiles. 

Alternatively, the tiny silica and molecular gas have no direct relation and are created by nonrelated processes. Volatiles can be delivered into the inner planetary region from scattered icy bodies originally stored in the outer reservoir. Erosion of a rocky body in a high temperature environment can produce tiny silica dust via the degradation of silicates. The keys for these alternatives are whether these processes can be efficient enough to account for the amount of silica and molecular gas observed in the system. Future observations to better determine the gas properties, search for other warm/cold gas and dust, assess the temporal change in both the dust composition and volatiles, and identify the shielding mechanisms against photodissociation will help to shed more light into the EDD phenomenon. 
 
%TC:ignore

\appendix 
\restartappendixnumbering

\section{Stellar Properties and Dust Emissivity Calculation}
\label{sec:starsubemi}

\subsection{Stellar Photosphere and Subtraction}
\label{sec:starsub}

For the stellar emission, we used a Kurucz model of $T_{\ast}$=6500 K and logg=4.5, normalized to the optical and $JHK_S$ photometry corrected by an interstellar extinction of $A_V$=0.1 \citep{roman-zuniga23_apogee2}. Given a distance of 139 pc, the integrated stellar luminosity is $\sim$3 $L_{\sun}$ and the stellar radius is 1.4 $R_{\sun}$. These values are all consistent with the values found in the literature for a 1.3 $M_\sun$ young star  \citep{rodriguez12_hd23514,fu22_lamost,roman-zuniga23_apogee2}. This stellar model was used for deriving IR excess, computing thermal equilibrium dust temperatures in SED modeling, and photodissociation rates of molecules. 

\begin{figure}
    \centering
    \includegraphics[width=\linewidth]{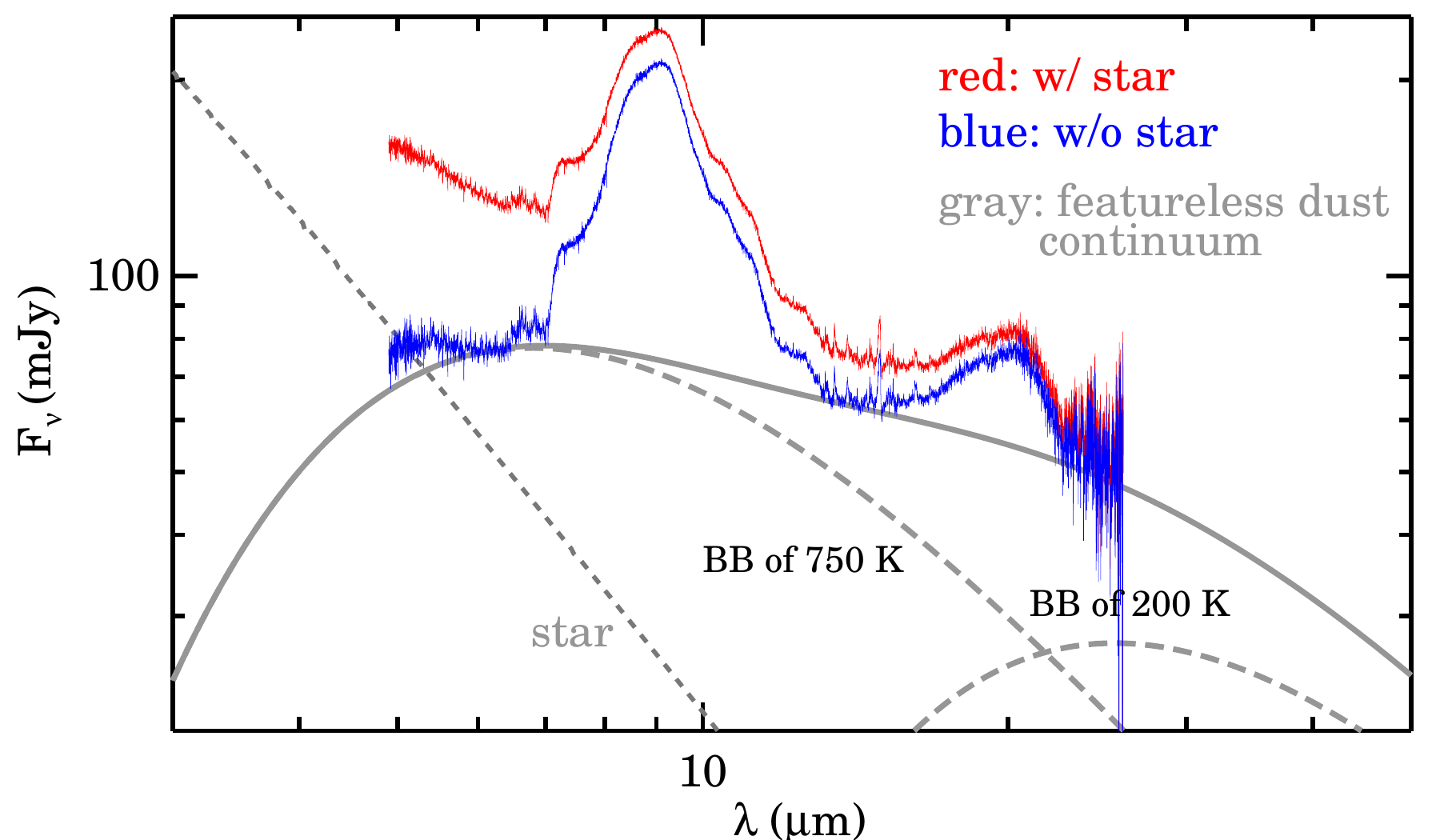}
    \caption{MIRI MRS data before (red) and after (blue) stellar photosphere subtraction. The gray line represents the featureless dust continuum with a combination of two blackbody emissions at 750 K and 200 K. This featureless dust continuum was used to compute the dust emissivity as shown in the right panel of Figure \ref{fig:mrs_result}.}
    \label{fig:starsubemi}
\end{figure}

The stellar emission contributes to the majority of the emission shortward of 2.2 \micron. Using the Kurucz model, we determine the star has an integrated flux of 153 and 99 mJy for the Spitzer/IRAC I1 and I2 bands, respectively, but 170 and 95 mJy for WISE W1 and W2 bands (due to the different filter profiles). \tgt has been monitored by the Kepler and TESS missions. Gaia DR3 catalog gives $T_{\rm eff}$, log\,$g$, [Fe/H], RV as 6319 K, 4.21, -0.24 and 6.32$\pm$0.44 km\,s$^{-1}$. \citet{Tsantaka22} lists its RV as 5.18$\pm$0.28 km\,s$^{-1}$ and \citet{mermilliod09} lists its $V$\,sin$i$ as 49.5$\pm$5 km\,s$^{-1}$. The system shows no flaring behavior \citep{ilin19_k2_flares}, nor $\delta$ Scuti-like pulsations \citep{bedding23_TESS_Pleiades} over the monitoring periods. Assuming the star is stable over time, Appendix \ref{sec:ir_monitoring} discusses the IR variability of the \tgt system using archival data from Spitzer and WISE. Figure \ref{fig:starsubemi} shows a comparison in the MRS spectrum before and after stellar subtraction, suggesting that the IR excess has a roughly equal contribution as the stellar photosphere at 5 \micron\ (the shortest wavelength of MRS). 

\subsection{Dust Emissivity and Silica Properties}
\label{sec:dustemissivity}

There are a few ways to determine the featureless dust continuum (termed ``pseudocontinuum"). \citet{watson09} used a fifth-order polynomial function anchored at the wavelength range where the dust emission feature is small for a large sample of PPDs observed by Spitzer/IRS. \citet{harker23_spitzer_comets} analyzed a large sample of solar system comets observed by Spitzer/IRS, and they used a scaled Planck function as the pseudocontinuum defined as spanning $\lambda=$7.3--7.7 and 12.3--13.0 \micron\ to study the 10 \micron\ feature complex. For the JWST/MIRI/MRS spectrum, we decided to use a combination of two Planck functions as the pseudocontinuum by anchoring the wavelength regions that have minimal contribution from known solid-state dust features. As shown in Figure \ref{fig:starsubemi}, the pseudocontinuum as represented by $\sim$750 and $\sim$200 K Planck functions captures the featureless dust continuum well (after stellar subtraction). The resultant dust emissivity is shown in the right panel of Figure \ref{fig:mrs_result}. 

To further validate our approach, we also generated dust emissivity functions for a young star called LkCa 15. The dust emissivity in LkCa 15 was classified as an ISM-like profile by \citet{watson09} where the emission is dominated by pristine silicate dust grains (i.e., amorphous and sub-micron). Therefore, we used its mid-IR archival spectra from Spitzer/IRS and JWST/MRS as the input and constructed the pseudocontinuum using the polynomial function for the former but a combination of two blackbodies for the latter. The resultant two dust emissivity functions agree well within a few percent in the 10 \micron\ complex (5--14 \micron), but in the 20 \micron\ complex the JWST one is slightly higher by $\sim$20--40\% likely due to the limited long-wavelength in determining the pseudocontinuum. The derived JWST LkCa 15 dust emissivity is shown in the right panel of Figure \ref{fig:mrs_result}. Because our focus is in deriving the crystalline mass fraction indices by referencing the ISM-like profile in the 10 \micron\ complex, the difference in the 20 \micron\ region has no impact on the result. We used the JWST LkCa 15 emissivity function to compute the indices ($P_{10}$ for pyroxene, $O_{10}$ for olivine and $S_{10}$ for silica) as defined by \citet{watson09}. Note that these indices are defined as the normalized ratio in dust emissivity at a certain wavelength range covering the narrow features, the normalization of the reference (ISM-like) emissivity (i.e., the blue curve in the left panel of Figure \ref{fig:mrs_result}) has no effects on the results. For HD\,23514, the derived indices are $P_{10}$=2.54, $Q_{10}$=0.91, and $S_{10}=$1.8, suggesting a high fraction of crystallinity preferentially in the pyroxene and silica mineralogy.  We also derived the dust emissivity using the Spitzer IRS spectrum, and found similar indices within the uncertainty (corroborating no change in the dust composition).

To quantitatively characterize the silica feature in HD\,23514, we also derived the dust emissivity using the
Spitzer IRS spectrum of T51 (a few Myr-old star). T51 is one of the five silica-rich PPDs characterized by \citet{sargen09_silia} who found that the system lacks submicron amorphous dust and the crystalline silicate mass fraction is $\sim$15\% including both pyroxene and olivine types, and the silica dust is best scribed as annealed silica although obsidian or amorphous quartz give equally good fits. Our derived indices for T51 are $P_{10}$=2.02, $Q_{10}$=1.50, and $S_{10}$=2.36, respectively, consistent with the finding for \tgt above. Although the 10 \micron\ feature in T51 and \tgt peaks at 9.1 \micron\ (Figure \ref{fig:mrs_result} right), the FWHM of the 10 \micron\ feature in T51 is 1.5 times larger, suggesting the dominant size in \tgt is smaller, likely submicron.

Silica is not abundant in the ISM. Its presence in circumstellar disks requires special conditions for formation like giant impacts as we discuss here. Space weathering, as discussed by \citet{sargen09_silia}, is one way to deposit silica by converting silicate minerals; however, it is unlikely to be the primary mechanism responsible for the large amount of sub-um size silica seen in HD\,23514. Space weathering is a long-term process. $\sim$\micron-size silicate grains in the inner region of \tgt have short lifetimes (blowout size is 1 \micron, and will be gone on orbital timescales), implying that converting small silicates, produced by collisional cascades, to silica is an inefficient process. For stars just emerge from the protoplanetary disk stage, massive flare events could speed up the process as suggested by \citet{badlisse2020}, however, \tgt is not very active at an old age of 150 Myr (Appendix \ref{sec:starsub}).  Comets could bring a large amount of fresh dust into the inner region during a period of high delivery rates like the late heavy bombardment, and create silica through space weathering. For solar system comets, the dust mass is $\sim$10$^{14}$\,g on average \citep{harker23_spitzer_comets}, meaning a total of 10 billion comets is needed to account for the dust observed in HD\,23514 ($\sim$10$^{24}$\,g derived in Appendix \ref{sec:sedmodel_details}). The most challenging problem in the comet delivery scenario is that the typical dust size is large, $\sim$100 \micron\ \citep{rinaldi17}, i.e., space weathering should produce large-size silica, not sub-$\micron$. In summary, the proposed giant-impact scenario is the most likely explanation for the large amount of submicron silica grains in HD\,23514.

\section{Other Mid-IR Data and 3--5 \micron\ Photometric Monitoring}
\label{sec:ir_monitoring}

\begin{figure}
    \centering
    \includegraphics[width=\linewidth]{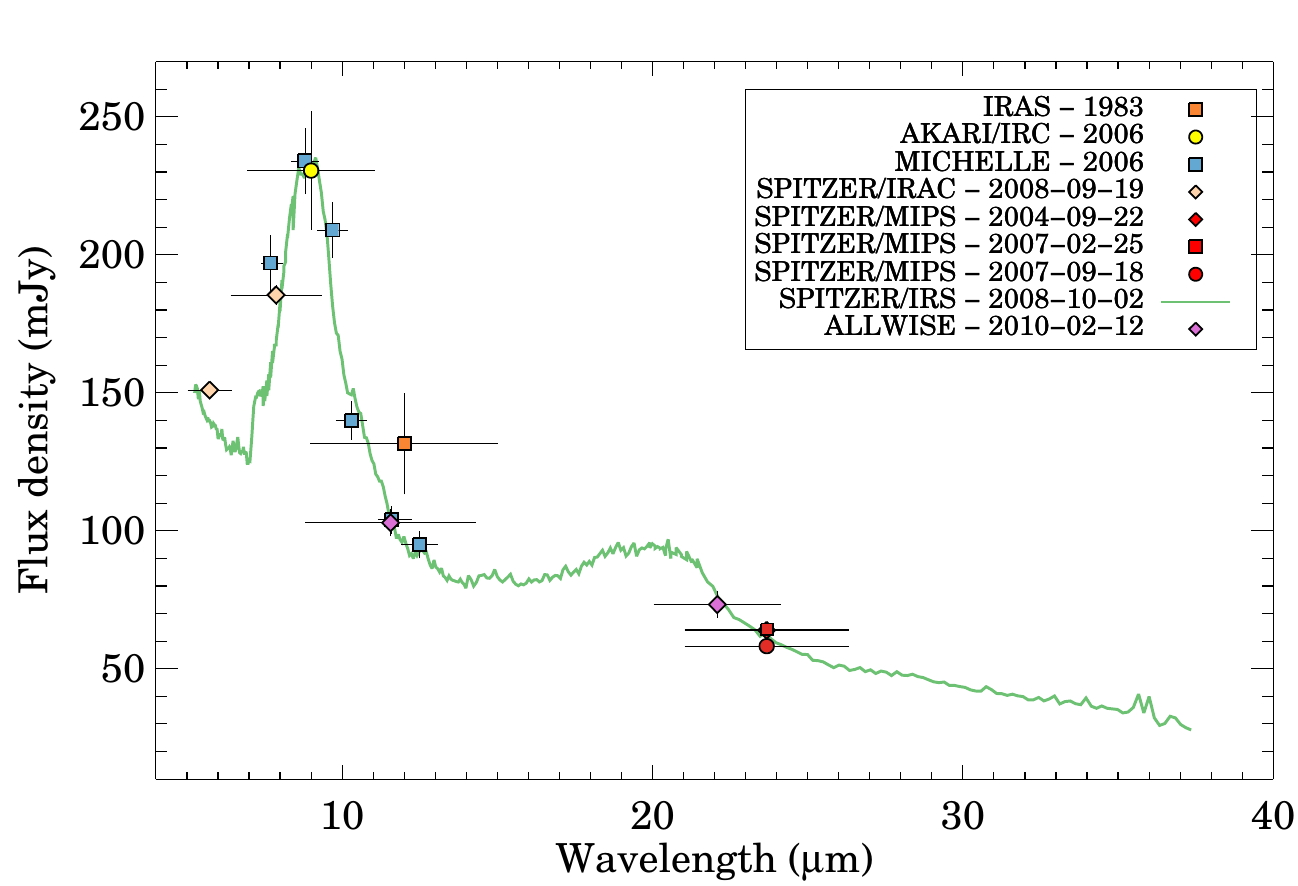}
    \caption{IR photometry of \tgt available in the literature/archive before 2013 in comparison with the Spitzer IRS spectrum. The horizontal bars show the bandpass of the measurement while the vertical ones for the 1$\sigma$ uncertainty. }
    \label{fig:mid-ir}
\end{figure}

\tgt has various space- and ground-based mid-IR photometry, with the earliest measurement from IRAS in 1983 \citep{irasfsc,akari_irc_catalog,rhee08,wright10_wisemission,meng12}. Figure \ref{fig:mid-ir} summarizes these photometric observations between 5 and 25 \micron. The plotted data have been color corrected using the Spitzer/IRS spectrum, assuming that it is a good representation of the SED shape in the given wavelength range over the period of interest (1983--2010). Because WISE 12 \micron\ band was selected to match approximately that of IRAS \citep{cutri12_wisemission}, they trace the same part of the 10\,{\micron} feature. The color corrected flux densities in these WISE and IRAS bands are 102.9$\pm$4.8\,mJy and 131.6$\pm$18.4\,mJy, respectively, i.e.  the two fluxes agree within the 1.5$\sigma$ photometric uncertainties.

\begin{figure*}
    \centering
    \includegraphics[width=\linewidth]{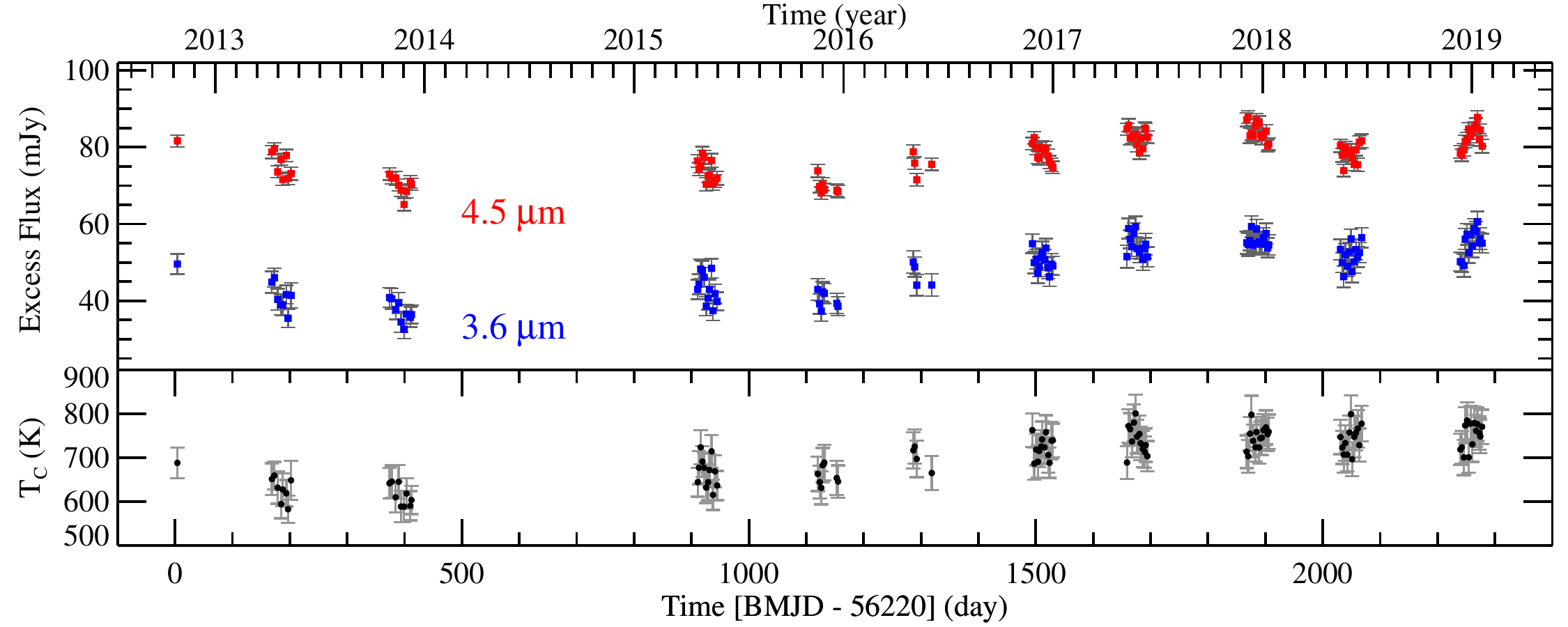}
    \caption{The IR variability around \tgt as monitored by warm Spitzer in $\sim$3 day cadence when available. The upper panel shows the excess fluxes obtained by the two shortest IRAC bands while the bottom depicts the color temperatures derived by the flux ratio. The system exhibits stochastic, weekly variability without obvious periodicity. There exists a weak, positive trend between the excess flux and color temperature. }
    \label{fig:var_irac}
\end{figure*}

\begin{figure}
    \centering
    \includegraphics[width=\linewidth]{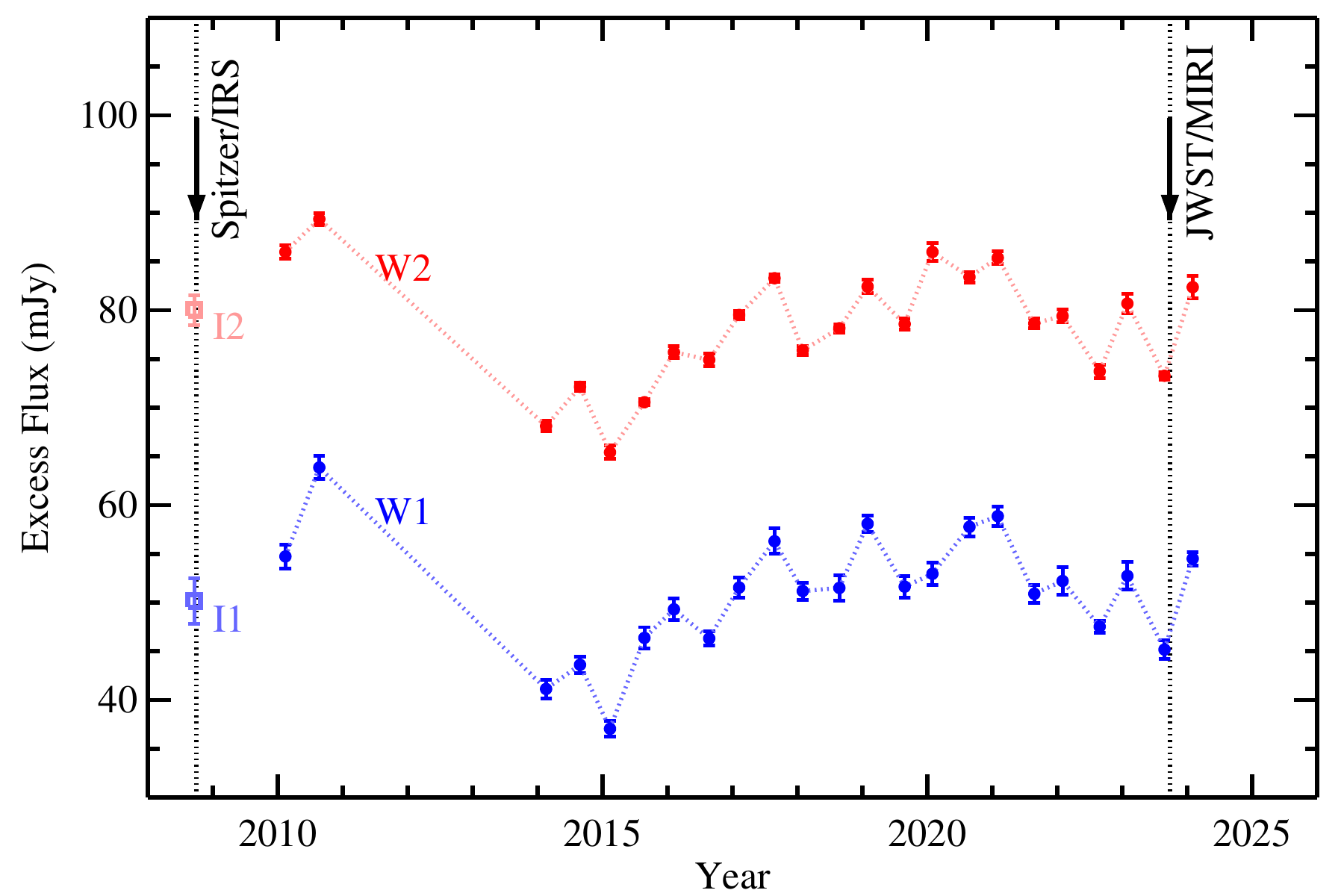}
    \caption{Similar to Figure \ref{fig:var_irac} but with WISE W1 and W2 data along with the Spitzer/IRAC I1 and I2 photometry taken in 2008, showing yearly IR variability. The lowest 3--5 \micron\ flux appeared to be in 2015. The two epochs of mid-IR spectra obtained by Spitzer and JWST are also indicated.  }
    \label{fig:var_wise}
\end{figure}

IR variability within $\sim$20--30\% levels, as an additional hallmark commonly found in the extreme debris disks \citep{moor21}, is known for HD 23514 \citep{meng12}. \tgt has been intensively monitoring by Spitzer/IRAC and WISE at 3--5 \micron\ with the former in $\sim$3--day cadence when available, while the latter in $\sim$6--month cadence. Following the analysis method used for another IR variable, the RZ Psc disk \citep{su23_rzpsc}, Figures \ref{fig:var_irac} and \ref{fig:var_wise} show the short- and long-term disk fluxes at 3--5 \micron\ between 2008 and 2024 where the two epochs of the IR spectra are also marked. 

Using these 3--5 \micron\ data, the \tgt disk exhibits stochastic, both weekly and yearly, IR variability with a weak, positive trend between the excess flux and color temperature. The inferred color temperatures are $\sim$700--800 K, consistent with the warmer dust temperature inferred from the featureless pseudocontinuum in Appendix \ref{sec:dustemissivity}. The typical peak-to-peak flux variation (Figure \ref{fig:var_irac}) is $\sim$15\% in both the 3.6 and 4.5 \micron\ bands, more than 10$\sigma$ significance compared to the photometry stability ($\sim$1\%) determined by field stars in the data. The 14-year long WISE data (Figure \ref{fig:var_wise}) show a similar behavior with the largest peak-to-peak flux difference of $\sim$25 mJy in the W2 band. Taking the typical ($\sim$10 mJy) and maximum ($\sim$25 mJy) 4.5 \micron\ flux change at a face value, the \textit{minimal} corresponding area change in dust cross section is (1--2.5)$\times10^{-3}$ au$^2$ by adopting a dust temperature of 800 K and a distance of 139 pc under optically thin assumption. It is interesting to note that this minimal area change in the dust cross section is similar to the hot CO$_2$ gas emitting area (Table \ref{tab:co2_parameters}). The cross section changes can also be translated to the \textit{minimal} change in the dust mass (assuming static state collisional cascades down to the radiation blowout size of $\sim$1 \micron). The very lower limit on the associated dust mass is on the order of $\sim$3--7$\times10^{21}$ g, corresponding to the total disruption of a $>$120--165 km-size asteroid (assuming a density of 3.5 g cm$^{-3}$). This mass/size is a lower limit because of our assumptions (optically thin and static state). 

\section{Gas Model and Other Related Calculations}
\label{sec:detailed_gas}

\subsection{Local Thermodynamic Equilibrium Slab Model}
\label{sec:gas_model}

For the LTE slab gas modeling, we first need to determine the gas contribution to the total emission by determining the dust emission, which includes both solid-state and featureless continuum. To fit the continuum of the spectrum, we applied a median smoothing and a second-order Savitzky–Golay filter, and used the procedure presented in \citet{pontoppidan24_jdisc} to apply a final offset based on line-free regions. We excluded the bad negative pixels in the continuum fitting using \texttt{scipy.signal.find\_peaks} with a prominence of 0.008. Because the visual inspection showed that CO$_2$ is the most dominant feature in the data, we also excluded the wavelength range around the CO$_2$ emission (14.78 \micron$<\lambda<$15.0 \micron) and the tentative NH$_3$ emission (10.32 \micron$<\lambda<$10.7 \micron). 
This assumption excludes the potential pseudo-continuum produced by optically thick molecular emission (visual inspection does not find such a emission), a limitation of modeling gas and dust separately \citep{liu19_click}. The resulting continuum fit is shown in the left panel of Figure \ref{fig:mrs_result}. 

To fit the line emission, we adopted the slab-modeling Python package \texttt{iris} \citep{iris_23} assuming the gas is in LTE, with line data from the HITRAN database \citep{hitran22}. The line width was estimated with thermal broadening based on the temperature of the gas. In \texttt{iris}, the fitting is performed with the Bayesian nested sampling Python package \texttt{dynesty} \citep{speagle20_dynesty}. We set the stopping criterion by the change in the remaining evidence (marginal likelihood $\mathcal{Z}$) when $\Delta \log \mathcal{Z} \leq 0.001 (n_{live} - 1) + 0.01 $, where $n_{live}$ = $10\times n_{dim}$ with $n_{dim}$ as the number of parameters in the model.

\begin{figure}
    \centering
    \includegraphics[width=\linewidth]{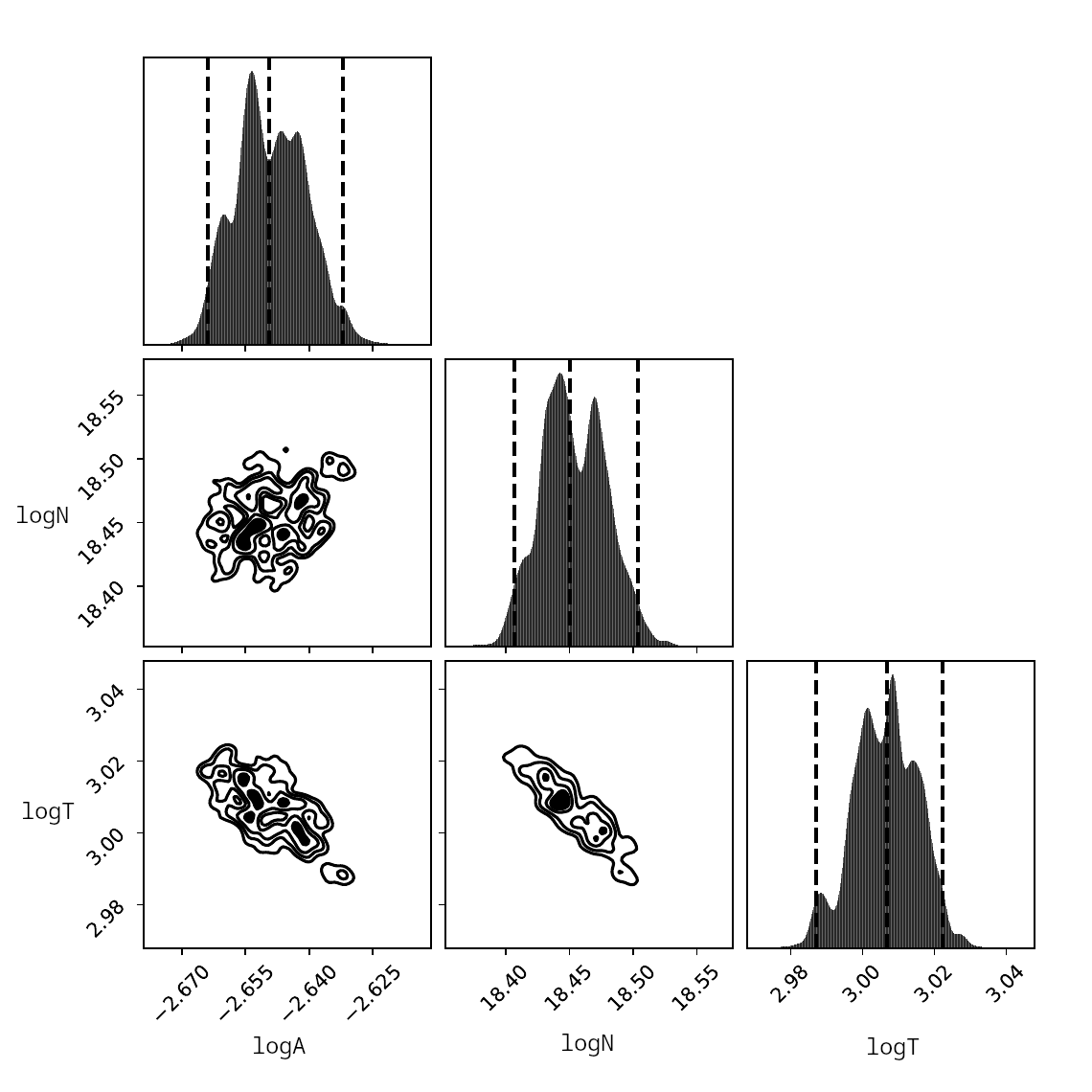}
    \caption{The posterior distributions for the CO$_2$ fitting using the three parameters: emitting area (log $A$ in units of au$^2$), column density (log $N$ in units of cm$^{-2}$), and gas temperatures (log $T$ in units of K). }
    \label{fig:co2_corner}
\end{figure}

We restricted the fitting within the wavelength range of 14.0--17.0 \micron\ for the emission of CO$_2$, and adopted uniform priors for emitting area ($A$) and column density ($N$), with $\log_{10}A =\mathcal{U} (-3, 3) \text{ au}^{2}$ and $\log_{10} N = \mathcal{U} (12, 22) \text{ cm}^{-2}$. The prior for the gas temperature ($T_{\rm gas}$) was set to be uniform initially, and later set as a normal distribution centered around 500\,K with a standard deviation of around 200\,K to speed up the search. We included $^{13}$CO$_{2}$ with CO$_2$ in our fitting, and assumed the emissions were from the same region ($T_{\rm gas}$ and $A$ are the same) and the $^{12}$C/$^{13}$C ratio was the same as ISM value ($N_{\rm CO_2}=68\times N_{^{13}\rm CO_2}$, e.g., \citealp{milam05}). The posterior distributions for the CO2 fitting are shown in Figure \ref{fig:co2_corner}. The 15.4 \micron\ $^{13}$CO$_{2}$ line is the only (i.e., strongest) isotopologue line present in the data. Excluding it produces a slightly poor fit at 15.4 \micron\ as shown in Figure \ref{fig:co2model}. The best fit CO$_2$ model parameters are summarized in Table \ref{tab:co2_parameters} by assuming two different turbulence velocities: 0 km\,s$^{-1}$ as the nominal and 2 km\,s$^{-1}$ as the upper limit value. The two cases make very little difference in the resulting fits visually, illustrating the associated uncertainties in the MCMC fits might be underestimated, and reflecting the limitations of using the simple LTE slab model. 

\begin{figure*}
    \includegraphics[width=\linewidth]{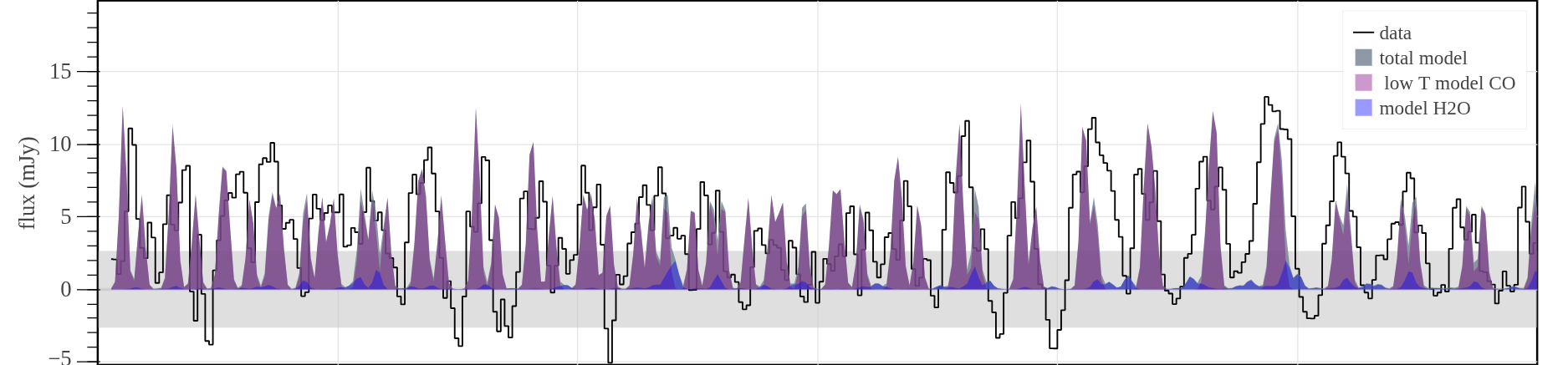}
    \includegraphics[width=\linewidth]{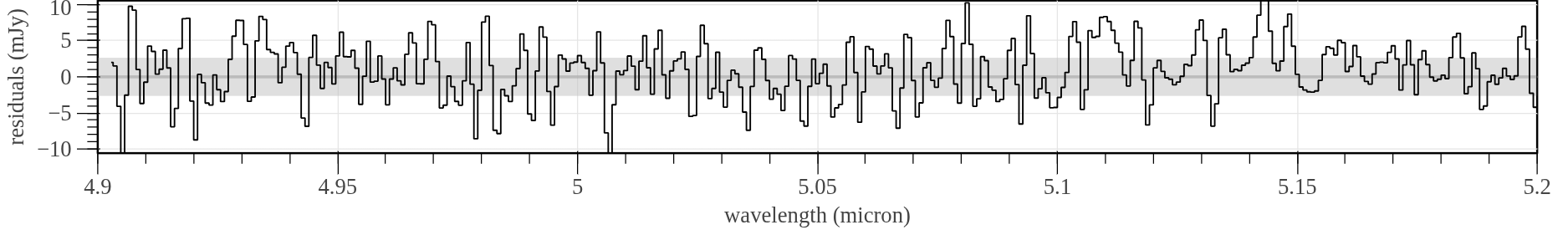}
    \includegraphics[width=\linewidth]{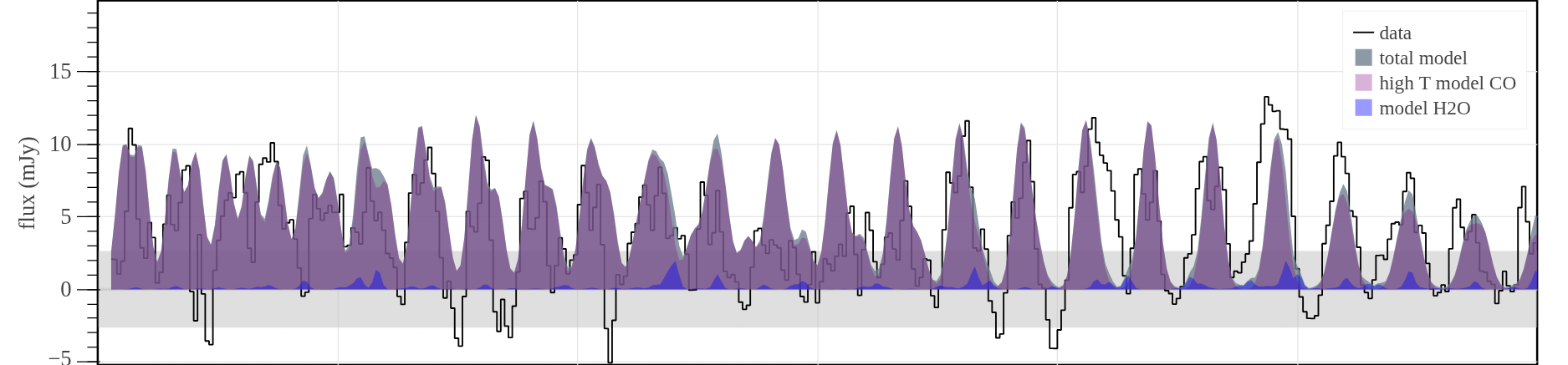}
    \includegraphics[width=\linewidth]{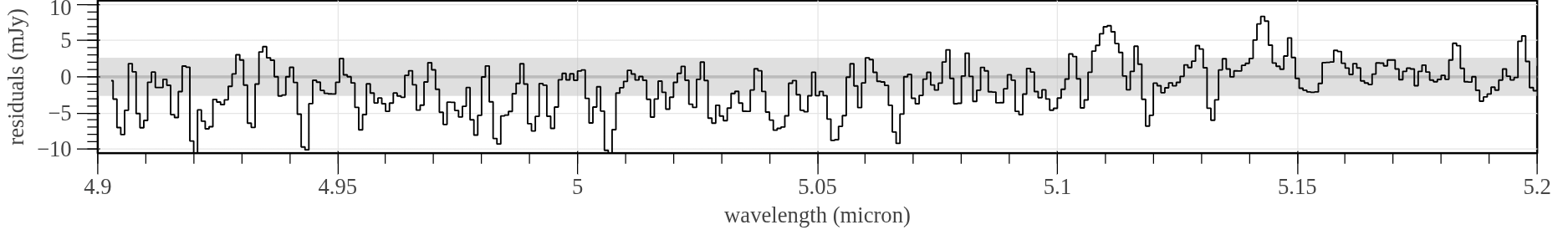}
    \caption{The continuum-subtracted spectrum of \tgt centered at the CO gas (upper for the low $T_{\rm gas}$ and lower for the high $T_{\rm gas}$ models; details see Appendix \ref{sec:detailed_gas}). The shaded color regions show the slab model predictions of CO (purple) and others (same as Figure \ref{fig:co2model}). For all the comparison plots between the data and model, the flux panels are shown starting from $-2$ rms to 1.5 times the peak flux in the displayed segment and the residual panels are all shown in the range of $\pm$4 rms.}
    \label{fig:comodel}
\end{figure*}

\begin{figure*}
   \centering   
   \includegraphics[width=0.95\linewidth]{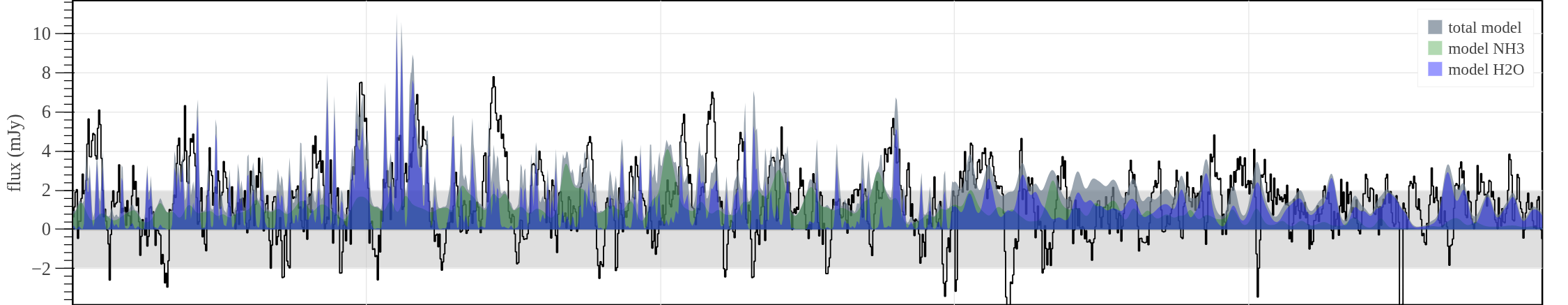}
    \includegraphics[width=0.95\linewidth]{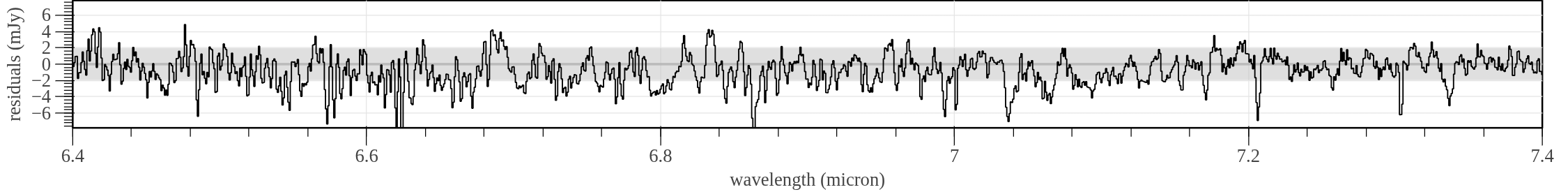}
    \includegraphics[width=0.95\linewidth]{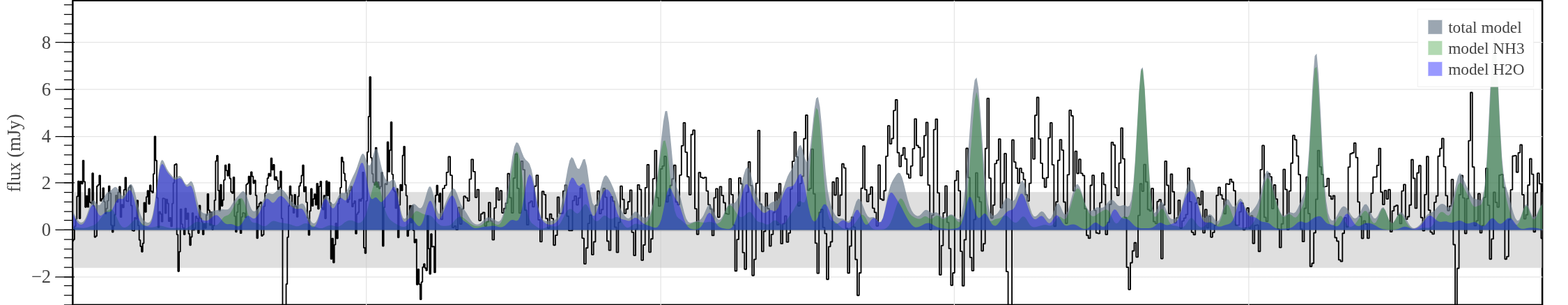}
    \includegraphics[width=0.95\linewidth]{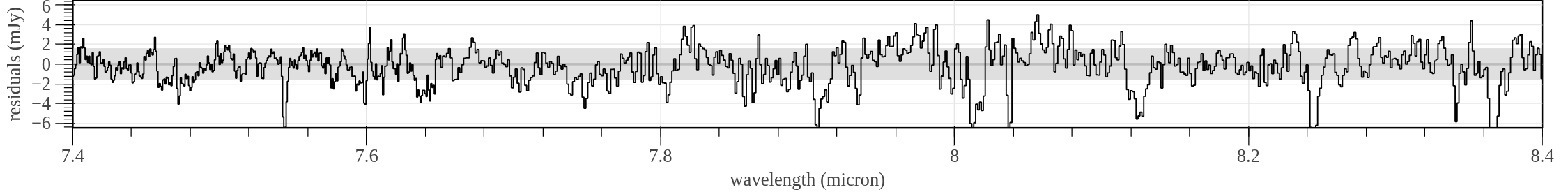}
    \includegraphics[width=0.95\linewidth]{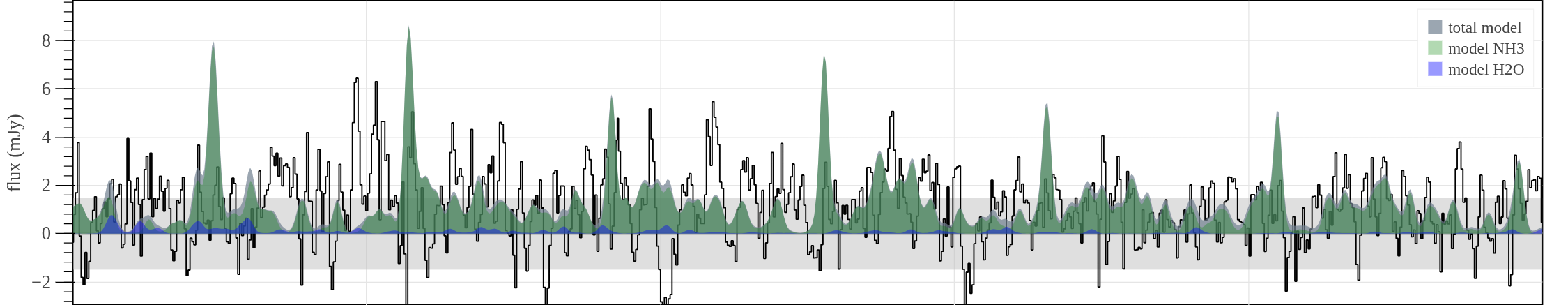}
    \includegraphics[width=0.95\linewidth]{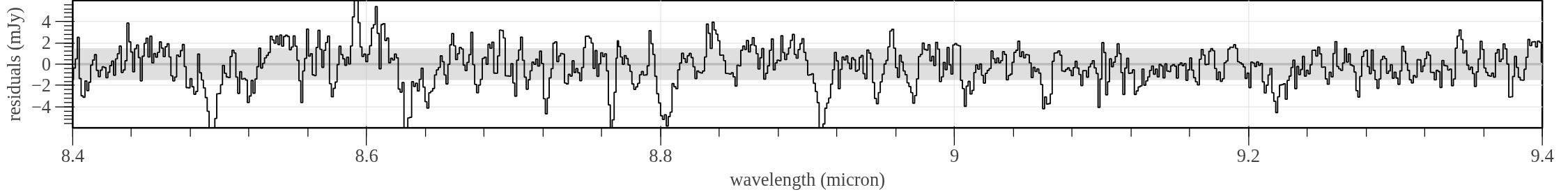}
    \includegraphics[width=0.95\linewidth]{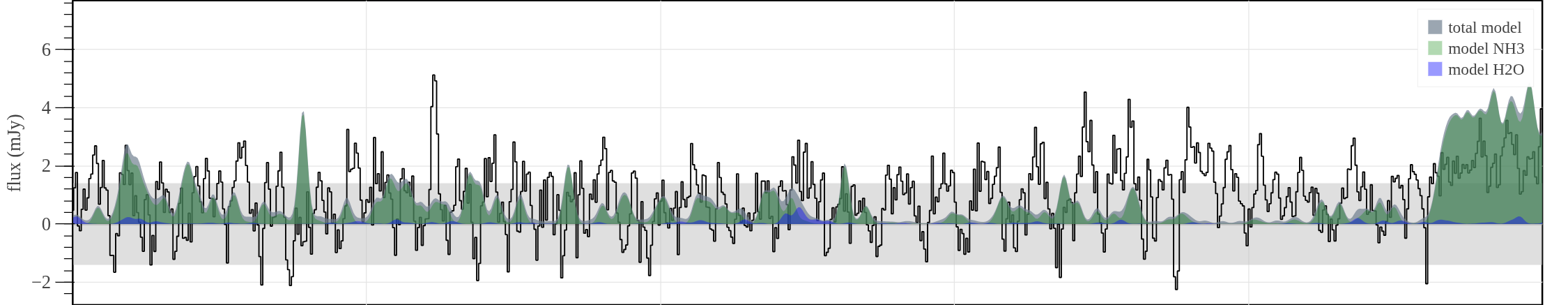}
    \includegraphics[width=0.95\linewidth]{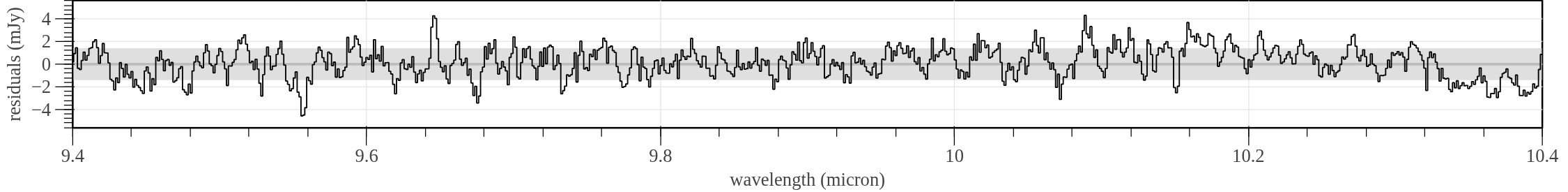}
    \caption{Same as Figure \ref{fig:comodel} for different spectral segments with different molecules using the same plotting rules. }
    \label{fig:hd23514_full_contsub}
\end{figure*}

\setcounter{figure}{6} 

\begin{figure*}
    \centering   
    \includegraphics[width=0.95\linewidth]{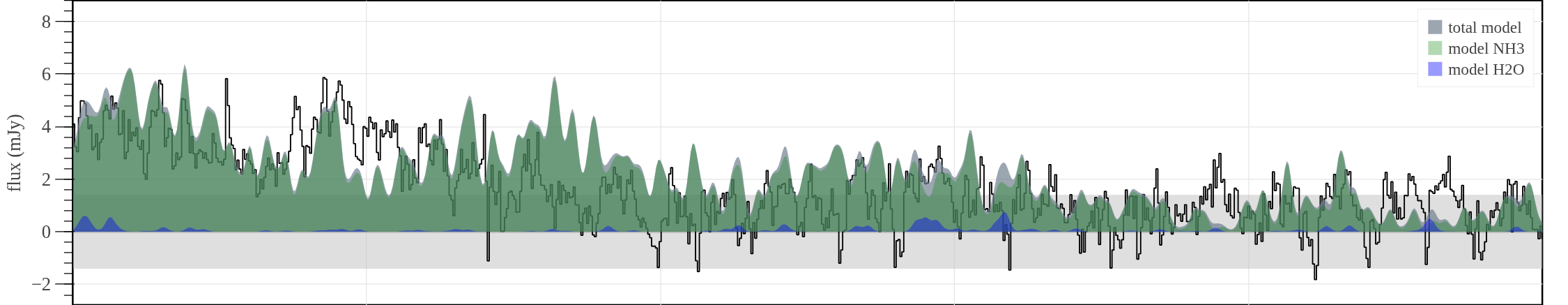}
    \includegraphics[width=0.95\linewidth]{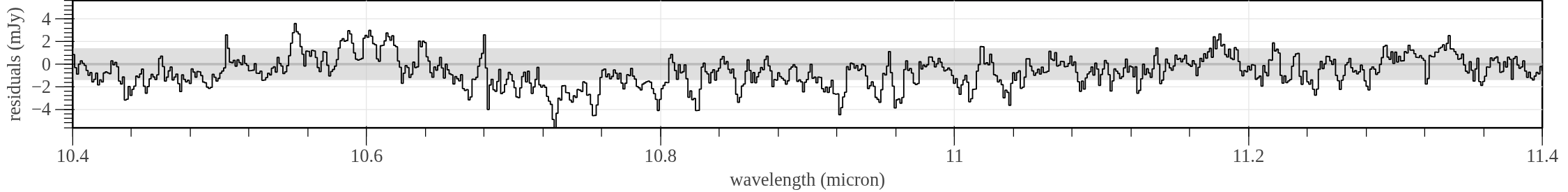}
    \includegraphics[width=0.95\linewidth]{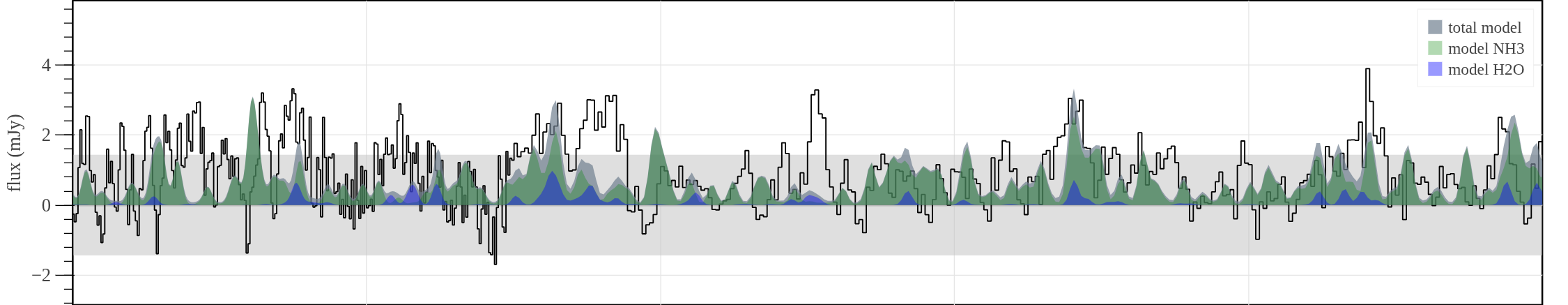}
    \includegraphics[width=0.95\linewidth]{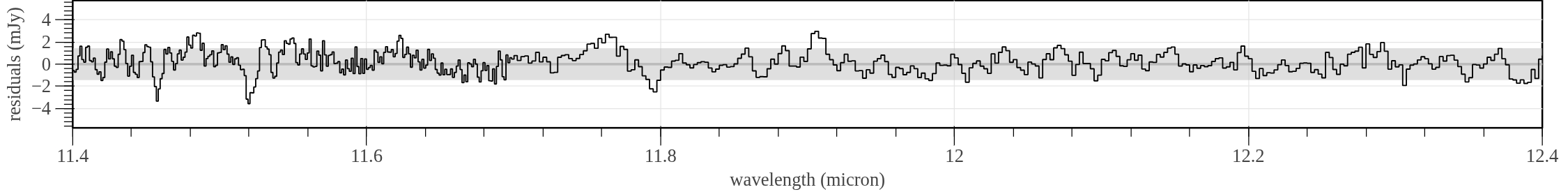}
    \includegraphics[width=0.95\linewidth]{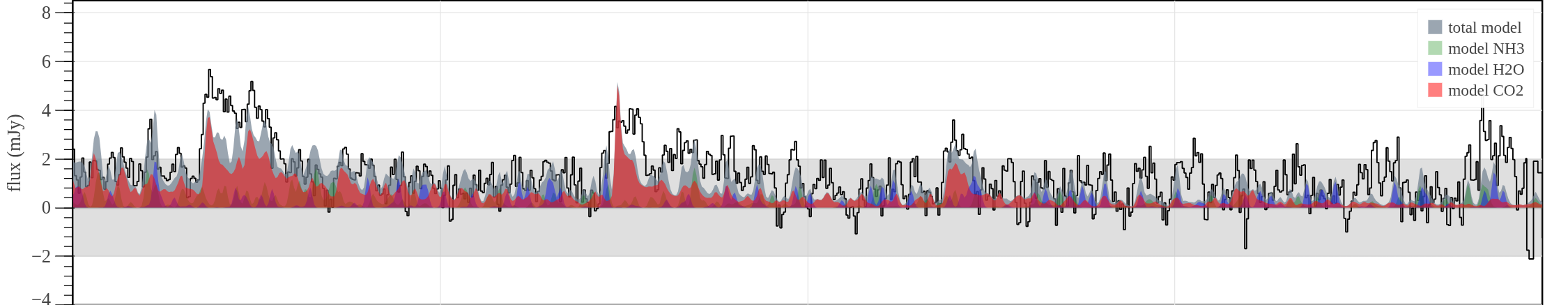}
    \includegraphics[width=0.95\linewidth]{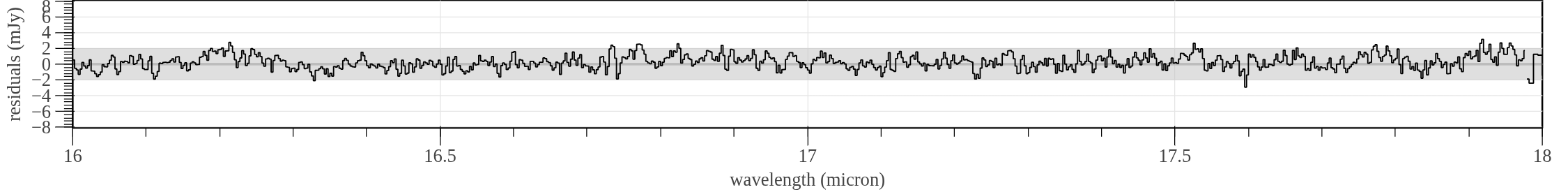}
    \includegraphics[width=0.95\linewidth]{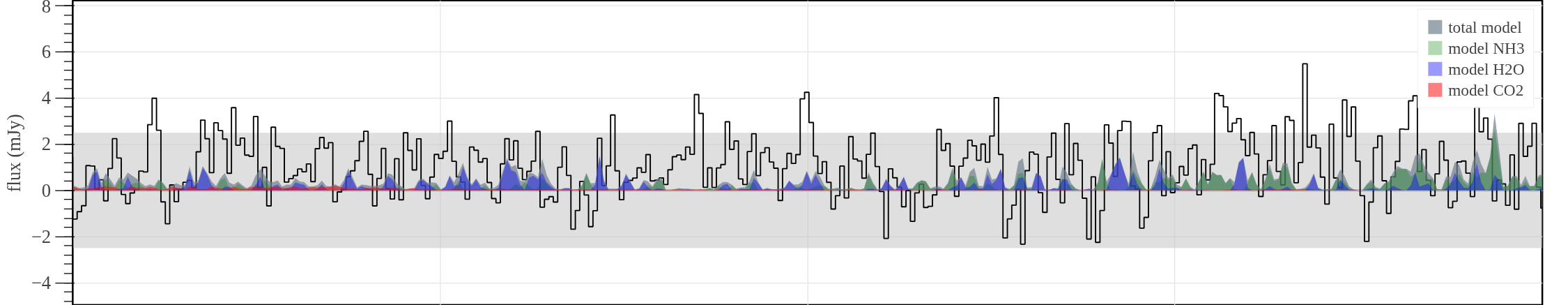}
    \includegraphics[width=0.95\linewidth]{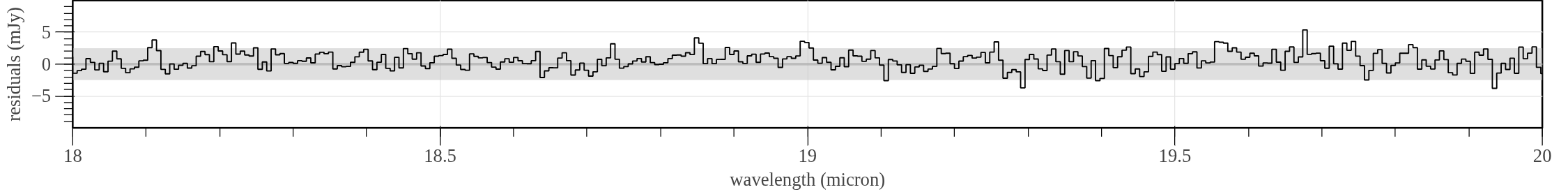}
    \caption{Continue Figure \ref{fig:hd23514_full_contsub}}
    \label{fig:hd23514_full_contsub2}
\end{figure*}

\begin{deluxetable}{lcc}
\tablewidth{0pt}
\tablecaption{Best fits CO$_2$ slab model parameters \label{tab:co2_parameters}}
\tablehead{
\colhead{parameter} & \colhead{$v_{\rm turb}$=0 km\ s$^{-1}$} & \colhead{$v_{\rm turb}$=2 km\ s$^{-1}$} 
}
\startdata
$\log_{10}A$ (au$^2$)      & $-$2.07$\pm$0.02 & $-$2.65$\pm$0.02 \\
$\log_{10}N$ (cm$^{-2}$)   & 18.14$\pm$0.06   & 18.45$\pm$0.06   \\
$\log_{10}T_{\rm gas}$ (K) & 2.95 $\pm$0.02   & 3.01 $\pm$0.02   \\  
\enddata
%\vspace{-0.5cm}
\tablecomments{The uncertainties given in the table are derived from MCMC, which are likely to be underestimated.}
\end{deluxetable}

Although there are hints of other molecules (CO, H$_2$O and NH$_3$) present in the continuum-subtracted spectra (Figure \ref{fig:hd23514_full_contsub}), we did not attempt to determine their precise parameters using the slab models due to high noise and weak lines. Instead, we assumed they share similar model parameters as CO$_2$, and adjust $N$ (and $A$, $T_{\rm gas}$ if necessary) to achieve reasonable fits visually. The mid-IR emission from molecules is due to ro-vibrational or vibrational transitions and is sensitive to their temperatures, i.e., the lines get significantly weaker as the temperature gets cooler. It is then meaningless to explore lower temperatures (or other parameters) without high S/N data. Keeping the same temperature for both H$_2$O and NH$_3$, the column density $N$ needs to be reduced by a factor of $\sim$2.2 and 1.4, respectively, in order to produce similar flux levels of the lines between 6.5 and 9.5 \micron. The low abundances in both H$_2$O and NH$_3$ are consistent with their very short photodissociation lifetimes (Figure \ref{fig:photodisociation_timescale}), by $\gtrsim$2--3 order of magnitudes compared to that of CO$_2$, without shielding. The model parameters for CO, which has the longest photodissociation lifetimes near HD\,23514, are very uncertain due to the limited wavelength coverage from MIRI/MRS. The slab model has great difficulty in reproducing the observed line width if CO has a similar $T_{\rm gas}$ as CO$_2$ (termed low $T$ case in the top panel of Figure \ref{fig:comodel}). Alternatively, the broad width of CO lines can roughly be reproduced by an extra broadening of 200 km\,s$^{-1}$ and increasing $T_{\rm gas}$ to $\sim$3000\,K (termed high $T$ case in the middle panel of Figure \ref{fig:comodel}). Such a high temperature can be reached at a stellocentric distance of 0.015 au around \tgt where its Keplerian velocity ($v_K$) is $\sim$277 km\,s$^{-1}$. This seems to suggest the additional broadening might be related to the Keplerian motion of the CO gas. If the gas is indeed in a disk form, resolving the double-peak profiles in the CO emission could help to constrain the inclination of the system, which should be explored in better future data.  For the low $T_{\rm gas}$ CO model, the associated log$N$ and log$A$ are, respectively, 20.5 cm$^{-2}$ and -2.0 au$^2$, but 17.5 cm$^{-2}$ and -2.5 au$^2$ for the high $T_{\rm gas}$ model, illustrating the sensitivity to the gas temperature. We note that the CO lines are optically thick in both cases, and that the slab model we used does include optical depth effects, i.e., the derived $N$ and $A$ are reliable for the assumed temperature.

We can also estimate the minimum gas mass by assuming the emission is optically thin. Taking the slab model parameters at $v_{\rm turb}$ = 0 km\ s$^{-1}$, the total gas mass for CO$_2$, H$_2$O, and NH$_3$ is 3.21$\times10^{-8}$, 6.05$\times10^{-9}$ and 5.04$\times10^{-9}$ $M_\oplus$, respectively. Note that the masses for H$_2$O and NH$_3$ should be considered as upper limits given the tentative detection. The total mass for CO is 5.54$\times10^{-6}$ $M_\oplus$ for the low $T_{\rm gas}$ case, but 1.75$\times10^{-9}$ $M_\oplus$ for the high $T_{\rm gas}$ case. The total gas mass ranges from 4.5 $\times10^{-8}$ to 5.6$\times10^{-6}$ $M_\oplus$ (spreading two orders of mag, depending on the CO mass). Using the Equ.\,(14) from \citet{birnstiel10}, we can estimate the Stokes number between the gas and small (0.5 \micron) dust by spreading the masses within 0.3 au for simplicity. The Stokes number between the gas and small dust is 0.44 (for the low $T_{\rm gas}$ CO model) or 55 (for the high $T_{\rm gas}$ CO model) assuming a grain density of 3 g\,cm${^{-3}}$. In the former case, the small dust and gas are coupled and the presence of gas would have an impact on dust dynamics, affecting the inward or outward draft rates depending on the gas surface density. In the later case, the tiny amount of gas has no strong effects on dust, i.e., radiation blowout operates on orbital timescales (less than a year in the sub-au region). 

\subsection{Upper Limits on Molecular Hydrogen Gas}
\label{h2_limits}

No H$_2$ lines are detected in the continuum-subtracted spectrum of HD\,23514. We used the local rms estimated around $\pm$50 spectral resolution elements around the potential lines to estimate the 3$\sigma$ upper limits on the integrated line flux assuming the lines are optically thin and unresolved.
Fixing the gas temperature at $T_{\rm gas} =$9 00\,K, we used all the upper limit line fluxes from H$_2$(S1) to H$_2$(S8) to estimate the gas mass and found that H$_2$(S3) line at 9.664 \micron\ (integrated line flux of 3.27$\times10^{-23}$ W\,cm$^{-2}$) gives the best 3-$\sigma$ upper limit mass, 2.25$\times10^{-3}M_{\oplus}$. The best estimated CO mass is 5.54$\times10^{-6}M_{\oplus}$ (taking from the low $T_{\rm gas}=$ 900\,K case), implying the CO-to-H$_2$ mass ratio is $>2.5\times10^{-3}$ which is slightly larger than the typical ISM value (an abundance of 10$^{-4}$ translates to a mass ratio of 1.4$\times10^{-3}$. Although the CO-to-H$_2$ mass ratio cannot unambiguously prove the gas is secondary, the old age of the system and the lack of HI and [Ne II] emission favor nonprimordial.

\subsection{Photodissociation Rate and Molecular Shielding Efficiency}
\label{sec:pf_shielding}

We derived the photodissociation lifetimes for CO, CO$_2$, H$_2$O, and NH$_3$ molecules as a function of stellocentric distance assuming that these molecules are exposed to ultraviolet (UV) photons from both the interstellar radiation field (ISRF) and the star without shielding. The Kurucz model described in Appendix \ref{sec:starsub} and the standard ISRF \citep{draine78,vandishoeck82} were adopted, and the relevant photodissociation cross sections were taken from the online database of the Leiden Observatory\footnote{ \url{https://home.strw.leidenuniv.nl/\~ewine/photo/data/photo\_data/}} \citep{visser09,LeidenLab_pd_calculation}. The result is shown in Figure \ref{fig:photodisociation_timescale}. For all molecules, the UV radiation from the star dominates the photodissociation processes within 10 au. We note that \tgt is located in the outskirts of the Pleiades cluster (far away from the central B-type stars), where the ISRF may be different from that of typical ISM, it does not significantly affect the results within 10 au unless its local ISRF is very different from the assumed one. 

For CO, CO$_2$, H$_2$O, and NH$_3$, UV photons with wavelengths shorter than $\sim$109, $\sim$115, $\sim$180, and $\sim$207 nm, respectively, are involved in the photodissociation of the molecules. Unfortunately, there are no UV observations available for \tgt that would assess the reliability of the adopted Kurucz model at these wavelengths. Since it is a relatively young F5-type star, it is suspected that it may show magnetic activity and thus probably exhibits excess emission in the UV regime. In the case of our Sun, the Kurucz model underestimates the observed emission increasingly toward the shortest wavelengths\footnote{see the measured spectrum from \url{https://home.strw.leidenuniv.nl/\~ewine/photo/data/photo_data/radiation\_fields/solar.dat}}. Therefore, our photodissociation lifetimes derived for HD 23514 should be considered as upper limits.

\begin{figure}
    \centering
    \includegraphics[width=\linewidth]{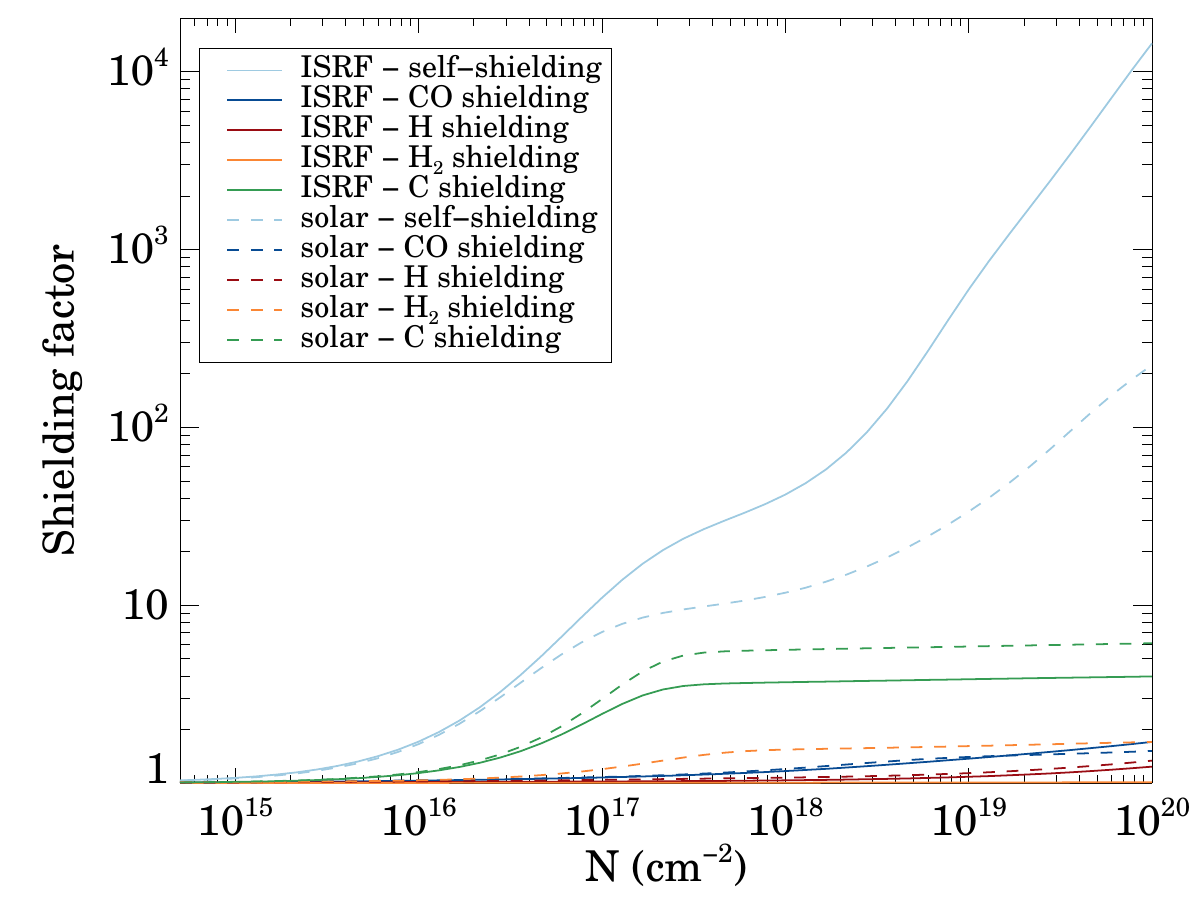}
    \caption{CO$_2$ shielding efficiency (in terms of factors to prolong the photodissociation lifetime) from different molecules (colored) against photodissociation as a function of column density. Solid lines used the ISRF as the radiation field while the dashed ones from the solar radiation.}
    \label{fig:selfshielding}
\end{figure}

UV shielding could occur by the molecule itself (self-shielding), by other molecules and atoms, and by the presence of dust. We investigated the 
shielding factor (efficiency) for CO$_2$ by various gases as a function of the column density of the given gases as shown in Figure \ref{fig:selfshielding}. Two different radiation fields (ISRF and solar) were used along with the molecular properties from the Leiden Observatory database (\url{https://home.strw.leidenuniv.nl/\~ewine/photo/}, \citealt{LeidenLab_pd_calculation}). Shielding from CO$_2$ itself is really efficient once the column density reaches $\sim$10$^{18}$cm$^{-2}$, which can increase its photodissociation lifetime by more than tenfold. The dust component of typical debris disks is so dilute that it cannot contribute significantly to the shielding of the molecules. However, in the case of HD\,23514, it is conceivable that some parts of the dust disk could be optically thick, which could slightly increase the photodissociation lifetime of the molecules. Nevertheless, the photodissociation lifetime of CO$_2$ with self-shielding is still very short, on the order of a few days at 0.05 au. Assuming the gas remains stable over 15 yr between the Spitzer and JWST observation, the shielding from tiny dust needs to be thousandfold more effective than molecular gas shielding.

\subsection{The Nature of the 15 \micron\ bump in the Spitzer/IRS Spectrum}
\label{sec:irs_exam}

We investigated whether the 15 \micron\ bump in the Spitzer spectrum as shown in Figure \ref{fig:co2_irs} could be due to data reduction artifacts by examining the CASSIS IRS spectra of a handful debris systems (e.g., HD\,172555, HD\,15407 and HD\,69830, and others), which are brighter or at a similar brightness as \tgt and observed in the same observing mode, using the CASSIS products. We found no similar bump near 15 \micron, suggesting that the 15 \micron\ bump is likely astrophysical in nature.

\section{Estimating the Dust Location, Blowout Sizes, and Various Timescales}
\label{sec:sedmodel_details}

\begin{figure*}
    \centering
    \includegraphics[width=0.49\linewidth]{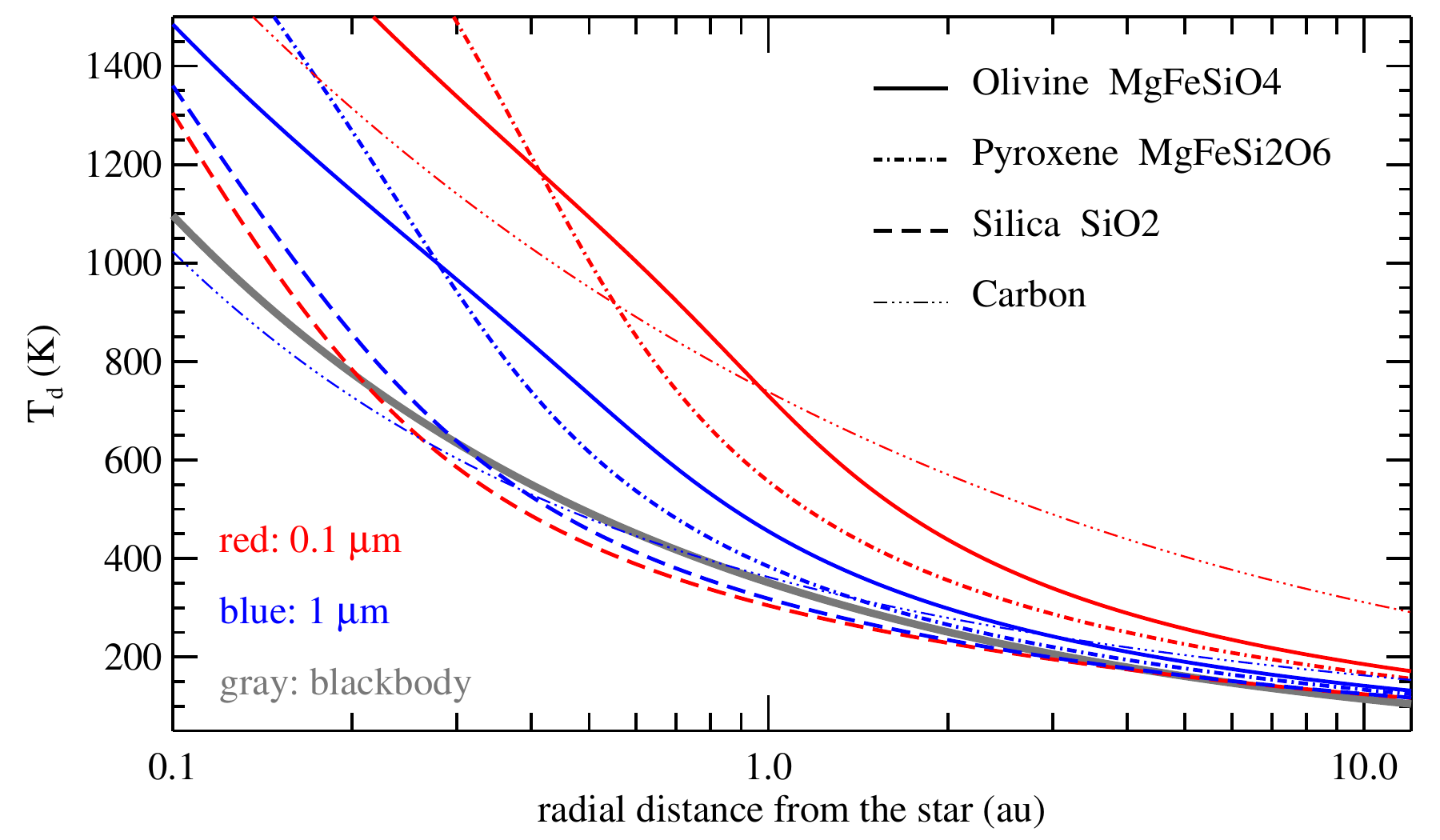}
    \includegraphics[width=0.49\linewidth]{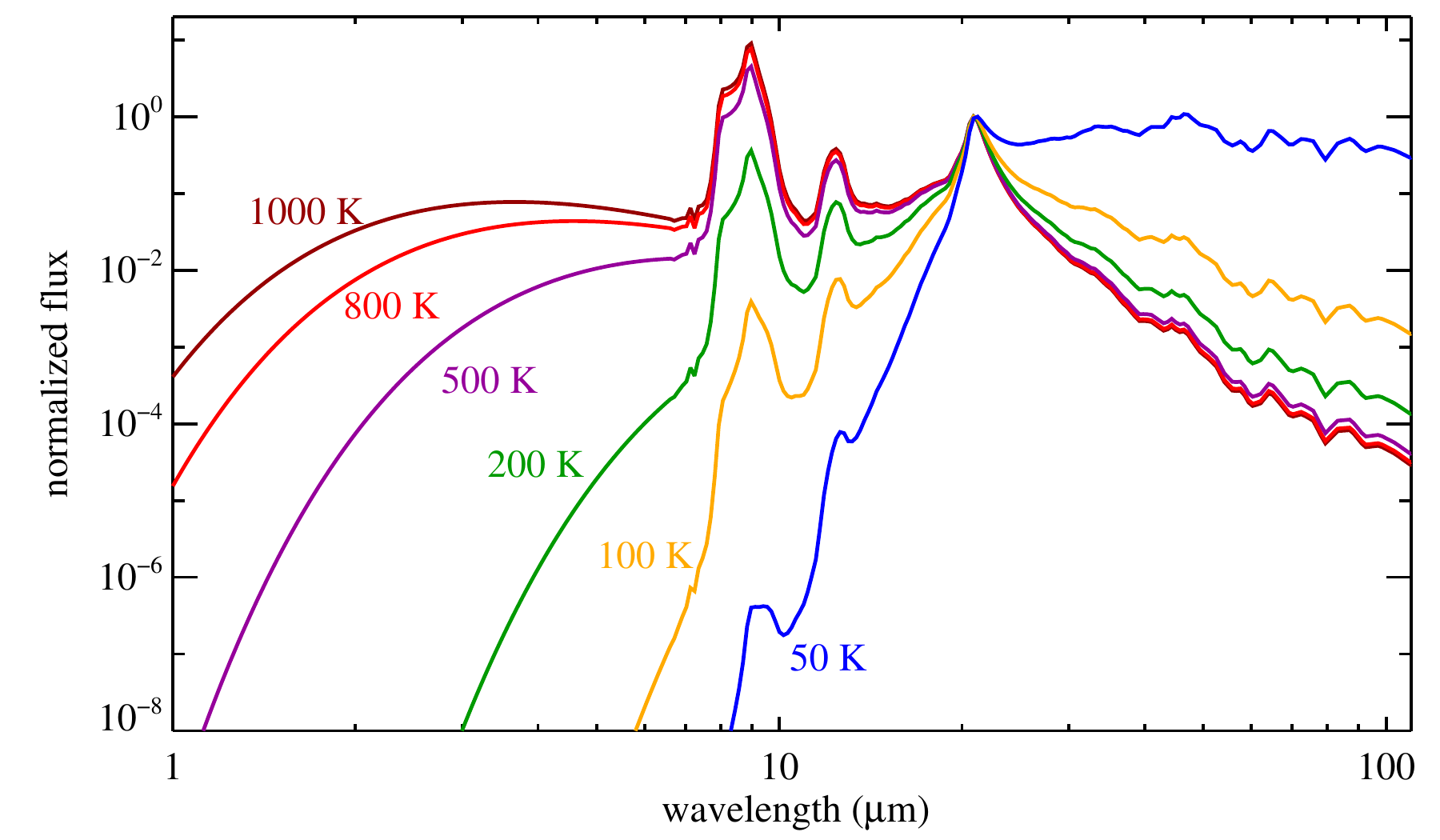}
    \caption{The left panel shows the estimated thermal equilibrium dust temperature as a function of stellocentric distance from \tgt under the optically thin condition. Line styles depict different compositions with red for 0.1 \micron\ and blue for 1 \micron\ grains. The thick gray line represents the temperature of a blackbody emitter. The right panel shows the expected SED for 0.1 \micron\ silica grains at different (colored) temperatures normalized at the 20 \micron\ peak, illustrating that the contrast/ratio between 9 \micron\ and 20 \micron\ is sensitive to the dust temperature.  }
    \label{fig:radialdistribution_dustTd}
\end{figure*}

\subsection{Spectral Energy Distribution Models}
\label{sec:sedmodels}

The purpose of the SED modeling is to estimate the disk extent and to illustrate that the bulk of the silica dust is colocated with the hot molecular gas (from the temperature argument). The derived parameters are then used to estimate various timescales that are relevant in HD\,23514. 

Given the number of assumed parameters, the SED models are extremely degenerate and sensitive to the disk density distribution, and only serve as a zero-order estimate without other constraints like resolved disk images. We stress that the SED modeling results are only applicable to the part of the disk where its bulk emission is fully captured by the SED measurements. The assumptions for our SED models are (1) the star is the only heating source and in an optically thin environment, (2) there is one single, flat (constant surface density), continuous disk with a radial range of inner ($r_{\rm in}$) and outer ($r_{\rm out}$) radii from the star, and (3) the dust grains are all in one uniform size distribution with a size power-law index of $-$3.5 and a maximum size of 1000 \micron. We explored several minimum grain sizes and found that a minimum size of 0.1--0.5 \micron\ silica- and silicate-like grains are needed in order to produce the prominent 10/20 \micron\ features, even though the typical radiation blowout size is $\sim$1 \micron\ given the stellar mass and luminosity (1.3 $M_{\sun}$ and $\sim$3 $L_{\sun}$, respectively). The dynamical stability of these small grains is subject to their exact optical properties in response to the stellar radiation, which is not explored/considered here. We consider that the impact of other parameters such as the maximum grain size ($a_{\rm max}$) and surface density power-law is minimal in a way that (1) $a_{\rm max}$ affects the total dust mass in a $\sqrt{a_{\rm max}}$ fashion under the assumed size distribution, and (2) a steeper surface density distribution would extend $r_{\rm out}$ so enough colder material can account for the total emission. 

A large uncertainty in our SED model resides in the grain composition. The dust temperatures depend sensitively on the grain absorption efficiency at the wavelengths where the star emits most its energy (i.e., UV and optical for HD\,23514). Submicron grains are generally hotter than their larger counterparts and the exact temperatures can be vastly different between different types of composition as shown in the left panel of Figure \ref{fig:radialdistribution_dustTd}. Because the carbonaceous and silica grains are less absorptive than the typical silicates in the UV and optical, they generally have lower temperatures than those of silicates at the same stellocentric distance from the star.

Determining detailed dust mineralogy to infer the exact dust composition is beyond the scope of this short paper. We only explored limited dust compositions that are commonly found in thermally processed dust in circumstellar environments. For the dust emission calculation, we adopted the grain properties computed by \texttt{optool} \citep{dominik21_optool} using the Distribution of Hollow Spheres (DHS, \citealt{min2005}) approach with an irregularity factor $f_{\rm max}$ of 0.8. We found that the system's SED can be well represented by three types of dust: silica (density $\rho$ = 2.65 g\,cm$^{-3}$, \citealt{silica_HM97}), pyroxene ($\rho$ = 3.2 g\,cm$^{-3}$, \citealt{silicates_dorschner95}) and amorphous carbon ($\rho$ = 1.85 g\,cm$^{-3}$, \citealt{zubko96}) where the first two give rise to the prominent dust features in the mid-IR while the amorphous carbon and a small fraction of Mg-rich pyroxene grains contribute the underlining featureless dust continuum as shown in Figure \ref{fig:sedmodel}. A single component disk with $r_{\rm in}$ = 0.1 au and  $r_{\rm out}$ $\sim$2--3 au gives a good fit to the measured SED except that the silica grains need to be restricted within 0.1--0.2 au to account for enough flux in the 3--5 \micron\ range without changing the contrast/ratio between the 10 and 20 \micron\ features (the right panel of Figure \ref{fig:radialdistribution_dustTd}). In other words, having silica grains beyond 1 au would shift the near-IR flux toward the far-IR because of its low thermal temperature outside 1 au, resulting in poor fits between 3 and 5 \micron. We note that the model produces a much narrower 10 \micron\ feature than the observed one, likely due to the mismatch in the various silica polymorphs \citep{koike13}. 

Based on the SED model, the system's IR fractional luminosity ($f_d$) is $\sim$1.76$\times10^{-2}$ with $\sim$70\% coming from the featureless dust continuum. The total dust mass is $\sim$$10^{24}$ g (1.67$\times 10^{-4} M_{\oplus}$) and the feature-producing grains account for $\sim$5\% of the mass. Because the SED models prefer a confined (sub-au) region for the silica grains, the total silica dust mass (integrated to 1 mm in size) is on the order of 1\% ($\sim$10$^{22}$ g) of the total dust mass.  In summary, the bulk of the dust in \tgt is within 3 au and the silica dust is mostly located inside the sub-au region, similar to the location of the hot molecular gas found in Appendix \ref{sec:gas_model}. \tgt was not detected by the ALMA 1.3 mm shallow survey of the Pleiades \citep{sullivan22_alma_pleiades}, suggesting lack of cold ($\lesssim$40 K) dust, although the ALMA data could not exclude a less dusty planetesimal belt like our current Kuiper belt. 
In conclusion, the \tgt system does not possess a massive (tens of au) Kuiper-belt-like analog that is found in $\sim$20\% of solar-like stars \citep{sierchio14,matra25_reasons}. 

\begin{figure}
    \centering
    \includegraphics[width=\linewidth]{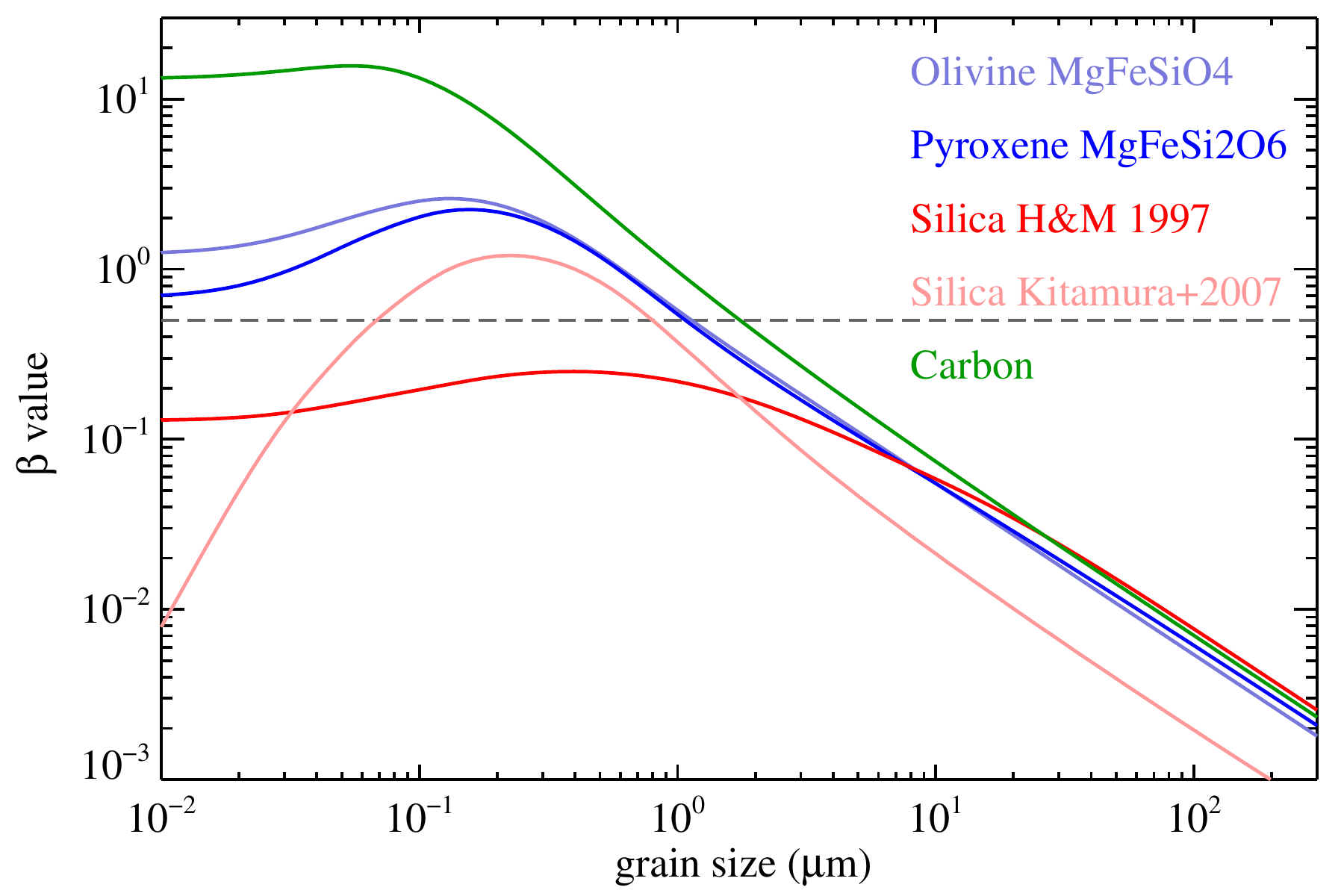}
    \caption{Computed $\beta$ values, the ratio between radiation pressure and gravitational forces, for different sizes and compositions of ideal dust grains around HD\,23514. The horizontal dashed line marks $\beta=0.5$ where the blowout size is defined and grains above the line are subject to radiation-pressure blowout (details see Appendix \ref{sec:betavalues}).}
    \label{fig:betvalues}
\end{figure}

\subsection{Radiation-pressure Blowout Sizes around HD\,23514 }
\label{sec:betavalues}

In gas-free debris disks, stellar radiation pressure is one of the most important forces acting on grains that counters with the star's gravity and dictates the dust dynamics, which can be characterized by the $\beta$ parameter defined as the ratio of radiation pressure to gravitational forces: 
\begin{equation}
    \beta = \frac{3 \langle Q_{\rm pr}\rangle L_{\ast} }{ 16 \pi G c M_{\ast} a \rho }
\end{equation} where $L_{\ast}$ and $M_{\ast}$ are, respectively, the luminosity and mass of the star, $a$ is the grain size, $\rho$ is the grain density, and $ \langle Q_{\rm pr}\rangle $ is the radiation-pressure efficiency averaged over stellar spectrum \citep{burns79}. The last three parameters have strong dependency on the exact grain properties (composition, size, density, and porosity, as detailed in \citealt{arnold19}). 
Grains with $\beta > 0.5$ get pushed into hyperbolic orbits and leave the system, while grains with $\beta < 0.5$ are on elliptical orbits and spiral inward due to Poynting-Robertson (PR) drag. The radiation-pressure blowout size is defined as $\beta = 0.5$, and the exact size depends on the type of star, and most importantly, the grain properties as illustrated in Figure \ref{fig:betvalues}. The $\beta$ values are calculated using the stellar Kurucz model described in Appendix \ref{sec:starsub}, the grain compositions adopted in Appendix \ref{sec:sedmodels} with two additions (olivine: $\rho=$ 3.71 from \citealt{silicates_dorschner95}, and silica: $\rho=$ 2.65 from \citealt{kitamura07}), and assuming compact (i.e., ideal with zero porosity) grain shape. For nominal silicate and carbonaceous compositions, grains smaller than the blowout size ($\sim$1 \micron\ for HD\,23514) are ejected at a terminal velocity of $\sqrt{2(\beta - \frac{1}{2}}) v_K$ \citep{su05}. While for silica composition, some (such as the one with the optical constant from \citealt{silica_HM97}) could have $\beta<$ 0.5 for all sizes, and some (from \citealt{kitamura07}) would have $\beta >$0.5 grains within two blowout sizes. This also illustrates that blowout size is not a strict number, highly depending on the exact composition.  
We stress the importance of grain optical properties in the UV and optical wavelengths where the star emits most of its energy in estimating the grain's $\beta$ values.

\subsection{Relevant Timescales}
\label{sec:tau_c}

For a dust-dominated debris disk, the dominant grain loss mechanisms are collisional grinding and PR/stellar wind drags. Once the grains generated by collisional cascades are smaller than the radiation blowout size ($\sim$1 \micron\ in \tgt when the ratio between the radiation pressure force and gravity, defined as $\beta$, equal to 0.5), they are ejected from the system by radiation pressure under normal conditions. The blowout timescale is on the order of the orbital period, which is relatively fast (1.3 week at 0.1 au, 0.8 yr at 1 au, and 4.5 yr at 3 au around an 1.3 $M_\sun$ star), although the the presence of overabundant submicron grains suggests that such rapid removal is not taking place in this system, likely due to the presence of gas. HD\,23514's chromospheric activity index \citep{fang18_lamostII} indicates that its stellar wind should not be exceptionally strong, suggesting the drag force is dominated by the PR drag with timescale $\tau_{PR} = c r^2/ (4 G M_{\ast} \beta) \sim 800\ {\rm yr}\ (r/{\rm au})^2 (M_{\sun}/M_{\ast})(0.5/\beta)$ where $c$ is the speed of light, $G$ is the gravitational constant, and $M_{\ast}$ is the stellar mass. Within 1 au, the PR timescale is a few 100s yr in\,HD 23514. 

Collisional grinding operates in a cascade fashion among different sizes of particles with a loss time that depends on the properties of the particle swarm (belt location and width, total mass, and the size of the particle) and stellar mass. Continuous grindings supply enough new grains against the loss from radiation blowout and PR drag, therefore sustaining the dust level over the collisional timescale ($\tau_c$). The collisional timescale can be estimated as formulated by Equation (13) in \citealt{wyatt07a} where the largest unknown properties are the total disk mass ($M_{\rm tot}$), the maximum size ($D_c$) of the planetesimals, and belt location ($r$). 

Based on the SED modeling described above, we can set $r\sim$3 au with a width $dr$ = 0.1\,$r$ for simplicity. If we take the SED-derived dust mass (up to 1 mm) at face value, $\tau_c$ is $\sim$500 yr for $D_c$ of 1 mm. However, if we extrapolate the dust mass to 1 km size planetesimals (i.e., 0.167 $M_\oplus$ as the disk mass), $\tau_c \sim$5$\times10^{5}$ yr in this case. The collisional timescale is inversely proportional to the total disk mass if extrapolated to larger sizes of planetesimals (by a factor of $\sqrt{a_{\rm max}}$), and is directly related to the sizes of colliding bodies. If the disk in HD\,23514 formed by a past giant impact at 3 au, it would remain IR bright for $\gtrsim$10$^{3-4}$ yr, consistent with previous studies \citep{jackson12}. 

An alternative way to support the giant-impact hypothesis for \tgt is to examine the mass-loss rate and backtrack $M_{\rm tot}$. From the SED model, we estimate the dust mass for the small grains, the ones that give rise to the solid-state feature, is $\sim5\times10^{22}$ g. These $\lesssim$1 \micron\ grains are expected to be ejected from the system on the blowout timescale, which is $\lesssim$1 yr in HD\,23514. Sustaining this high mass-loss rate through collisional cascades would requires a total planetesimal mass of $\sim$1260 $M_\oplus$ over the age of the star (150 Myr) without counting the loss from PR drag. This would require that the HD\,23514 system is 35 times more massive than the MMSN at birth (assuming 100:1 gas-to-dust mass ratio). It is then unlikely such a high rate has been sustained over the age of the system, i.e., the high mass-loss rate is transient, consistent with the giant-impact hypothesis.

\facilities{JWST (MIRI), Spitzer, WISE}

\software{\texttt{astropy} \citep{astropy}, \texttt{scipy} \citep{scipy}}

\begin{acknowledgments}
Work on this paper was supported by grants 80NSSC18K0555 from NASA Goddard Space Flight Center to the University of Arizona, and by 80NSSC20K1002 under the NASA ADAP program. A.M. is supported by the Hungarian National Research, Development and Innovation Ofﬁce \'Elvonal grant KKP-143986. C.X. and I.P. acknowledge partial support by NASA STScI GO grant JWST-GO-02970.010. A.K. acknowledges the support by the NKFIH NKKP grant ADVANCED 149943 and the NKFIH excellence grant TKP2021-NKTA-64. Project no. 149943 has been implemented with the support provided by the Ministry of Culture and Innovation of Hungary from the National Research, Development and Innovation Fund, financed under the NKKP ADVANCED funding scheme. P.A. is partially supported by the Hungarian NKFIH grant K-147380. L.M. and Z.R. acknowledges funding by the European Union through the E-BEANS ERC project (grant number 100117693). Views and opinions expressed are however those of the author(s) only and do not necessarily reflect those of the European Union or the European Research Council Executive Agency. Neither the European Union nor the granting authority can be held responsible for them.

This research has made use of the NASA/IPAC Infrared
Science Archive, which is funded by the National Aeronautics
and Space Administration and operated by the California
Institute of Technology.

The data presented in this article were obtained from the
Mikulski Archive for Space Telescopes (MAST) at the Space
Telescope Science Institute. These observations are associated
with JWST program 1206. The standard pipeline-processed products can be accessed via doi:10.17909/qwmm-1c59.

\end{acknowledgments}

\bibliography{ksuref}{}

\begin{thebibliography}{}
\expandafter\ifx\csname natexlab\endcsname\relax\def\natexlab#1{#1}\fi
\providecommand{\url}[1]{\href{#1}{#1}}
\providecommand{\dodoi}[1]{doi:~\href{http://doi.org/#1}{\nolinkurl{#1}}}
\providecommand{\doeprint}[1]{\href{http://ascl.net/#1}{\nolinkurl{http://ascl.net/#1}}}
\providecommand{\doarXiv}[1]{\href{https://arxiv.org/abs/#1}{\nolinkurl{https://arxiv.org/abs/#1}}}

\bibitem[{{Abe} \& {Matsui}(1985)}]{abe_matsui85}
{Abe}, Y., \& {Matsui}, T. 1985, Lunar and Planetary Science Conference
  Proceedings, 90, C545

\bibitem[{{Arnold} {et~al.}(2019){Arnold}, {Weinberger}, {Videen}, \&
  {Zubko}}]{arnold19}
{Arnold}, J.~A., {Weinberger}, A.~J., {Videen}, G., \& {Zubko}, E.~S. 2019,
  \aj, 157, 157, \dodoi{10.3847/1538-3881/ab095e}

\bibitem[{{Artymowicz}(1988)}]{artymowicz88}
{Artymowicz}, P. 1988, \apjl, 335, L79, \dodoi{10.1086/185344}

\bibitem[{{Astropy Collaboration} {et~al.}(2022){Astropy Collaboration},
  {Price-Whelan}, {Lim}, {Earl}, {Starkman}, {Bradley}, {Shupe}, {Patil},
  {Corrales}, {Brasseur}, {N{\"o}the}, {Donath}, {Tollerud}, {Morris},
  {Ginsburg}, {Vaher}, {Weaver}, {Tocknell}, {Jamieson}, {van Kerkwijk},
  {Robitaille}, {Merry}, {Bachetti}, {G{\"u}nther}, {Aldcroft},
  {Alvarado-Montes}, {Archibald}, {B{\'o}di}, {Bapat}, {Barentsen},
  {Baz{\'a}n}, {Biswas}, {Boquien}, {Burke}, {Cara}, {Cara}, {Conroy},
  {Conseil}, {Craig}, {Cross}, {Cruz}, {D'Eugenio}, {Dencheva}, {Devillepoix},
  {Dietrich}, {Eigenbrot}, {Erben}, {Ferreira}, {Foreman-Mackey}, {Fox},
  {Freij}, {Garg}, {Geda}, {Glattly}, {Gondhalekar}, {Gordon}, {Grant},
  {Greenfield}, {Groener}, {Guest}, {Gurovich}, {Handberg}, {Hart},
  {Hatfield-Dodds}, {Homeier}, {Hosseinzadeh}, {Jenness}, {Jones}, {Joseph},
  {Kalmbach}, {Karamehmetoglu}, {Ka{\l}uszy{\'n}ski}, {Kelley}, {Kern},
  {Kerzendorf}, {Koch}, {Kulumani}, {Lee}, {Ly}, {Ma}, {MacBride}, {Maljaars},
  {Muna}, {Murphy}, {Norman}, {O'Steen}, {Oman}, {Pacifici}, {Pascual},
  {Pascual-Granado}, {Patil}, {Perren}, {Pickering}, {Rastogi}, {Roulston},
  {Ryan}, {Rykoff}, {Sabater}, {Sakurikar}, {Salgado}, {Sanghi}, {Saunders},
  {Savchenko}, {Schwardt}, {Seifert-Eckert}, {Shih}, {Jain}, {Shukla}, {Sick},
  {Simpson}, {Singanamalla}, {Singer}, {Singhal}, {Sinha}, {Sip{\H{o}}cz},
  {Spitler}, {Stansby}, {Streicher}, {{\v{S}}umak}, {Swinbank}, {Taranu},
  {Tewary}, {Tremblay}, {de Val-Borro}, {Van Kooten}, {Vasovi{\'c}}, {Verma},
  {de Miranda Cardoso}, {Williams}, {Wilson}, {Winkel}, {Wood-Vasey}, {Xue},
  {Yoachim}, {Zhang}, {Zonca}, \& {Astropy Project Contributors}}]{astropy}
{Astropy Collaboration}, {Price-Whelan}, A.~M., {Lim}, P.~L., {et~al.} 2022,
  \apj, 935, 167, \dodoi{10.3847/1538-4357/ac7c74}

\bibitem[{{Bedding} {et~al.}(2023){Bedding}, {Murphy}, {Crawford}, {Hey},
  {Huber}, {Kjeldsen}, {Li}, {Mann}, {Torres}, {White}, \&
  {Zhou}}]{bedding23_TESS_Pleiades}
{Bedding}, T.~R., {Murphy}, S.~J., {Crawford}, C., {et~al.} 2023, \apjl, 946,
  L10, \dodoi{10.3847/2041-8213/acc17a}

\bibitem[{{Beust} {et~al.}(1996){Beust}, {Lagrange}, {Plazy}, \&
  {Mouillet}}]{beust96_betapic_FEBs}
{Beust}, H., {Lagrange}, A.~M., {Plazy}, F., \& {Mouillet}, D. 1996, \aap, 310,
  181

\bibitem[{{Birnstiel} {et~al.}(2010){Birnstiel}, {Dullemond}, \&
  {Brauer}}]{birnstiel10}
{Birnstiel}, T., {Dullemond}, C.~P., \& {Brauer}, F. 2010, \aap, 513, A79,
  \dodoi{10.1051/0004-6361/200913731}

\bibitem[{{Bonsor} {et~al.}(2023){Bonsor}, {Wyatt}, {Marino}, {Davidsson},
  {Kral}, \& {Thebault}}]{bonsor23_debrisgas}
{Bonsor}, A., {Wyatt}, M.~C., {Marino}, S., {et~al.} 2023, \mnras, 526, 3115,
  \dodoi{10.1093/mnras/stad2912}

\bibitem[{{Bosman} {et~al.}(2017){Bosman}, {Bruderer}, \& {van
  Dishoeck}}]{bosman17}
{Bosman}, A.~D., {Bruderer}, S., \& {van Dishoeck}, E.~F. 2017, \aap, 601, A36,
  \dodoi{10.1051/0004-6361/201629946}

\bibitem[{{Burns} {et~al.}(1979){Burns}, {Lamy}, \& {Soter}}]{burns79}
{Burns}, J.~A., {Lamy}, P.~L., \& {Soter}, S. 1979, \icarus, 40, 1,
  \dodoi{10.1016/0019-1035(79)90050-2}

\bibitem[{{Bushouse} {et~al.}(2024){Bushouse}, {Eisenhamer}, {Dencheva},
  {Davies}, {Greenfield}, {Morrison}, {Hodge}, {Simon}, {Grumm}, {Droettboom},
  {Slavich}, {Sosey}, {Pauly}, {Miller}, {Jedrzejewski}, {Hack}, {Davis},
  {Crawford}, {Law}, {Gordon}, {Regan}, {Cara}, {MacDonald}, {Bradley},
  {Shanahan}, {Jamieson}, {Teodoro}, {Williams}, \&
  {Pena-Guerrero}}]{bushouse24_jwstpipeline}
{Bushouse}, H., {Eisenhamer}, J., {Dencheva}, N., {et~al.} 2024, {JWST
  Calibration Pipeline}, 1.15.1,  Zenodo, \dodoi{10.5281/zenodo.12692459}

\bibitem[{{Canup}(2004)}]{canup04_simulations}
{Canup}, R.~M. 2004, \icarus, 168, 433, \dodoi{10.1016/j.icarus.2003.09.028}

\bibitem[{{Cutri} {et~al.}(2012){Cutri}, {Wright}, {Conrow}, {Bauer},
  {Benford}, {Brandenburg}, {Dailey}, {Eisenhardt}, {Evans}, {Fajardo-Acosta},
  {Fowler}, {Gelino}, {Grillmair}, {Harbut}, {Hoffman}, {Jarrett},
  {Kirkpatrick}, {Leisawitz}, {Liu}, {Mainzer}, {Marsh}, {Masci}, {McCallon},
  {Padgett}, {Ressler}, {Royer}, {Skrutskie}, {Stanford}, {Wyatt}, {Tholen},
  {Tsai}, {Wachter}, {Wheelock}, {Yan}, {Alles}, {Beck}, {Grav}, {Masiero},
  {McCollum}, {McGehee}, {Papin}, \& {Wittman}}]{cutri12_wisemission}
{Cutri}, R.~M., {Wright}, E.~L., {Conrow}, T., {et~al.} 2012, {Explanatory
  Supplement to the WISE All-Sky Data Release Products}, Explanatory Supplement
  to the WISE All-Sky Data Release Products

\bibitem[{{de Wit} {et~al.}(2013){de Wit}, {Grinin}, {Potravnov},
  {Shakhovskoi}, {M{\"u}ller}, \& {Moerchen}}]{dewit13_rzpsc}
{de Wit}, W.~J., {Grinin}, V.~P., {Potravnov}, I.~S., {et~al.} 2013, \aap, 553,
  L1, \dodoi{10.1051/0004-6361/201220715}

\bibitem[{{Dent} {et~al.}(2014){Dent}, {Wyatt}, {Roberge}, {Augereau},
  {Casassus}, {Corder}, {Greaves}, {de Gregorio-Monsalvo}, {Hales}, {Jackson},
  {Hughes}, {Lagrange}, {Matthews}, \& {Wilner}}]{dent14_betapic}
{Dent}, W.~R.~F., {Wyatt}, M.~C., {Roberge}, A., {et~al.} 2014, Science, 343,
  1490, \dodoi{10.1126/science.1248726}

\bibitem[{{Dominik} {et~al.}(2021){Dominik}, {Min}, \&
  {Tazaki}}]{dominik21_optool}
{Dominik}, C., {Min}, M., \& {Tazaki}, R. 2021, {OpTool: Command-line driven
  tool for creating complex dust opacities}, Astrophysics Source Code Library,
  record ascl:2104.010.
\newblock \doeprint{2104.010}

\bibitem[{{Dorschner} {et~al.}(1995){Dorschner}, {Begemann}, {Henning},
  {Jaeger}, \& {Mutschke}}]{silicates_dorschner95}
{Dorschner}, J., {Begemann}, B., {Henning}, T., {Jaeger}, C., \& {Mutschke}, H.
  1995, \aap, 300, 503

\bibitem[{{Draine}(1978)}]{draine78}
{Draine}, B.~T. 1978, \apjs, 36, 595, \dodoi{10.1086/190513}

\bibitem[{{Fang} {et~al.}(2018){Fang}, {Zhao}, {Zhao}, \& {Bharat
  Kumar}}]{fang18_lamostII}
{Fang}, X.-S., {Zhao}, G., {Zhao}, J.-K., \& {Bharat Kumar}, Y. 2018, \mnras,
  476, 908, \dodoi{10.1093/mnras/sty212}

\bibitem[{{Fu} {et~al.}(2022){Fu}, {Bragaglia}, {Liu}, {Zhang}, {Xu}, {Wang},
  {Zhang}, {Zhong}, {Chang}, {Li}, {Chen}, {Chen}, {Wang}, {Gjergo}, {Wang},
  {Yue}, \& {Zhang}}]{fu22_lamost}
{Fu}, X., {Bragaglia}, A., {Liu}, C., {et~al.} 2022, \aap, 668, A4,
  \dodoi{10.1051/0004-6361/202243590}

\bibitem[{{Fujiwara} {et~al.}(2012){Fujiwara}, {Onaka}, {Yamashita},
  {Ishihara}, {Kataza}, {Fukagawa}, {Takeda}, \& {Murakami}}]{fujiwara12}
{Fujiwara}, H., {Onaka}, T., {Yamashita}, T., {et~al.} 2012, \apjl, 749, L29,
  \dodoi{10.1088/2041-8205/749/2/L29}

\bibitem[{{Furukawa} {et~al.}(2007){Furukawa}, {Nakazawa}, {Sekine}, \&
  {Kakegawa}}]{furukawa07_UPs}
{Furukawa}, Y., {Nakazawa}, H., {Sekine}, T., \& {Kakegawa}, T. 2007, Earth and
  Planetary Science Letters, 258, 543, \dodoi{10.1016/j.epsl.2007.04.014}

\bibitem[{{Gaia Collaboration} {et~al.}(2021){Gaia Collaboration}, {Brown},
  {Vallenari}, {Prusti}, {de Bruijne}, {Babusiaux}, {Biermann}, {Creevey},
  {Evans}, {Eyer}, {Hutton}, {Jansen}, {Jordi}, {Klioner}, {Lammers},
  {Lindegren}, {Luri}, {Mignard}, {Panem}, {Pourbaix}, {Randich}, {Sartoretti},
  {Soubiran}, {Walton}, {Arenou}, {Bailer-Jones}, {Bastian}, {Cropper},
  {Drimmel}, {Katz}, {Lattanzi}, {van Leeuwen}, {Bakker}, {Cacciari},
  {Casta{\~n}eda}, {De Angeli}, {Ducourant}, {Fabricius}, {Fouesneau},
  {Fr{\'e}mat}, {Guerra}, {Guerrier}, {Guiraud}, {Jean-Antoine Piccolo},
  {Masana}, {Messineo}, {Mowlavi}, {Nicolas}, {Nienartowicz}, {Pailler},
  {Panuzzo}, {Riclet}, {Roux}, {Seabroke}, {Sordo}, {Tanga}, {Th{\'e}venin},
  {Gracia-Abril}, {Portell}, {Teyssier}, {Altmann}, {Andrae}, {Bellas-Velidis},
  {Benson}, {Berthier}, {Blomme}, {Brugaletta}, {Burgess}, {Busso}, {Carry},
  {Cellino}, {Cheek}, {Clementini}, {Damerdji}, {Davidson}, {Delchambre},
  {Dell'Oro}, {Fern{\'a}ndez-Hern{\'a}ndez}, {Galluccio}, {Garc{\'\i}a-Lario},
  {Garcia-Reinaldos}, {Gonz{\'a}lez-N{\'u}{\~n}ez}, {Gosset}, {Haigron},
  {Halbwachs}, {Hambly}, {Harrison}, {Hatzidimitriou}, {Heiter},
  {Hern{\'a}ndez}, {Hestroffer}, {Hodgkin}, {Holl}, {Jan{\ss}en}, {Jevardat de
  Fombelle}, {Jordan}, {Krone-Martins}, {Lanzafame}, {L{\"o}ffler}, {Lorca},
  {Manteiga}, {Marchal}, {Marrese}, {Moitinho}, {Mora}, {Muinonen}, {Osborne},
  {Pancino}, {Pauwels}, {Petit}, {Recio-Blanco}, {Richards}, {Riello},
  {Rimoldini}, {Robin}, {Roegiers}, {Rybizki}, {Sarro}, {Siopis}, {Smith},
  {Sozzetti}, {Ulla}, {Utrilla}, {van Leeuwen}, {van Reeven}, {Abbas}, {Abreu
  Aramburu}, {Accart}, {Aerts}, {Aguado}, {Ajaj}, {Altavilla}, {{\'A}lvarez},
  {{\'A}lvarez Cid-Fuentes}, {Alves}, {Anderson}, {Anglada Varela}, {Antoja},
  {Audard}, {Baines}, {Baker}, {Balaguer-N{\'u}{\~n}ez}, {Balbinot}, {Balog},
  {Barache}, {Barbato}, {Barros}, {Barstow}, {Bartolom{\'e}}, {Bassilana},
  {Bauchet}, {Baudesson-Stella}, {Becciani}, {Bellazzini}, {Bernet}, {Bertone},
  {Bianchi}, {Blanco-Cuaresma}, {Boch}, {Bombrun}, {Bossini}, {Bouquillon},
  {Bragaglia}, {Bramante}, {Breedt}, {Bressan}, {Brouillet}, {Bucciarelli},
  {Burlacu}, {Busonero}, {Butkevich}, {Buzzi}, {Caffau}, {Cancelliere},
  {C{\'a}novas}, {Cantat-Gaudin}, {Carballo}, {Carlucci}, {Carnerero},
  {Carrasco}, {Casamiquela}, {Castellani}, {Castro-Ginard}, {Castro Sampol},
  {Chaoul}, {Charlot}, {Chemin}, {Chiavassa}, {Cioni}, {Comoretto}, {Cooper},
  {Cornez}, {Cowell}, {Crifo}, {Crosta}, {Crowley}, {Dafonte}, {Dapergolas},
  {David}, {David}, {de Laverny}, {De Luise}, {De March}, {De Ridder}, {de
  Souza}, {de Teodoro}, {de Torres}, {del Peloso}, {del Pozo}, {Delbo},
  {Delgado}, {Delgado}, {Delisle}, {Di Matteo}, {Diakite}, {Diener},
  {Distefano}, {Dolding}, {Eappachen}, {Edvardsson}, {Enke}, {Esquej}, {Fabre},
  {Fabrizio}, {Faigler}, {Fedorets}, {Fernique}, {Fienga}, {Figueras},
  {Fouron}, {Fragkoudi}, {Fraile}, {Franke}, {Gai}, {Garabato},
  {Garcia-Gutierrez}, {Garc{\'\i}a-Torres}, {Garofalo}, {Gavras}, {Gerlach},
  {Geyer}, {Giacobbe}, {Gilmore}, {Girona}, {Giuffrida}, {Gomel}, {Gomez},
  {Gonzalez-Santamaria}, {Gonz{\'a}lez-Vidal}, {Granvik},
  {Guti{\'e}rrez-S{\'a}nchez}, {Guy}, {Hauser}, {Haywood}, {Helmi}, {Hidalgo},
  {Hilger}, {H{\l}adczuk}, {Hobbs}, {Holland}, {Huckle}, {Jasniewicz},
  {Jonker}, {Juaristi Campillo}, {Julbe}, {Karbevska}, {Kervella}, {Khanna},
  {Kochoska}, {Kontizas}, {Kordopatis}, {Korn}, {Kostrzewa-Rutkowska},
  {Kruszy{\'n}ska}, {Lambert}, {Lanza}, {Lasne}, {Le Campion}, {Le Fustec},
  {Lebreton}, {Lebzelter}, {Leccia}, {Leclerc}, {Lecoeur-Taibi}, {Liao},
  {Licata}, {Lindstr{\o}m}, {Lister}, {Livanou}, {Lobel}, {Madrero Pardo},
  {Managau}, {Mann}, {Marchant}, {Marconi}, {Marcos Santos}, {Marinoni},
  {Marocco}, {Marshall}, {Martin Polo}, {Mart{\'\i}n-Fleitas}, {Masip},
  {Massari}, {Mastrobuono-Battisti}, {Mazeh}, {McMillan}, {Messina},
  {Michalik}, {Millar}, {Mints}, {Molina}, {Molinaro}, {Moln{\'a}r},
  {Montegriffo}, {Mor}, {Morbidelli}, {Morel}, {Morris}, {Mulone}, {Munoz},
  {Muraveva}, {Murphy}, {Musella}, {Noval}, {Ord{\'e}novic}, {Orr{\`u}},
  {Osinde}, {Pagani}, {Pagano}, {Palaversa}, {Palicio}, {Panahi}, {Pawlak},
  {Pe{\~n}alosa Esteller}, {Penttil{\"a}}, {Piersimoni}, {Pineau}, {Plachy},
  {Plum}, {Poggio}, {Poretti}, {Poujoulet}, {Pr{\v{s}}a}, {Pulone}, {Racero},
  {Ragaini}, {Rainer}, {Raiteri}, {Rambaux}, {Ramos}, {Ramos-Lerate}, {Re
  Fiorentin}, {Regibo}, {Reyl{\'e}}, {Ripepi}, {Riva}, {Rixon}, {Robichon},
  {Robin}, {Roelens}, {Rohrbasser}, {Romero-G{\'o}mez}, {Rowell}, {Royer},
  {Rybicki}, {Sadowski}, {Sagrist{\`a} Sell{\'e}s}, {Sahlmann}, {Salgado},
  {Salguero}, {Samaras}, {Sanchez Gimenez}, {Sanna}, {Santove{\~n}a},
  {Sarasso}, {Schultheis}, {Sciacca}, {Segol}, {Segovia}, {S{\'e}gransan},
  {Semeux}, {Shahaf}, {Siddiqui}, {Siebert}, {Siltala}, {Slezak}, {Smart},
  {Solano}, {Solitro}, {Souami}, {Souchay}, {Spagna}, {Spoto}, {Steele},
  {Steidelm{\"u}ller}, {Stephenson}, {S{\"u}veges}, {Szabados}, {Szegedi-Elek},
  {Taris}, {Tauran}, {Taylor}, {Teixeira}, {Thuillot}, {Tonello}, {Torra},
  {Torra}, {Turon}, {Unger}, {Vaillant}, {van Dillen}, {Vanel}, {Vecchiato},
  {Viala}, {Vicente}, {Voutsinas}, {Weiler}, {Wevers}, {Wyrzykowski}, {Yoldas},
  {Yvard}, {Zhao}, {Zorec}, {Zucker}, {Zurbach}, \& {Zwitter}}]{gaia_edr3}
{Gaia Collaboration}, {Brown}, A.~G.~A., {Vallenari}, A., {et~al.} 2021, \aap,
  649, A1, \dodoi{10.1051/0004-6361/202039657}

\bibitem[{{Gaidos} {et~al.}(2019){Gaidos}, {Jacobs}, {LaCourse}, {Vanderburg},
  {Rappaport}, {Berger}, {Pearce}, {Mann}, {Weiss}, {Fulton}, {Behmard},
  {Howard}, {Ansdell}, {Ricker}, {Vanderspek}, {Latham}, {Seager}, {Winn}, \&
  {Jenkins}}]{gaidos19_hd240779}
{Gaidos}, E., {Jacobs}, T., {LaCourse}, D., {et~al.} 2019, \mnras, 488, 4465,
  \dodoi{10.1093/mnras/stz1942}

\bibitem[{{Gordon} {et~al.}(2022){Gordon}, {Rothman}, {Hargreaves}, {Hashemi},
  {Karlovets}, {Skinner}, {Conway}, {Hill}, {Kochanov}, {Tan}, {Wcis{\l}o},
  {Finenko}, {Nelson}, {Bernath}, {Birk}, {Boudon}, {Campargue}, {Chance},
  {Coustenis}, {Drouin}, {Flaud}, {Gamache}, {Hodges}, {Jacquemart}, {Mlawer},
  {Nikitin}, {Perevalov}, {Rotger}, {Tennyson}, {Toon}, {Tran}, {Tyuterev},
  {Adkins}, {Baker}, {Barbe}, {Can{\`e}}, {Cs{\'a}sz{\'a}r}, {Dudaryonok},
  {Egorov}, {Fleisher}, {Fleurbaey}, {Foltynowicz}, {Furtenbacher}, {Harrison},
  {Hartmann}, {Horneman}, {Huang}, {Karman}, {Karns}, {Kassi}, {Kleiner},
  {Kofman}, {Kwabia-Tchana}, {Lavrentieva}, {Lee}, {Long}, {Lukashevskaya},
  {Lyulin}, {Makhnev}, {Matt}, {Massie}, {Melosso}, {Mikhailenko}, {Mondelain},
  {M{\"u}ller}, {Naumenko}, {Perrin}, {Polyansky}, {Raddaoui}, {Raston},
  {Reed}, {Rey}, {Richard}, {T{\'o}bi{\'a}s}, {Sadiek}, {Schwenke},
  {Starikova}, {Sung}, {Tamassia}, {Tashkun}, {Vander Auwera}, {Vasilenko},
  {Vigasin}, {Villanueva}, {Vispoel}, {Wagner}, {Yachmenev}, \&
  {Yurchenko}}]{hitran22}
{Gordon}, I.~E., {Rothman}, L.~S., {Hargreaves}, R.~J., {et~al.} 2022, \jqsrt,
  277, 107949, \dodoi{10.1016/j.jqsrt.2021.107949}

\bibitem[{{Harker} {et~al.}(2023){Harker}, {Wooden}, {Kelley}, \&
  {Woodward}}]{harker23_spitzer_comets}
{Harker}, D.~E., {Wooden}, D.~H., {Kelley}, M. S.~P., \& {Woodward}, C.~E.
  2023, \psj, 4, 242, \dodoi{10.3847/PSJ/ad0382}

\bibitem[{{Heays} {et~al.}(2017){Heays}, {Bosman}, \& {van
  Dishoeck}}]{LeidenLab_pd_calculation}
{Heays}, A.~N., {Bosman}, A.~D., \& {van Dishoeck}, E.~F. 2017, \aap, 602,
  A105, \dodoi{10.1051/0004-6361/201628742}

\bibitem[{{Henning}(2010)}]{henning10_astromineralogy}
{Henning}, T. 2010, {Astromineralogy}, Vol. 815,
  \dodoi{10.1007/978-3-642-13259-9}

\bibitem[{{Henning} \& {Mutschke}(1997)}]{silica_HM97}
{Henning}, T., \& {Mutschke}, H. 1997, \aap, 327, 743

\bibitem[{{Ilin} {et~al.}(2019){Ilin}, {Schmidt}, {Davenport}, \&
  {Strassmeier}}]{ilin19_k2_flares}
{Ilin}, E., {Schmidt}, S.~J., {Davenport}, J. R.~A., \& {Strassmeier}, K.~G.
  2019, \aap, 622, A133, \dodoi{10.1051/0004-6361/201834400}

\bibitem[{{Ishihara} {et~al.}(2010){Ishihara}, {Onaka}, {Kataza}, {Salama},
  {Alfageme}, {Cassatella}, {Cox}, {Garc{\'\i}a-Lario}, {Stephenson}, {Cohen},
  {Fujishiro}, {Fujiwara}, {Hasegawa}, {Ita}, {Kim}, {Matsuhara}, {Murakami},
  {M{\"u}ller}, {Nakagawa}, {Ohyama}, {Oyabu}, {Pyo}, {Sakon}, {Shibai},
  {Takita}, {Tanab{\'e}}, {Uemizu}, {Ueno}, {Usui}, {Wada}, {Watarai},
  {Yamamura}, \& {Yamauchi}}]{akari_irc_catalog}
{Ishihara}, D., {Onaka}, T., {Kataza}, H., {et~al.} 2010, \aap, 514, A1,
  \dodoi{10.1051/0004-6361/200913811}

\bibitem[{{Jackson} \& {Wyatt}(2012)}]{jackson12}
{Jackson}, A.~P., \& {Wyatt}, M.~C. 2012, \mnras, 425, 657,
  \dodoi{10.1111/j.1365-2966.2012.21546.x}

\bibitem[{{Johnson} \& {Melosh}(2012)}]{johnson12a}
{Johnson}, B.~C., \& {Melosh}, H.~J. 2012, Icarus, 217, 416,
  \dodoi{10.1016/j.icarus.2011.11.020}

\bibitem[{{Johnson} {et~al.}(2012){Johnson}, {Lisse}, {Chen}, {Melosh},
  {Wyatt}, {Thebault}, {Henning}, {Gaidos}, {Elkins-Tanton}, {Bridges}, \&
  {Morlok}}]{johnson12b}
{Johnson}, B.~C., {Lisse}, C.~M., {Chen}, C.~H., {et~al.} 2012, \apj, 761, 45,
  \dodoi{10.1088/0004-637X/761/1/45}

\bibitem[{{Juh{\'a}sz} {et~al.}(2010){Juh{\'a}sz}, {Bouwman}, {Henning},
  {Acke}, {van den Ancker}, {Meeus}, {Dominik}, {Min}, {Tielens}, \&
  {Waters}}]{juhasz10_herbig}
{Juh{\'a}sz}, A., {Bouwman}, J., {Henning}, T., {et~al.} 2010, \apj, 721, 431,
  \dodoi{10.1088/0004-637X/721/1/431}

\bibitem[{{Kennedy} \& {Wyatt}(2013)}]{kennedy_wyatt13}
{Kennedy}, G.~M., \& {Wyatt}, M.~C. 2013, \mnras, 433, 2334,
  \dodoi{10.1093/mnras/stt900}

\bibitem[{{Kenyon} {et~al.}(2016){Kenyon}, {Najita}, \&
  {Bromley}}]{kenyon_najita_bromley2016}
{Kenyon}, S.~J., {Najita}, J.~R., \& {Bromley}, B.~C. 2016, \apj, 831, 8,
  \dodoi{10.3847/0004-637X/831/1/8}

\bibitem[{{Kitamura} {et~al.}(2007){Kitamura}, {Pilon}, \&
  {Jonasz}}]{kitamura07}
{Kitamura}, R., {Pilon}, L., \& {Jonasz}, M. 2007, \ao, 46, 8118,
  \dodoi{10.1364/AO.46.008118}

\bibitem[{{Koike} {et~al.}(2013){Koike}, {Noguchi}, {Chihara}, {Suto},
  {Ohtaka}, {Imai}, {Matsumoto}, \& {Tsuchiyama}}]{koike13}
{Koike}, C., {Noguchi}, R., {Chihara}, H., {et~al.} 2013, \apj, 778, 60,
  \dodoi{10.1088/0004-637X/778/1/60}

\bibitem[{{Lange} \& {Ahrens}(1982)}]{lange_ahrens82}
{Lange}, M.~A., \& {Ahrens}, T.~J. 1982, \icarus, 51, 96,
  \dodoi{10.1016/0019-1035(82)90031-8}

\bibitem[{{Law} {et~al.}(2025){Law}, {Argyriou}, {Gordon}, {Sloan}, {Gasman},
  {Glasse}, {Larson}, {Fletcher}, {Labiano}, \&
  {Noriega-Crespo}}]{law25_mrscalibration}
{Law}, D.~R., {Argyriou}, I., {Gordon}, K.~D., {et~al.} 2025, \aj, 169, 67,
  \dodoi{10.3847/1538-3881/ad9685}

\bibitem[{{Lebouteiller} {et~al.}(2011){Lebouteiller}, {Barry}, {Spoon},
  {Bernard-Salas}, {Sloan}, {Houck}, \& {Weedman}}]{cassis_ref}
{Lebouteiller}, V., {Barry}, D.~J., {Spoon}, H.~W.~W., {et~al.} 2011, \apjs,
  196, 8, \dodoi{10.1088/0067-0049/196/1/8}

\bibitem[{{Lebreton} {et~al.}(2013){Lebreton}, {van Lieshout}, {Augereau},
  {Absil}, {Mennesson}, {Kama}, {Dominik}, {Bonsor}, {Vandeportal}, {Beust},
  {Defr{\`e}re}, {Ertel}, {Faramaz}, {Hinz}, {Kral}, {Lagrange}, {Liu}, \&
  {Th{\'e}bault}}]{lebreton13}
{Lebreton}, J., {van Lieshout}, R., {Augereau}, J.~C., {et~al.} 2013, \aap,
  555, A146, \dodoi{10.1051/0004-6361/201321415}

\bibitem[{{Lisse} {et~al.}(2009){Lisse}, {Chen}, {Wyatt}, {Morlok}, {Song},
  {Bryden}, \& {Sheehan}}]{lisse09}
{Lisse}, C.~M., {Chen}, C.~H., {Wyatt}, M.~C., {et~al.} 2009, \apj, 701, 2019,
  \dodoi{10.1088/0004-637X/701/2/2019}

\bibitem[{{Lisse} {et~al.}(2020){Lisse}, {Meng}, {Sitko}, {Morlok}, {Johnson},
  {Jackson}, {Vervack}, {Chen}, {Wolk}, {Lucas}, {Marengo}, \&
  {Britt}}]{badlisse2020}
{Lisse}, C.~M., {Meng}, H.~Y.~A., {Sitko}, M.~L., {et~al.} 2020, \apj, 894,
  116, \dodoi{10.3847/1538-4357/ab7b80}

\bibitem[{{Liu} {et~al.}(2019){Liu}, {Pascucci}, \& {Henning}}]{liu19_click}
{Liu}, Y., {Pascucci}, I., \& {Henning}, T. 2019, \aap, 623, A106,
  \dodoi{10.1051/0004-6361/201834418}

\bibitem[{{Lodieu} {et~al.}(2019){Lodieu}, {P{\'e}rez-Garrido}, {Smart}, \&
  {Silvotti}}]{lodieu19_Pleiades_Praesepe_alphaPer}
{Lodieu}, N., {P{\'e}rez-Garrido}, A., {Smart}, R.~L., \& {Silvotti}, R. 2019,
  \aap, 628, A66, \dodoi{10.1051/0004-6361/201935533}

\bibitem[{{Marino} {et~al.}(2017){Marino}, {Wyatt}, {Pani{\'c}}, {Matr{\`a}},
  {Kennedy}, {Bonsor}, {Kral}, {Dent}, {Duchene}, {Wilner}, {Lisse},
  {Lestrade}, \& {Matthews}}]{marino17_etaCrv}
{Marino}, S., {Wyatt}, M.~C., {Pani{\'c}}, O., {et~al.} 2017, \mnras, 465,
  2595, \dodoi{10.1093/mnras/stw2867}

\bibitem[{{Matr{\`a}} {et~al.}(2019){Matr{\`a}}, {Wyatt}, {Wilner}, {Dent},
  {Marino}, {Kennedy}, \& {Milli}}]{matra19b}
{Matr{\`a}}, L., {Wyatt}, M.~C., {Wilner}, D.~J., {et~al.} 2019, \aj, 157, 135,
  \dodoi{10.3847/1538-3881/ab06c0}

\bibitem[{{Matr{\`a}} {et~al.}(2025){Matr{\`a}}, {Marino}, {Wilner}, {Kennedy},
  {Booth}, {Krivov}, {Williams}, {Hughes}, {del Burgo}, {Carpenter}, {Davies},
  {Ertel}, {Kral}, {Lestrade}, {Marshall}, {Milli}, {{\"O}berg}, {Pawellek},
  {Sepulveda}, {Wyatt}, {Matthews}, \& {MacGregor}}]{matra25_reasons}
{Matr{\`a}}, L., {Marino}, S., {Wilner}, D.~J., {et~al.} 2025, \aap, 693, A151,
  \dodoi{10.1051/0004-6361/202451397}

\bibitem[{{Melis} {et~al.}(2021){Melis}, {Olofsson}, {Song}, {Sarkis},
  {Weinberger}, {Kennedy}, \& {Krumpe}}]{melis21}
{Melis}, C., {Olofsson}, J., {Song}, I., {et~al.} 2021, arXiv e-prints,
  arXiv:2104.06448.
\newblock \doarXiv{2104.06448}

\bibitem[{{Meng} {et~al.}(2012){Meng}, {Rieke}, {Su}, {Ivanov}, {Vanzi}, \&
  {Rujopakarn}}]{meng12}
{Meng}, H.~Y.~A., {Rieke}, G.~H., {Su}, K.~Y.~L., {et~al.} 2012, \apjl, 751,
  L17, \dodoi{10.1088/2041-8205/751/1/L17}

\bibitem[{{Meng} {et~al.}(2014){Meng}, {Su}, {Rieke}, {Stevenson}, {Plavchan},
  {Rujopakarn}, {Lisse}, {Poshyachinda}, \& {Reichart}}]{meng14}
{Meng}, H.~Y.~A., {Su}, K.~Y.~L., {Rieke}, G.~H., {et~al.} 2014, Science, 345,
  1032, \dodoi{10.1126/science.1255153}

\bibitem[{{Mermilliod} {et~al.}(2009){Mermilliod}, {Mayor}, \&
  {Udry}}]{mermilliod09}
{Mermilliod}, J.~C., {Mayor}, M., \& {Udry}, S. 2009, \aap, 498, 949,
  \dodoi{10.1051/0004-6361/200810244}

\bibitem[{{Milam} {et~al.}(2005){Milam}, {Savage}, {Brewster}, {Ziurys}, \&
  {Wyckoff}}]{milam05}
{Milam}, S.~N., {Savage}, C., {Brewster}, M.~A., {Ziurys}, L.~M., \& {Wyckoff},
  S. 2005, \apj, 634, 1126, \dodoi{10.1086/497123}

\bibitem[{{Min} {et~al.}(2005){Min}, {Hovenier}, \& {de Koter}}]{min2005}
{Min}, M., {Hovenier}, J.~W., \& {de Koter}, A. 2005, \aap, 432, 909,
  \dodoi{10.1051/0004-6361:20041920}

\bibitem[{{Mo{\'o}r} {et~al.}(2017){Mo{\'o}r}, {Cur{\'e}}, {K{\'o}sp{\'a}l},
  {{\'A}brah{\'a}m}, {Csengeri}, {Eiroa}, {Gunawan}, {Henning}, {Hughes},
  {Juh{\'a}sz}, {Pawellek}, \& {Wyatt}}]{moor17}
{Mo{\'o}r}, A., {Cur{\'e}}, M., {K{\'o}sp{\'a}l}, {\'A}., {et~al.} 2017, \apj,
  849, 123, \dodoi{10.3847/1538-4357/aa8e4e}

\bibitem[{{Mo{\'o}r} {et~al.}(2021){Mo{\'o}r}, {{\'A}brah{\'a}m}, {Szab{\'o}},
  {Vida}, {Cataldi}, {Derekas}, {Henning}, {Kinemuchi}, {K{\'o}sp{\'a}l},
  {Kov{\'a}cs}, {P{\'a}l}, {Sarkis}, {Seli}, {Szab{\'o}}, \&
  {Tak{\'a}ts}}]{moor21}
{Mo{\'o}r}, A., {{\'A}brah{\'a}m}, P., {Szab{\'o}}, G., {et~al.} 2021, \apj,
  910, 27, \dodoi{10.3847/1538-4357/abdc26}

\bibitem[{{Mo{\'o}r} {et~al.}(2024){Mo{\'o}r}, {{\'A}brah{\'a}m}, {Su},
  {Henning}, {Marino}, {Chen}, {K{\'o}sp{\'a}l}, {Pawellek}, {Varga}, \&
  {Vida}}]{moor24_edds_visir}
{Mo{\'o}r}, A., {{\'A}brah{\'a}m}, P., {Su}, K. Y.~L., {et~al.} 2024, \mnras,
  528, 4528, \dodoi{10.1093/mnras/stae155}

\bibitem[{{Morlok} {et~al.}(2016){Morlok}, {Stojic}, {Weber}, {Hiesinger},
  {Zanetti}, \& {Helbert}}]{morlok16_reflectance_impactglasses}
{Morlok}, A., {Stojic}, A., {Weber}, I., {et~al.} 2016, \icarus, 278, 162,
  \dodoi{10.1016/j.icarus.2016.06.013}

\bibitem[{{Moshir} {et~al.}(1992){Moshir}, {Kopman}, \& {Conrow}}]{irasfsc}
{Moshir}, M., {Kopman}, G., \& {Conrow}, T.~A.~O. 1992, {IRAS Faint Source
  Survey, Explanatory supplement version 2}

\bibitem[{{Munoz-Romero} {et~al.}(2023){Munoz-Romero}, {Banzatti}, \&
  {{\"O}berg}}]{iris_23}
{Munoz-Romero}, C.~E., {Banzatti}, A., \& {{\"O}berg}, K.~I. 2023, {iris
  (InfraRed Isothermal Slabs)},  Zenodo, \dodoi{10.5281/zenodo.10369000}

\bibitem[{{Naoz}(2016)}]{naoz16review_kozai}
{Naoz}, S. 2016, \araa, 54, 441, \dodoi{10.1146/annurev-astro-081915-023315}

\bibitem[{{Pascucci} {et~al.}(2007){Pascucci}, {Hollenbach}, {Najita},
  {Muzerolle}, {Gorti}, {Herczeg}, {Hillenbrand}, {Kim}, {Carpenter}, {Meyer},
  {Mamajek}, \& {Bouwman}}]{pascucci07}
{Pascucci}, I., {Hollenbach}, D., {Najita}, J., {et~al.} 2007, \apj, 663, 383,
  \dodoi{10.1086/518535}

\bibitem[{{Pearce} {et~al.}(2020){Pearce}, {Krivov}, \&
  {Booth}}]{pearce20_gastrapping}
{Pearce}, T.~D., {Krivov}, A.~V., \& {Booth}, M. 2020, \mnras, 498, 2798,
  \dodoi{10.1093/mnras/staa2514}

\bibitem[{{Pontoppidan} {et~al.}(2024){Pontoppidan}, {Salyk}, {Banzatti},
  {Zhang}, {Pascucci}, {{\"O}berg}, {Long}, {Mu{\~n}oz-Romero}, {Carr},
  {Najita}, {Blake}, {Arulanantham}, {Andrews}, {Ballering}, {Bergin},
  {Calahan}, {Cobb}, {Colmenares}, {Dickson-Vandervelde}, {Dignan}, {Green},
  {Heretz}, {Herczeg}, {Kalyaan}, {Krijt}, {Pauly}, {Pinilla}, {Trapman}, \&
  {Xie}}]{pontoppidan24_jdisc}
{Pontoppidan}, K.~M., {Salyk}, C., {Banzatti}, A., {et~al.} 2024, \apj, 963,
  158, \dodoi{10.3847/1538-4357/ad20f0}

\bibitem[{{Quintana} {et~al.}(2016){Quintana}, {Barclay}, {Borucki}, {Rowe}, \&
  {Chambers}}]{quintana16}
{Quintana}, E.~V., {Barclay}, T., {Borucki}, W.~J., {Rowe}, J.~F., \&
  {Chambers}, J.~E. 2016, \apj, 821, 126, \dodoi{10.3847/0004-637X/821/2/126}

\bibitem[{{Rhee} {et~al.}(2008){Rhee}, {Song}, \& {Zuckerman}}]{rhee08}
{Rhee}, J.~H., {Song}, I., \& {Zuckerman}, B. 2008, \apj, 675, 777,
  \dodoi{10.1086/524935}

\bibitem[{{Rieke} {et~al.}(2016){Rieke}, {G{\'a}sp{\'a}r}, \&
  {Ballering}}]{rieke16}
{Rieke}, G.~H., {G{\'a}sp{\'a}r}, A., \& {Ballering}, N.~P. 2016, \apj, 816,
  50, \dodoi{10.3847/0004-637X/816/2/50}

\bibitem[{{Rieke} {et~al.}(2021){Rieke}, {Su}, {Melis}, \&
  {G{\'a}sp{\'a}r}}]{rieke21_v488per}
{Rieke}, G.~H., {Su}, K.~Y.~L., {Melis}, C., \& {G{\'a}sp{\'a}r}, A. 2021,
  \apj, 918, 71, \dodoi{10.3847/1538-4357/ac0dc4}

\bibitem[{{Rinaldi} {et~al.}(2017){Rinaldi}, {Della Corte}, {Fulle},
  {Capaccioni}, {Rotundi}, {Ivanovski}, {Bockel{\'e}e-Morvan}, {Filacchione},
  {D'Aversa}, {Capria}, {Tozzi}, {Erard}, {Leyrat}, {Palomba}, {Longobardo},
  {Ciarniello}, {Taylor}, {Mottola}, \& {Salatti}}]{rinaldi17}
{Rinaldi}, G., {Della Corte}, V., {Fulle}, M., {et~al.} 2017, \mnras, 469,
  S598, \dodoi{10.1093/mnras/stx1873}

\bibitem[{{Roberge} {et~al.}(2000){Roberge}, {Feldman}, {Lagrange},
  {Vidal-Madjar}, {Ferlet}, {Jolly}, {Lemaire}, \&
  {Rostas}}]{roberge00_betapic_gas}
{Roberge}, A., {Feldman}, P.~D., {Lagrange}, A.~M., {et~al.} 2000, \apj, 538,
  904, \dodoi{10.1086/309157}

\bibitem[{{Rodriguez} {et~al.}(2012){Rodriguez}, {Marois}, {Zuckerman},
  {Macintosh}, \& {Melis}}]{rodriguez12_hd23514}
{Rodriguez}, D.~R., {Marois}, C., {Zuckerman}, B., {Macintosh}, B., \& {Melis},
  C. 2012, \apj, 748, 30, \dodoi{10.1088/0004-637X/748/1/30}

\bibitem[{{Rom{\'a}n-Z{\'u}{\~n}iga} {et~al.}(2023){Rom{\'a}n-Z{\'u}{\~n}iga},
  {Kounkel}, {Hern{\'a}ndez}, {Pe{\~n}a Ram{\'\i}rez}, {L{\'o}pez-Valdivia},
  {Covey}, {Stutz}, {Roman-Lopes}, {Campbell}, {Khilfeh}, {Tapia},
  {Stringfellow}, {Downes}, {Stassun}, {Minniti}, {Bayo}, {Kim}, {Su{\'a}rez},
  {Ybarra}, {Fern{\'a}ndez-Trincado}, {Longa-Pe{\~n}a},
  {Ram{\'\i}rez-Preciado}, {Serna}, {Lane}, {Garc{\'\i}a-Hern{\'a}ndez},
  {Beaton}, {Bizyaev}, \& {Pan}}]{roman-zuniga23_apogee2}
{Rom{\'a}n-Z{\'u}{\~n}iga}, C.~G., {Kounkel}, M., {Hern{\'a}ndez}, J., {et~al.}
  2023, \aj, 165, 51, \dodoi{10.3847/1538-3881/aca3a4}

\bibitem[{{Samland} {et~al.}(2025){Samland}, {Henning}, {Garatti}, {Giannini},
  {Bouwman}, {Tabone}, {Arabhavi}, {Olofsson}, {G{\"u}del}, {Pawellek}, {Kamp},
  {Waters}, {Semenov}, {van Dishoeck}, {Absil}, {Barrado}, {Boccaletti},
  {Christiaens}, {Gasman}, {Grant}, {Jang}, {Kaeufer}, {Kanwar}, {Perotti},
  {Schwarz}, \& {Temmink}}]{samland25_hd172555}
{Samland}, M., {Henning}, T., {Garatti}, A. C.~o., {et~al.} 2025, arXiv
  e-prints, arXiv:2506.09976, \dodoi{10.48550/arXiv.2506.09976}

\bibitem[{{Sargent} {et~al.}(2006){Sargent}, {Forrest}, {D'Alessio}, {Li},
  {Najita}, {Watson}, {Calvet}, {Furlan}, {Green}, {Kim}, {Sloan}, {Chen},
  {Hartmann}, \& {Houck}}]{sargent06}
{Sargent}, B., {Forrest}, W.~J., {D'Alessio}, P., {et~al.} 2006, \apj, 645,
  395, \dodoi{10.1086/504283}

\bibitem[{{Sargent} {et~al.}(2009){Sargent}, {Forrest}, {Tayrien}, {McClure},
  {Li}, {Basu}, {Manoj}, {Watson}, {Bohac}, {Furlan}, {Kim}, {Green}, \&
  {Sloan}}]{sargen09_silia}
{Sargent}, B.~A., {Forrest}, W.~J., {Tayrien}, C., {et~al.} 2009, \apj, 690,
  1193, \dodoi{10.1088/0004-637X/690/2/1193}

\bibitem[{{Schaefer} \& {Fegley}(2010)}]{schaefer2010}
{Schaefer}, L., \& {Fegley}, B. 2010, \icarus, 208, 438,
  \dodoi{10.1016/j.icarus.2010.01.026}

\bibitem[{{Schneiderman} {et~al.}(2021){Schneiderman}, {Matr{\`a}}, {Jackson},
  {Kennedy}, {Kral}, {Marino}, {{\"O}berg}, {Su}, {Wilner}, \&
  {Wyatt}}]{schneiderman21_hd172555}
{Schneiderman}, T., {Matr{\`a}}, L., {Jackson}, A.~P., {et~al.} 2021, \nat,
  598, 425, \dodoi{10.1038/s41586-021-03872-x}

\bibitem[{{Sierchio} {et~al.}(2014){Sierchio}, {Rieke}, {Su}, \&
  {G{\'a}sp{\'a}r}}]{sierchio14}
{Sierchio}, J.~M., {Rieke}, G.~H., {Su}, K.~Y.~L., \& {G{\'a}sp{\'a}r}, A.
  2014, \apj, 785, 33, \dodoi{10.1088/0004-637X/785/1/33}

\bibitem[{{Speagle}(2020)}]{speagle20_dynesty}
{Speagle}, J.~S. 2020, \mnras, 493, 3132, \dodoi{10.1093/mnras/staa278}

\bibitem[{{Su} {et~al.}(2022){Su}, {Kennedy}, {Schlawin}, {Jackson}, \&
  {Rieke}}]{su22_hd166}
{Su}, K. Y.~L., {Kennedy}, G.~M., {Schlawin}, E., {Jackson}, A.~P., \& {Rieke},
  G.~H. 2022, \apj, 927, 135, \dodoi{10.3847/1538-4357/ac4bbb}

\bibitem[{{Su} {et~al.}(2020){Su}, {Rieke}, {Melis}, {Jackson}, {Smith},
  {Meng}, \& {G{\'a}sp{\'a}r}}]{su20}
{Su}, K. Y.~L., {Rieke}, G.~H., {Melis}, C., {et~al.} 2020, \apj, 898, 21,
  \dodoi{10.3847/1538-4357/ab9c9b}

\bibitem[{{Su} {et~al.}(2005){Su}, {Rieke}, {Misselt}, {Stansberry},
  {Moro-Martin}, {Stapelfeldt}, {Werner}, {Trilling}, {Bendo}, {Gordon},
  {Hines}, {Wyatt}, {Holland}, {Marengo}, {Megeath}, \& {Fazio}}]{su05}
{Su}, K.~Y.~L., {Rieke}, G.~H., {Misselt}, K.~A., {et~al.} 2005, \apj, 628,
  487, \dodoi{10.1086/430819}

\bibitem[{{Su} {et~al.}(2019){Su}, {Jackson}, {G{\'a}sp{\'a}r}, {Rieke},
  {Dong}, {Olofsson}, {Kennedy}, {Leinhardt}, {Malhotra}, {Hammer}, {Meng},
  {Rujopakarn}, {Rodriguez}, {Pepper}, {Reichart}, {James}, \&
  {Stassun}}]{su19}
{Su}, K.~Y.~L., {Jackson}, A.~P., {G{\'a}sp{\'a}r}, A., {et~al.} 2019, AJ, 157,
  202, \dodoi{10.3847/1538-3881/ab1260}

\bibitem[{{Su} {et~al.}(2023){Su}, {Kennedy}, {Rieke}, {Hughes}, {Lin},
  {Kittling}, {Jackson}, {Anche}, \& {Liu}}]{su23_rzpsc}
{Su}, K. Y.~L., {Kennedy}, G.~M., {Rieke}, G.~H., {et~al.} 2023, \apj, 959, 43,
  \dodoi{10.3847/1538-4357/ad04d9}

\bibitem[{{Sullivan} {et~al.}(2022){Sullivan}, {Wilner}, {Matr{\`a}}, {Wyatt},
  {Andrews}, {MacGregor}, \& {Matthews}}]{sullivan22_alma_pleiades}
{Sullivan}, D., {Wilner}, D.~J., {Matr{\`a}}, L., {et~al.} 2022, \aj, 164, 100,
  \dodoi{10.3847/1538-3881/ac80c5}

\bibitem[{{Takarada} {et~al.}(2020){Takarada}, {Sato}, {Omiya}, {Hori}, \&
  {Fujii}}]{takarada20_RV_Pleiades}
{Takarada}, T., {Sato}, B., {Omiya}, M., {Hori}, Y., \& {Fujii}, M.~S. 2020,
  \pasj, 72, 104, \dodoi{10.1093/pasj/psaa105}

\bibitem[{{Takeuchi} \& {Artymowicz}(2001)}]{takeuchi01}
{Takeuchi}, T., \& {Artymowicz}, P. 2001, \apj, 557, 990,
  \dodoi{10.1086/322252}

\bibitem[{{Thebault} \& {Kral}(2019)}]{thebault19}
{Thebault}, P., \& {Kral}, Q. 2019, \aap, 626, A24,
  \dodoi{10.1051/0004-6361/201935341}

\bibitem[{{Thompson} {et~al.}(2021){Thompson}, {Telus}, {Schaefer}, {Fortney},
  {Joshi}, \& {Lederman}}]{thompson21_outgassingexperiments}
{Thompson}, M.~A., {Telus}, M., {Schaefer}, L., {et~al.} 2021, Nature
  Astronomy, 5, 575, \dodoi{10.1038/s41550-021-01338-8}

\bibitem[{{Tsantaki} {et~al.}(2022){Tsantaki}, {Pancino}, {Marrese},
  {Marinoni}, {Rainer}, {Sanna}, {Turchi}, {Randich}, {Gallart}, {Battaglia},
  \& {Masseron}}]{Tsantaka22}
{Tsantaki}, M., {Pancino}, E., {Marrese}, P., {et~al.} 2022, \aap, 659, A95,
  \dodoi{10.1051/0004-6361/202141702}

\bibitem[{{van Dishoeck} \& {Black}(1982)}]{vandishoeck82}
{van Dishoeck}, E.~F., \& {Black}, J.~H. 1982, \apj, 258, 533,
  \dodoi{10.1086/160104}

\bibitem[{{Vican} {et~al.}(2016){Vican}, {Schneider}, {Bryden}, {Melis},
  {Zuckerman}, {Rhee}, \& {Song}}]{vican16}
{Vican}, L., {Schneider}, A., {Bryden}, G., {et~al.} 2016, \apj, 833, 263,
  \dodoi{10.3847/1538-4357/833/2/263}

\bibitem[{{Virtanen} {et~al.}(2020){Virtanen}, {Gommers}, {Oliphant},
  {Haberland}, {Reddy}, {Cournapeau}, {Burovski}, {Peterson}, {Weckesser},
  {Bright}, {van der Walt}, {Brett}, {Wilson}, {Millman}, {Mayorov}, {Nelson},
  {Jones}, {Kern}, {Larson}, {Carey}, {Polat}, {Feng}, {Moore}, {VanderPlas},
  {Laxalde}, {Perktold}, {Cimrman}, {Henriksen}, {Quintero}, {Harris},
  {Archibald}, {Ribeiro}, {Pedregosa}, {van Mulbregt}, \& {SciPy 1. 0
  Contributors}}]{scipy}
{Virtanen}, P., {Gommers}, R., {Oliphant}, T.~E., {et~al.} 2020, Nature
  Methods, 17, 261, \dodoi{10.1038/s41592-019-0686-2}

\bibitem[{{Visser} {et~al.}(2009){Visser}, {van Dishoeck}, \&
  {Black}}]{visser09}
{Visser}, R., {van Dishoeck}, E.~F., \& {Black}, J.~H. 2009, \aap, 503, 323,
  \dodoi{10.1051/0004-6361/200912129}

\bibitem[{{Watson} {et~al.}(2009){Watson}, {Leisenring}, {Furlan}, {Bohac},
  {Sargent}, {Forrest}, {Calvet}, {Hartmann}, {Nordhaus}, {Green}, {Kim},
  {Sloan}, {Chen}, {Keller}, {d'Alessio}, {Najita}, {Uchida}, \&
  {Houck}}]{watson09}
{Watson}, D.~M., {Leisenring}, J.~M., {Furlan}, E., {et~al.} 2009, \apjs, 180,
  84, \dodoi{10.1088/0067-0049/180/1/84}

\bibitem[{{Watt} {et~al.}(2024){Watt}, {Leinhardt}, \&
  {Carter}}]{watt24_postimpact_edd_evolution}
{Watt}, L., {Leinhardt}, Z.~M., \& {Carter}, P.~J. 2024, \mnras, 527, 7749,
  \dodoi{10.1093/mnras/stad3606}

\bibitem[{{Wright} {et~al.}(2010){Wright}, {Eisenhardt}, {Mainzer}, {Ressler},
  {Cutri}, {Jarrett}, {Kirkpatrick}, {Padgett}, {McMillan}, {Skrutskie},
  {Stanford}, {Cohen}, {Walker}, {Mather}, {Leisawitz}, {Gautier}, {McLean},
  {Benford}, {Lonsdale}, {Blain}, {Mendez}, {Irace}, {Duval}, {Liu}, {Royer},
  {Heinrichsen}, {Howard}, {Shannon}, {Kendall}, {Walsh}, {Larsen}, {Cardon},
  {Schick}, {Schwalm}, {Abid}, {Fabinsky}, {Naes}, \&
  {Tsai}}]{wright10_wisemission}
{Wright}, E.~L., {Eisenhardt}, P. R.~M., {Mainzer}, A.~K., {et~al.} 2010, \aj,
  140, 1868, \dodoi{10.1088/0004-6256/140/6/1868}

\bibitem[{{Wyatt} {et~al.}(2007){Wyatt}, {Smith}, {Greaves}, {Beichman},
  {Bryden}, \& {Lisse}}]{wyatt07a}
{Wyatt}, M.~C., {Smith}, R., {Greaves}, J.~S., {et~al.} 2007, \apj, 658, 569,
  \dodoi{10.1086/510999}

\bibitem[{{Young} \& {Wyatt}(2024)}]{young_wyatt24}
{Young}, S.~D., \& {Wyatt}, M.~C. 2024, \mnras, 527, 5244,
  \dodoi{10.1093/mnras/stad2963}

\bibitem[{{Zubko} {et~al.}(1996){Zubko}, {Mennella}, {Colangeli}, \&
  {Bussoletti}}]{zubko96}
{Zubko}, V.~G., {Mennella}, V., {Colangeli}, L., \& {Bussoletti}, E. 1996,
  \mnras, 282, 1321

\bibitem[{{Zuckerman}(2015)}]{zuckerman15}
{Zuckerman}, B. 2015, \apj, 798, 86, \dodoi{10.1088/0004-637X/798/2/86}

\end{thebibliography}
\bibliographystyle{aasjournal}

%TC:endignore

\end{document}